\documentclass[11pt]{cernrep}
\usepackage{epsfig}
\usepackage{rotating}

\usepackage{subfigure}
\usepackage{psfrag}
\usepackage{rotating}
\usepackage{axodraw}

\renewcommand\thesection{\arabic{section}}

\renewcommand\thesubsection{\thesection.\arabic{subsection}}

\newcommand{\RE}{{\rm Re}}
\newcommand{\IM}{{\rm Im}}
\newcommand{\vcb}{|V_{cb}|}
\newcommand{\vtd}{|V_{td}|}
\newcommand{\vub}{|V_{ub}/V_{cb}|}
\newcommand{\vts}{|V_{ts}|}
\newcommand{\vus}{|V_{us}|}

\newcommand{\caladir}{{\cal A}_{\rm CP}^{\rm dir}}

\newcommand{\calamix}{{\cal A}_{\rm CP}^{\rm mix}}

\newcommand{\calaCP}{{\cal A}_{\rm CP}}

\def\R1{\varepsilon_1}
\def\E8{\varepsilon_8}

\def\eps{\varepsilon}
\def\epe{\varepsilon'/\varepsilon}
\def\as{\alpha_s}
\newcommand{\eqn}{\ref}

\newcommand{\mt}{m_{\rm t}}
\newcommand{\mtb}{\overline{m}_{\rm t}}

\newcommand{\mc}{m_{\rm c}}

\newcommand{\mb}{m_{\rm b}}
\newcommand{\mw}{M_{\rm W}}
\newcommand{\mz}{M_{\rm Z}}
\newcommand{\gev}{\, {\rm GeV}}
\newcommand{\mev}{\, {\rm MeV}}

\newcommand{\bea}{\begin{eqnarray}}
\newcommand{\eea}{\end{eqnarray}}
\newcommand{\bd}{\begin{displaymath}}
\newcommand{\ed}{\end{displaymath}}

\newcommand{\beq}{\begin{equation}}
\newcommand{\eeq}{\end{equation}}
\newcommand{\be}{\begin{equation}}
\newcommand{\ee}{\end{equation}}
\newcommand{\bi}{\begin{itemize}}
\newcommand{\ei}{\end{itemize}}
\newcommand{\ord}{{\cal O}}

\newcommand{\f}{\frac}

\def\kpn{K^+\rightarrow\pi^+\nu\bar\nu}
\def\klpn{K_{\rm L}\rightarrow\pi^0\nu\bar\nu}

\newcommand{\kmm}{K_{\rm L} \to \mu^+ \mu^-}

\newcommand{\imlt}{\IM\lambda_t}
\newcommand{\relt}{\RE\lambda_t}

\renewcommand{\baselinestretch}{1.1}

\begin{document}
\pagestyle{plain}

\thispagestyle{empty}
\phantom{xxx}
\vskip1truecm
\begin{flushright}
TUM-HEP-590/05 \\
May 2005
\end{flushright}
\vskip1.8truecm
\centerline{\LARGE\bf Flavour Physics and
 CP Violation }
   \vskip1truecm
\centerline{\Large\bf Andrzej J. Buras}
\bigskip
\centerline{\sl Technische Universit{\"a}t M{\"u}nchen}
\centerline{\sl Physik Department} 
\centerline{\sl D-85748 Garching, Germany}
\vskip1truecm
\centerline{\bf Abstract}
After listing basic properties of the Standard Model (SM) that play the crucial
role in the field of flavour and CP violation, we discuss the following
topics: 1) CKM matrix and the unitarity triangle. 2) Theoretical framework in
a non-technical manner, classifying various extentions of the SM.
3) Particle-Antiparticle mixing and various types of CP violation. 4)
Standard analysis of the unitarity triangle. 5) Strategies for the
determination of the angles $\alpha$, $\beta$ and $\gamma$ in 
non-leptonic $B$ decays. 
6) The rare decays $\kpn$ and $\klpn$. 7) Models with minimal flavour
violation (MFV). 8) Models with new complex phases, addressing in
particular possible signals of new physics in the $B\to\pi K$ 
data and their implications
for rare $K$ and $B$ decays. A personal shopping list for the rest of this
decade closes these lectures.
\vskip1.52truecm

\centerline{\it Lectures given at the European CERN School}
\centerline{\bf Saint Feliu de Guixols}
\centerline{\it  June, 2004}

\newpage

\pagenumbering{roman}

\tableofcontents

\newpage

\pagenumbering{arabic}


\title{FLAVOUR PHYSICS AND CP VIOLATION }
\author{Andrzej J. Buras}
\institute{Technische Universit{\"a}t M{\"u}nchen, Physik Department\\ 
D-85748 Garching, Germany}
\maketitle
\begin{abstract}
After listing basic properties of the Standard Model (SM) that play the crucial
role in the field of flavour and CP violation, we discuss the following
topics: 1) CKM matrix and the unitarity triangle. 2) Theoretical framework in
a non-technical manner, classifying various extentions of the SM.
3) Particle-Antiparticle mixing and various types of CP violation. 4)
Standard analysis of the unitarity triangle. 5) Strategies for the
determination of the angles $\alpha$, $\beta$ and $\gamma$ in 
non-leptonic $B$ decays. 
6) The rare decays $\kpn$ and $\klpn$. 7) Models with minimal flavour
violation (MFV). 8) Models with new complex phases, addressing in
particular possible signals of new physics in the $B\to\pi K$ 
data and their implications
for rare $K$ and $B$ decays. A personal shopping list for the rest of this
decade closes these lectures.
\end{abstract}

\section{Introduction}\label{INTRO}
\setcounter{equation}{0}
Flavour physics and CP violation in $K$ and $B$ meson decays
are among the central topics in particle physics. In particular
particle--antiparticle
 mixing and CP violation in $K\to \pi\pi$ decays have been  of fundamental
importance for the construction and testing of the Standard Model (SM).
They have also proven often to be 
undefeatable challenges for suggested extensions of this model.

In this context a very important role is played by the
Glashow-Iliopoulos-Maiani (GIM) mechanism \cite{GIM} for the suppression 
of flavour
changing neutral current (FCNC) processes that in turn proceed first at the 
one--loop level and are consequently sensitive to the short distance
structure of the SM and its possible extensions.
In particular
from the calculation of the
$K_{\rm L}-K_{\rm S}$ mass difference, Gaillard and Lee \cite{GALE} 
were able to estimate the
value of the charm quark mass before charm discovery. On the
other hand the first measurement of the size of the $B_d^0-\bar B_d^0$ mixing
\cite{ARGUS} gave the first 
indication of a large top quark mass.

The pattern of flavour and CP violations, both in the charged current
and FCNC processes, is governed by
the Cabibbo-Kobayashi-Maskawa (CKM) matrix \cite{CAB,KM}
that parametrizes the weak charged current interactions of quarks.
In particular this matrix contains the single CP-violating phase that within
the SM and its simplest extensions is supposed to describe all CP-violating 
processes.

One of the important questions still to be answered, is whether the 
CKM matrix is capable to describe with its four parameters all weak 
decays that include in addition to tree level 
decays mediated by $W^\pm$-bosons, a vast number of FCNC processes 
in which the so-called penguin and box diagrams play the central role.
This sector of the SM has not yet been sufficiently tested and one should be 
prepared for surprises, in spite of the fact that the last three years 
of experimental and theoretical investigations indicate that the CKM matrix
is likely to be the dominant source of flavour and CP violation. The present
and future studies of CP violation in $B$ decays, of theoretically clean rare
decays $\kpn$ and $\klpn$, and of a number of rare decays such as
$K_L\to\pi^0l^+l^-$, $B\to X_{s,d}\gamma$, $B\to X_{s,d}l^+l^-$, 
$B\to X_{s,d}\nu\bar\nu$ and $B_{s,d}\to\mu^+\mu^-$ should give decisive
answers already in this decade. 
These studies will be complemented by the investigations of flavour violation
in the lepton sector, its possible relation to flavour violation in the quark
sector, studies of $D^0-\bar D^0$ mixing and of the electric dipole moments.

It should be emphasized that all these efforts
are very challenging because the relevant rare 
and CP-violating decays have small branching ratios and consequently are 
 very 
difficult to measure. Moreover, as mesons are bound states of quarks and
antiquarks, the determination of the CKM parameters requires in many cases 
a quantitative control over QCD effects at long distances where the 
existing non-perturbative methods are not yet satisfactory. 

In spite of these difficulties, we strongly believe that 
the picture of flavour and CP violation in the quark sector will be much
clearer at the end of this decade and certainly in ten years from now with 
the studies of these phenomena in the leptonic sector probably requiring more
time.
This belief is based on an 
impressive 
progress in the experimental measurements in this field and on a 
similar progress made by theorists in  perturbative and 
to a lesser extend non-perturbative QCD calculations. The development of 
various strategies for the determination of the CKM parameters, 
that are essentially free from hadronic uncertainties, is also an important 
ingredient in this progress.
The last account of these joined efforts by experimentalists and theorists 
appeared  in \cite{CERNCKM} and certainly other accounts of this type will 
follow in the coming years.

These lecture notes provide a rather non-technical up to date description 
of flavour and CP violation in the SM and in its simplest extensions. 
In particular we will discuss the decays
that are best suited for the determination of the CKM matrix. 
There is unavoidably an overlap with our Les Houches  
\cite{AJBLH}, Lake Louise \cite{AJBLAKE}, Erice \cite{Erice}, 
Zacatecas \cite{MEX01}, Schladming \cite{Schladming} and Zakopane
\cite{Zakopane}  lectures 
and with the 
reviews \cite{BBL}, \cite{BF97} and \cite{BSU}.
On the other hand new developments until the appearance of this article
 have been taken
into account, as far as the space allowed for it, and all numerical
results have been  updated. 
In particular with respect to the Schladming lectures we present an 
extended discussion of $B\to\pi\pi$, $B\to\pi K$ and $K\to\pi\nu\bar\nu$ 
decays. Moreover, the discussion of the physics beyond the SM is
significantly extended.

The material of these lectures is organized as follows. In
Section~\ref{Facts}  we
recall those facts about the SM that are fundamental for the topics discussed
here. In Section~\ref{CKMM} we describe the CKM matrix and 
the Unitarity Triangle.
Section~\ref{THF}
 summarizes briefly the general aspects of the theoretical framework
for weak decays. Here it will turn out to be useful to make a classification
of various extensions of the SM. 
In Section~\ref{PPCP} the  particle--antiparticle mixing and
various types of CP violation are presented. In Section~\ref{UT-Det}
 we describe 
the so--called standard analysis of the unitarity triangle, present the 
shape of the 2004 unitarity triangle and comment briefly on the ratio $\epe$.

In Section~\ref{STRATEGIES} a number of strategies for the 
determination of the angles 
$\alpha$, $\beta$ and $\gamma$ in non-leptonic $B$ decays
 are discussed in some detail. 
In Section~\ref{KPNN} the
rare decays $\kpn$ and $\klpn$ are reviewed with particular emphasis put 
on their important virtues with respect to a clean determination of the
parameters of the CKM matrix. In Section~\ref{MFVM}
 we discuss briefly the models with
minimal flavour violation (MFV). 
Section~\ref{NPHASE} discusses 
 new developments in $B\to\pi\pi$ and $B\to\pi K$ decays, with the latter 
possibly indicating some new physics not only beyond the SM but more
generally beyond the MFV framework. This new physics implies spectacular
effects in rare $K$ and $B$ decays. A shopping list closes our lectures.

We hope that these lecture notes
will be helpful in following the new developments in this exciting field. 
In this respect
the books \cite{BULIND,Branco,Bigi,Kleinknecht}, the working group reports
\cite{CERNCKM,BABAR,SUPERB,LHCB,FERMILAB} 
and the reviews \cite{REV1,REV2,REV3,REV4}
are 
also strongly recommended.

\section{Basic Facts about the Standard Model}
\label{Facts}
\setcounter{equation}{0}
In the first part of these lectures we will dominantly work 
in the context of the SM with three generations of 
quarks and leptons and
the interactions described by the gauge group 
$ SU(3)_C\otimes SU(2)_L\otimes U(1)_Y$ spontaneously broken to
$SU(3)_C\otimes U(1)_Q$.
There are many text books on the dynamics of the SM. At this 
school  excellent lectures have been given by Toni Pich \cite{PICH}.

Here we will only collect those ingredients of the SM which
are fundamental for the subject of these lectures.

\begin{figure}[thb]
\centerline{
\epsfysize=9.5in
\epsffile{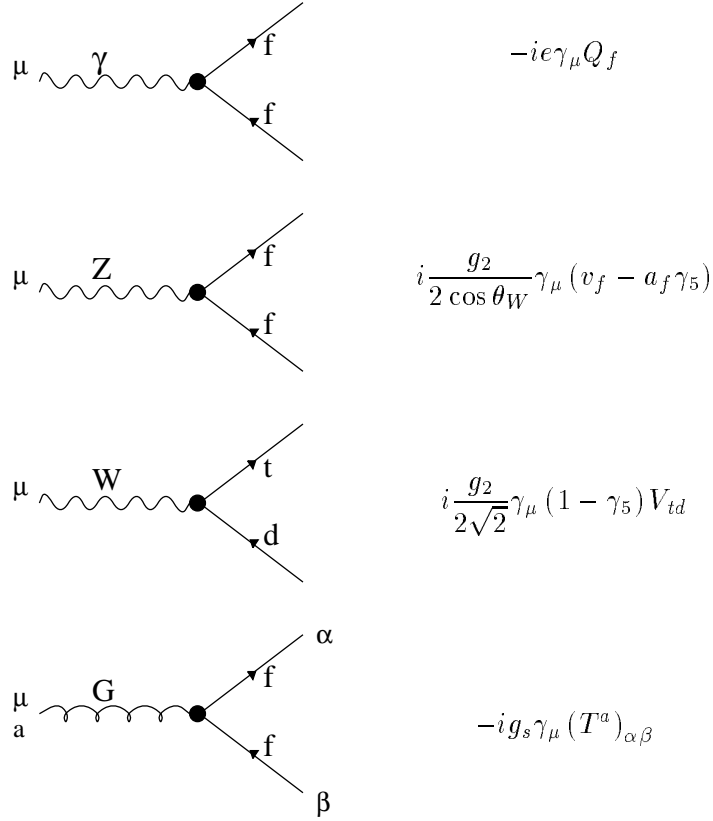}
}
\vspace{-4.95in}
\caption[]{
Feynman Rules for the Vertices
\label{fig:3}}
\end{figure}

\bi
\item
The strong interactions are mediated by eight gluons $G_a$, the
electroweak interactions by $W^{\pm}$, $Z^0$ and $\gamma$.
\item
Concerning {\it Electroweak Interactions}, the left-handed leptons and
quarks are put into $SU(2)_L$ doublets:
\begin{equation}\label{2.31}
\left(\begin{array}{c}
\nu_e \\
e^-
\end{array}\right)_L\qquad
\left(\begin{array}{c}
\nu_\mu \\
\mu^-
\end{array}\right)_L\qquad
\left(\begin{array}{c}
\nu_\tau \\
\tau^-
\end{array}\right)_L
\end{equation}
\begin{equation}\label{2.66}
\left(\begin{array}{c}
u \\
d^\prime
\end{array}\right)_L\qquad
\left(\begin{array}{c}
c \\
s^\prime
\end{array}\right)_L\qquad
\left(\begin{array}{c}
t \\
b^\prime
\end{array}\right)_L       
\end{equation}
with the corresponding right-handed fields transforming as singlets
under $ SU(2)_L $. The primes in (\ref{2.66}) indicate that the
down quark fields  $(d^\prime,s^\prime,b^\prime)$ placed in these doublets 
are weak (flavour) eigenstates. They differ from the 
mass eigenstates $(d,s,b)$. Without loss of generality one can set 
$u=u^\prime$, $c=c^\prime$, $t=t^\prime$. Similarly one can set 
the weak and the mass eigenstates for $(e^-,\mu^-,\tau^-)$ to be equal 
to each other. Then the flavour eigenstates $(\nu_e,\nu_\mu,\nu_\tau)$ 
must differ from the corresponding mass eigenstates $(\nu_1,\nu_2,\nu_2)$ 
because of the observed neutrino oscillations. 
The latter topic is beyond the scope of these lectures.
\item
The charged current processes mediated by $W^{\pm}$ are
flavour violating with the strength of violation given by
the gauge coupling $g_2$  and effectively at low energies 
by the Fermi constant (see Section \ref{THF})
\begin{equation}\label{2.100}
\frac{G_{\rm F}}{\sqrt{2}}=\frac{g^2_2}{8 \mw^2}
\end{equation}
and a {\it unitary} $3\times3$
{\rm CKM} matrix. 
\item
The {\rm CKM} matrix \cite{CAB,KM} connects the {\it weak
eigenstates} $(d^\prime,s^\prime,b^\prime)$ and the corresponding {\it mass 
eigenstates} $d,s,b$ through
\begin{equation}\label{2.67}
\left(\begin{array}{c}
d^\prime \\ s^\prime \\ b^\prime
\end{array}\right)=
\left(\begin{array}{ccc}
V_{ud}&V_{us}&V_{ub}\\
V_{cd}&V_{cs}&V_{cb}\\
V_{td}&V_{ts}&V_{tb}
\end{array}\right)
\left(\begin{array}{c}
d \\ s \\ b
\end{array}\right)\equiv\hat V_{\rm CKM}\left(\begin{array}{c}
d \\ s \\ b
\end{array}\right).
\end{equation}
In the leptonic sector the analogous mixing matrix is  
the MNS matrix \cite{MNS}, but due to the possibility of 
neutrinos being Majorana
particles, two additional complex phases could be present.
\item
The unitarity of the CKM matrix assures the absence of
flavour changing neutral current transitions at the tree level.
This means that the
elementary vertices involving neutral gauge bosons ($G_a$, $Z^0$,
$\gamma$) and the neutral Higgs are flavour conserving.
This property is known under the name of GIM mechanism \cite{GIM}.
\item
The fact that the $V_{ij}$'s can a priori be complex
numbers allows  CP violation in the SM \cite{KM}. 
\item
Feynman rules for charged current and neutral current vertices are given in 
Fig.~\ref{fig:3}, where $T^a$ are  colour matrices and
\begin{equation}\label{9}
v_f=T^f_3-2 Q_f \sin^2\theta_{W},
\qquad
a_f=T^f_3.
\end{equation}
Here $Q_f$ and $T^f_3$ denote the charge and the third component of the
weak isospin of the left-handed fermion $f_L$, respectively. These
electroweak charges are collected in Table~\ref{tab:ewcharges}.
\begin{table}[htb]
\caption[]{Electroweak Quantum Numbers.
\label{tab:ewcharges}}
\begin{center}
\begin{tabular}{@{}|llllllll|@{}}
\hline
 & $\nu^e_L$ & $ e^-_L$ & $ e^-_R$ & $u_L$ & $d_L$ & $u_R$ & $d_R$ \\
\hline
$Q$ & 0 & $-1$ & $-1$ & 2/3 & $-1/3$ & 2/3 & $-1/3$ \\
\hline
$T_3$ &1/2 & $-1/2$ & 0 & 1/2 & $-1/2$ & 0 & 0 \\
\hline
$Y$ & $-1$ & $-1$ & $-2$ & 1/3 & 1/3 & 4/3 & $-2/3$ \\
\hline
\end{tabular}
\end{center}
\end{table}
It should be noted that
the photonic and gluonic
vertices are vectorlike (V), the $W^{\pm}$ vertices are purely $V-A$
and the $Z^0$ vertices involve both $V-A$ and
$V+A$ structures.
\item
An important property of the strong interactions described by
Quantum Chromodynamics (QCD) is {\it the asymptotic freedom}
\cite{ASYM}.
This property implies that at short distance scales $\mu >\ord(1~\gev)$
the strong interaction effects in weak decays can be evaluated
by means of perturbative methods with the expansion parameter
$\alpha_{\overline{MS}}(\mu)\equiv\alpha_s(\mu)$ \cite{BBDM}. 
The existing analyses
of high energy processes give  
$\alpha_s(\mz)=0.1187\pm 0.0020$ \cite{PDG,Bethke}.
The value of $\alpha_s(\mu)$ for $\mu\not=\mz$ can be calculated by means of
($\alpha_s=g^2_s/4\pi$)
\be\label{alphaNLL}
\as(\mu) = \frac{\as(M_Z)}{v(\mu)} \left[1 - \f{\beta_1}{\beta_0} 
           \frac{\as(M_Z)}{4 \pi}    \f{\ln v(\mu)}{v(\mu)} \right],
\ee
where 
\be\label{v(mu)}
v(\mu) = 1 - \beta_0 \frac{\as(M_Z)}{2 \pi} 
\ln \left( \frac{M_Z}{\mu} \right).
\ee
Here
\be
\beta_0=11 -\frac{2}{3}f ,\qquad \beta_1=102 -\frac{38}{3}f
\ee
with $f$ denoting the number of flavours.
\item
At long distances, corresponding to $\mu<\ord( 1\gev)$, 
$\alpha_s(\mu)$
becomes  large and QCD effects in weak decays relevant
to these scales can only be evaluated by means of non-perturbative methods. 
As we will see in the course of these lectures, this is the main difficulty 
in the description of weak decays of mesons. We will address this problem
in Section \ref{THF}.
\ei

\section{CKM Matrix and the Unitarity Triangle}\label{CKMM}
\setcounter{equation}{0}
\subsection{Preliminaries}
Many parametrizations of the CKM
matrix have been proposed in the literature. The classification of different 
parametrizations can be found in \cite{FX1}. While the so called 
standard parametrization \cite{CHAU} 
\begin{equation}\label{2.72}
\hat V_{\rm CKM}=
\left(\begin{array}{ccc}
c_{12}c_{13}&s_{12}c_{13}&s_{13}e^{-i\delta}\\ -s_{12}c_{23}
-c_{12}s_{23}s_{13}e^{i\delta}&c_{12}c_{23}-s_{12}s_{23}s_{13}e^{i\delta}&
s_{23}c_{13}\\ s_{12}s_{23}-c_{12}c_{23}s_{13}e^{i\delta}&-s_{23}c_{12}
-s_{12}c_{23}s_{13}e^{i\delta}&c_{23}c_{13}
\end{array}\right)\,,
\end{equation}
with
$c_{ij}=\cos\theta_{ij}$ and $s_{ij}=\sin\theta_{ij}$ 
($i,j=1,2,3$) and the complex phase $\delta$ necessary for {\rm CP} violation,
should be recommended \cite{PDG} 
for any numerical 
analysis, a generalization of the Wolfenstein parametrization \cite{WO}, as 
presented in \cite{BLO}, is more suitable for these lectures. 
On the one hand it is more transparent than the standard parametrization and 
on the other hand it  satisfies the unitarity 
of the CKM matrix to higher accuracy  than the original parametrization 
in \cite{WO}. Let us then discuss it in some details. 

\subsection{Generalized Wolfenstein Parametrization}
In order to find this parametrization we make the following change of 
variables in
the standard parametrization (\ref{2.72}) 
\cite{BLO,schubert}
\begin{equation}\label{2.77} 
s_{12}=\lambda\,,
\qquad
s_{23}=A \lambda^2\,,
\qquad
s_{13} e^{-i\delta}=A \lambda^3 (\varrho-i \eta)~,
\end{equation}
where
\begin{equation}\label{2.76}
\lambda, \qquad A, \qquad \varrho, \qquad \eta \, 
\end{equation}
are the Wolfenstein parameters
with $\lambda\approx 0.22$ being an expansion parameter. We find then 
\be\label{f1}
V_{ud}=1-\frac{1}{2}\lambda^2-\frac{1}{8}\lambda^4, \qquad
V_{cs}= 1-\frac{1}{2}\lambda^2-\frac{1}{8}\lambda^4(1+4 A^2),
\ee
\be
V_{tb}=1-\frac{1}{2} A^2\lambda^4, \qquad
V_{cd}=-\lambda+\frac{1}{2} A^2\lambda^5 [1-2 (\varrho+i \eta)],
\ee
\be\label{VUS}
V_{us}=\lambda+\ord(\lambda^7),\qquad 
V_{ub}=A \lambda^3 (\varrho-i \eta), \qquad 
V_{cb}=A\lambda^2+\ord(\lambda^8),
\ee
\begin{equation}\label{2.83d}
 V_{ts}= -A\lambda^2+\frac{1}{2}A\lambda^4[1-2 (\varrho+i\eta)],
\qquad V_{td}=A\lambda^3(1-\bar\varrho-i\bar\eta)~,
\end{equation}
where terms 
$\ord(\lambda^6)$ and higher order terms have been neglected.
A non-vanishing $\eta$ is responsible for CP violation in the SM. It plays 
the role of $\delta$ in the standard parametrization.
Finally, the barred variables in (\ref{2.83d}) are given by
\cite{BLO}
\begin{equation}\label{2.88d}
\bar\varrho=\varrho (1-\frac{\lambda^2}{2}),
\qquad
\bar\eta=\eta (1-\frac{\lambda^2}{2}).
\end{equation}

The advantage of this generalization of the Wolfenstein parametrization
over other generalizations found in the literature is the absence of
relevant corrections to $V_{us}$, $V_{cb}$ and $V_{ub}$ and an elegant
change in $V_{td}$ which allows a simple generalization of the 
so-called unitarity triangle beyond LO. For these reasons this
generalization of the Wolfenstein parametrization has been adopted
by most authors in the literature.

Finally let us collect useful approximate analytic expressions
for $\lambda_i=V_{id}V^*_{is}$ with $i=c,t$:
\begin{equation}\label{2.51}
 \IM\lambda_t= -\IM\lambda_c=\eta A^2\lambda^5=
\mid V_{ub}\mid \mid V_{cb} \mid \sin\delta~, 
\end{equation}
\begin{equation}\label{2.52}
 \RE\lambda_c=-\lambda (1-\frac{\lambda^2}{2})~,
\end{equation}
\begin{equation}\label{2.53}
 \RE\lambda_t= -(1-\frac{\lambda^2}{2}) A^2\lambda^5 (1-\bar\varrho) \,.
\end{equation}
Expressions (\ref{2.51}) and (\ref{2.52}) represent to an accuracy of
0.2\% the exact formulae obtained using (\ref{2.72}). The expression
(\ref{2.53}) deviates by at most $0.5\%$ from the exact formula in the
full range of parameters considered.  
After inserting the expressions (\ref{2.51})--(\ref{2.53}) in the exact
formulae for quantities of interest, a further expansion in $\lambda$
should not be made. 

\subsection{Unitarity Triangle}
The unitarity of the CKM-matrix implies various relations between its
elements. In particular, we have
\begin{equation}\label{2.87h}
V_{ud}^{}V_{ub}^* + V_{cd}^{}V_{cb}^* + V_{td}^{}V_{tb}^* =0.
\end{equation}
Phenomenologically this relation is very interesting as it involves
simultaneously the elements $V_{ub}$, $V_{cb}$ and $V_{td}$ which are
under extensive discussion at present.

The relation (\ref{2.87h})  can be
represented as a ``unitarity'' triangle in the complex 
$(\bar\varrho,\bar\eta)$ plane. 
The invariance of (\ref{2.87h})  under any phase-transformations
implies that the  corresponding triangle
is rotated in the $(\bar\varrho,\bar\eta)$  plane under such transformations. 
Since the angles and the sides
(given by the moduli of the elements of the
mixing matrix)  in this triangle remain unchanged, they
 are phase convention independent and are  physical observables.
Consequently they can be measured directly in suitable experiments.  
One can construct additional five unitarity triangles corresponding
to other orthogonality relations, like the one in (\ref{2.87h}).
They are discussed in \cite{Kayser}. Some of them should be useful
when LHC-B experiment will provide data.
The areas of all unitarity triangles are equal and related to the measure of 
CP violation 
$J_{\rm CP}$ \cite{CJ}:
\begin{equation}
\mid J_{\rm CP} \mid = 2\cdot A_{\Delta},
\end{equation}
where $A_{\Delta}$ denotes the area of the unitarity triangle.

\begin{figure}[hbt]
\vspace{0.10in}
\centerline{
\epsfysize=2.1in
\epsffile{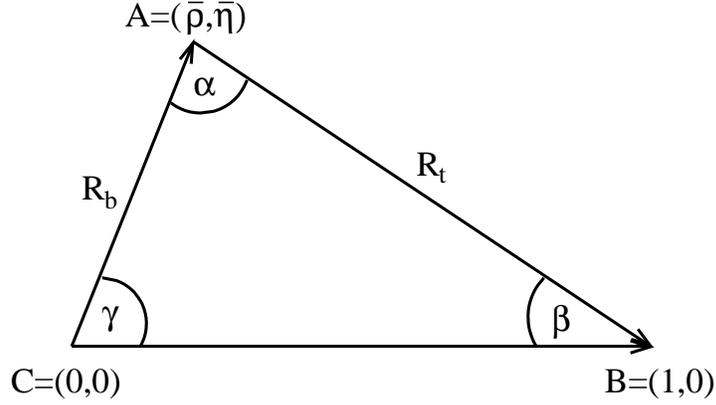}
}
\vspace{0.08in}
\caption{Unitarity Triangle.}\label{fig:utriangle}
\end{figure}

The construction of the unitarity triangle proceeds as follows:

\bi
\item
We note first that
\begin{equation}\label{2.88a}
V_{cd}^{}V_{cb}^*=-A\lambda^3+\ord(\lambda^7).
\end{equation}
Thus to an excellent accuracy $V_{cd}^{}V_{cb}^*$ is real with
$| V_{cd}^{}V_{cb}^*|=A\lambda^3$.
\item
Keeping $\ord(\lambda^5)$ corrections and rescaling all terms in
(\ref{2.87h})
by $A \lambda^3$ 
we find
\begin{equation}\label{2.88b}
 \frac{1}{A\lambda^3}V_{ud}^{}V_{ub}^*
=\bar\varrho+i\bar\eta,
\qquad
\qquad
 \frac{1}{A\lambda^3}V_{td}^{}V_{tb}^*
=1-(\bar\varrho+i\bar\eta)
\end{equation}
with $\bar\varrho$ and $\bar\eta$ defined in (\ref{2.88d}). 
\item
Thus we can represent (\ref{2.87h}) as the unitarity triangle 
in the complex $(\bar\varrho,\bar\eta)$ plane 
as shown in Fig. \ref{fig:utriangle}.
\ei

Let us collect useful formulae related 
to this triangle:
\bi
\item
We can express the angles $\alpha, \beta, \gamma$ 
in terms of $(\bar\varrho,\bar\eta)$. In particular:
\begin{equation}\label{2.90}
\sin(2\beta)=\frac{2\bar\eta(1-\bar\varrho)}{(1-\bar\varrho)^2 + \bar\eta^2}.
\end{equation}
\item
The lengths $CA$ and $BA$ are given respectively by
\begin{equation}\label{2.94}
R_b \equiv \frac{| V_{ud}^{}V^*_{ub}|}{| V_{cd}^{}V^*_{cb}|}
= \sqrt{\bar\varrho^2 +\bar\eta^2}
= (1-\frac{\lambda^2}{2})\frac{1}{\lambda}
\left| \frac{V_{ub}}{V_{cb}} \right|,
\end{equation}
\begin{equation}\label{2.95}
R_t \equiv \frac{| V_{td}^{}V^*_{tb}|}{| V_{cd}^{}V^*_{cb}|} =
 \sqrt{(1-\bar\varrho)^2 +\bar\eta^2}
=\frac{1}{\lambda} \left| \frac{V_{td}}{V_{cb}} \right|.
\end{equation}
\item
The angles $\beta$ and $\gamma=\delta$ of the unitarity triangle 
are related
directly to the complex phases of the CKM-elements $V_{td}$ and
$V_{ub}$, respectively, through
\beq\label{e417}
V_{td}=|V_{td}|e^{-i\beta},\quad V_{ub}=|V_{ub}|e^{-i\gamma}.
\eeq
\item
The unitarity relation (\ref{2.87h}) can be rewritten as
\be\label{RbRt}
R_b e^{i\gamma} +R_t e^{-i\beta}=1~.
\ee
\item
The angle $\alpha$ can be obtained through the relation
\beq\label{e419}
\alpha+\beta+\gamma=180^\circ~.
\eeq
\ei

Formula (\ref{RbRt}) shows transparently that the knowledge of
$(R_t,\beta)$ allows to determine $(R_b,\gamma)$ through 
\be\label{VUBG}
R_b=\sqrt{1+R_t^2-2 R_t\cos\beta},\qquad
\cot\gamma=\frac{1-R_t\cos\beta}{R_t\sin\beta}.
\ee
Similarly, $(R_t,\beta)$ can be expressed through $(R_b,\gamma)$:
\be\label{VTDG}
R_t=\sqrt{1+R_b^2-2 R_b\cos\gamma},\qquad
\cot\beta=\frac{1-R_b\cos\gamma}{R_b\sin\gamma}.
\ee
These relations are remarkable. They imply that the knowledge 
of the coupling $V_{td}$ between $t$ and $d$ quarks allows to deduce the 
strength of the corresponding coupling $V_{ub}$ between $u$ and $b$ quarks 
and vice versa.

The triangle depicted in Fig.~\ref{fig:utriangle}, $|V_{us}|$ 
and $\vcb$ give the full description of the CKM matrix. 
Looking at the expressions for $R_b$ and $R_t$, we observe that within
the SM the measurements of four CP
{\it conserving } decays sensitive to $|V_{us}|$, $|V_{ub}|$,   
$|V_{cb}|$ and $|V_{td}|$ can tell us whether CP violation
($\bar\eta \not= 0$ or $\gamma \not=0,\pi$) is predicted in the SM. 
This fact is often used to determine
the angles of the unitarity triangle without the study of CP-violating
quantities. 

\subsection{A Different Parametrization}
Parallel to the use of the Wolfenstein parameters  it is sometimes useful 
to express the CKM elements $V_{td}$ and $V_{ts}$ 
as follows \cite{BFRS-BIG}
\be\label{VTDVTS}
V_{td}=A R_t\lambda^3 e^{-i\beta},\quad V_{ts}=-|V_{ts}|e^{-i\beta_s},
\ee
with $\tan\beta_s\approx -\lambda^2 \bar\eta$.
The smallness of $\beta_s$ follows from the CKM phase conventions and the 
unitarity of the CKM matrix. Consequently it is valid beyond the SM
if three generation unitarity is assumed.

We have then
\be\label{LAMT}
\lambda_t\equiv V_{ts}^*V_{td}=-\tilde r \lambda\vcb^2 R_t e^{-i\beta} 
e^{i\beta_s} \quad\mbox{with}\quad 
\tilde r=\left|\frac{V_{ts}}{V_{cb}}\right|
=\sqrt{1+\lambda^2(2\bar\varrho-1)}
\approx 0.985,
\ee
where in order to avoid high powers of $\lambda$ we expressed the parameter
 $A$ through $\vcb$. 
Consequently 
\be\label{IMRE}
\imlt=\tilde r \lambda\vcb^2 R_t\sin(\beta_{\rm eff}), \qquad 
\relt=-\tilde r \lambda\vcb^2 R_t\cos(\beta_{\rm eff})
\ee
with $\beta_{\rm eff}=\beta-\beta_s$.

\boldmath
\subsection{Leading Strategies for $(\bar\varrho,\bar\eta)$}
\unboldmath
Next, we have the following useful 
relations, that correspond to the best strategies for the determination of 
$(\bar\varrho,\bar\eta)$ considered in \cite{BUPAST}:

{\bf \boldmath{$(R_t,\beta)$} Strategy}:

\be\label{S1}
\bar\varrho=1-R_t\cos\beta, \qquad \bar\eta=R_t\sin\beta
\ee
with $R_t$ determined through (\ref{Rt}) below and $\beta$ through 
the CP asymmetry $\calamix(\psi K_S)$ as discussed in Section 6.
In this strategy, $R_b$ and $\gamma$ are given by
(\ref{VUBG}).

{\bf \boldmath{($R_b,\gamma)$} Strategy}:

\be\label{S2}
\bar\varrho=R_b\cos\gamma, \qquad \bar\eta=R_b\sin\gamma
\ee
with $\gamma$ (see Fig.~\ref{fig:utriangle}), determined through clean 
strategies in tree dominated $B$-decays \cite{BABAR,SUPERB,LHCB,FERMILAB}.
In this strategy, $R_t$ and $\beta$ are given by
(\ref{VTDG}).

{\bf \boldmath{$(\beta,\gamma)$} Strategy}:

Formulae in (\ref{S1}) and
\be\label{S3}
R_t=\frac{\sin\gamma}{\sin(\beta+\gamma)}
\ee
with $\beta$ and $\gamma$ determined through $\calamix(\psi K_S)$ and clean 
strategies for $\gamma$ as in (\ref{S2}).
In this strategy, the length $R_b$ and $\vub$ can be determined through
\be\label{RTVUB2}
R_b=\frac{\sin\beta}{\sin(\beta+\gamma)},\qquad
\left|\frac{V_{ub}}{V_{cb}}\right|=\left(\frac{\lambda}{1-\lambda^2/2}
\right) R_b.
\ee

{\bf \boldmath{$(\bar\eta,\gamma)$} Strategy}:

\be\label{S4}
\bar\varrho=\frac{\bar\eta}{\tan\gamma}
\ee
with $\bar\eta$ determined for instance through $Br(\klpn)$ as discussed 
in Section \ref{KPNN} and $\gamma$ as in the two strategies above.

As demonstrated in \cite{BUPAST}, the $(R_t,\beta)$ strategy will be very useful
as soon as the $B^0_{s}-\bar B^0_{s}$ mixing mass difference
$\Delta M_s$ has been measured. However, the remaining three strategies turn 
out to be more efficient in determining $(\bar\varrho,\bar\eta)$. 
The strategies $(\beta,\gamma)$ and $(\bar\eta,\gamma)$
are theoretically cleanest as $\beta$ and $\gamma$  
can be measured 
precisely in two body $B$ decays one day and $\bar\eta$ can be extracted 
from $Br(\klpn)$ subject only to uncertainty in $\vcb$. Combining these
two strategies offers a precise determination of the CKM matrix 
including $\vcb$ and $|V_{ub}|$ \cite{AJB94,BSU}. 
On the other hand, these two strategies 
are subject to uncertainties
coming from new physics that can enter through $\beta$ and $\bar\eta$. 
The angle $\gamma$, the phase of $V_{ub}$, can be determined in principle 
without 
these uncertainties. 

The strategy $(R_b,\gamma)$, on the other hand, while
subject to hadronic uncertainties in the determination of $R_b$, is not
polluted by new physics contributions as, in addition to $\gamma$, also 
 $R_b$ can be determined from tree level decays. This strategy results 
in the so-called {\it reference unitarity triangle} as proposed and discussed
in \cite{refut}. We will return to some of these strategies in the course of 
our lectures.

\boldmath
\subsection{The Special Role of {$|V_{us}|$}, {$|V_{ub}|$}
and {$|V_{cb}|$}}
\unboldmath
What do we know about the CKM matrix and the unitarity triangle on the
basis of {\it tree level} decays? 
Here the semi-leptonic $K$ and $B$ decays play the decisive role. 
The present situation can be summarized by \cite{CERNCKM,BSU} 
\begin{equation}\label{vcb}
|V_{us}| = \lambda =  0.2240 \pm 0.0036\,
\quad\quad
\vcb=(41.5\pm0.8)\cdot 10^{-3},
\end{equation}
\begin{equation}\label{v13}
\frac{|V_{ub}|}{\vcb}=0.092\pm0.012, \quad\quad
|V_{ub}|=(3.81\pm0.46)\cdot 10^{-3}.
\end{equation}
implying
\be
 A=0.83\pm0.02,\qquad R_b=0.40\pm 0.06~.
\ee
There is an impressive work done by theorists and experimentalists hidden
behind these numbers. We refer to \cite{CERNCKM} for details.
See also \cite{PDG} and the recent improved determinations of $V_{us}$ 
\cite{VUSNEW}.

\begin{figure}[hbt]
\vspace{-0.10in}
\centerline{
\epsfysize=2.0in
\epsffile{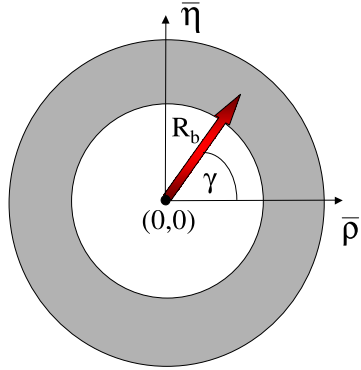}
}
\vspace{0.02in}
\caption[]{``Unitarity Clock".
\label{L:2}}
\end{figure}

The information given above tells us only that the apex $A$ of the unitarity 
triangle lies in the band shown in Fig.~\ref{L:2}. 
While this information appears at first sight to be rather limited, 
it is very important for the following reason. As $|V_{us}|$, $\vcb$, 
 $|V_{ub}|$ and consequently $R_b$ are determined here from tree level 
decays, their
values given above are to an excellent accuracy independent of any 
new physics contributions. They are universal fundamental 
constants valid in any extention of the SM. Therefore their precise 
determinations are of utmost importance. 
In order to answer the question where
the apex $A$ lies on the ``unitarity clock'' in Fig.~\ref{L:2} we have to 
look at other decays. Most promising in this respect are the so-called 
``loop induced'' decays and transitions and CP-violating $B$ decays. 
These decays are sensitive to the angles $\beta$ and $\gamma$ as well as 
to the length $R_t$ and measuring only one of these three quantities allows to 
find the unitarity triangle provided the universal $R_b$ is known.

Of course any pair among $(R_t,\beta,\gamma)$ is sufficient to construct 
the UT without any knowledge of $R_b$. Yet the special role of $R_b$ among
these variables lies in its universality, whereas the other three variables 
are generally sensitive functions of possible new physics contributions.
This means that assuming three generation unitarity 
of the CKM matrix and that the SM is a part of a grander 
theory, the apex of the unitarity triangle has to be eventually placed
on the unitarity clock with the radius $R_b$ obtained from tree level decays.
That is even 
if using SM expressions for loop induced processes, $(\bar\varrho,\bar\eta)$
would be found outside the unitarity clock, the corresponding expressions 
of the grander theory must include appropriate new contributions so that 
the apex of the unitarity triangle is shifted back to the band in  
Fig.~\ref{L:2}. In the case of CP asymmetries this could be achieved by 
realizing that the measured angles $\alpha$, $\beta$ and $\gamma$ are not
the true angles of the unitarity triangle but sums of the true angles and 
new complex phases present in extentions of the SM. The better $R_b$ is known,
the thinner the band in Fig.~\ref{L:2} will be, 
selecting in this manner efficiently the correct theory. On the other hand 
as 
the branching ratios for rare and CP-violating decays depend sensitively
on the parameter $A$, the precise knowledge of $\vcb$ is also very important.

\section{Theoretical Framework}\label{THF}
\setcounter{equation}{0}
\subsection{General View}
The basic starting point for any serious phenomenology of weak decays of
hadrons is the effective weak Hamiltonian which has the following generic
structure
\be\label{b1}
{\cal H}_{eff}=\frac{G_F}{\sqrt{2}}\sum_i V^i_{\rm CKM}C_i(\mu)Q_i~.
\ee
Here $G_F$ is the Fermi constant and $Q_i$ are the relevant local
operators which govern the decays in question. The Cabibbo-Kobayashi-Maskawa
factors $V^i_{CKM}$ \cite{CAB,KM} 
and the Wilson Coefficients $C_i$ \cite{OPE,ZIMM} describe the 
strength with which a given operator enters the Hamiltonian.
We will soon give a more intuitive names to $C_i$ and $Q_i$.

\begin{figure}[hbt]
\vspace{0.10in}
\centerline{
\epsfysize=1.9in
\epsffile{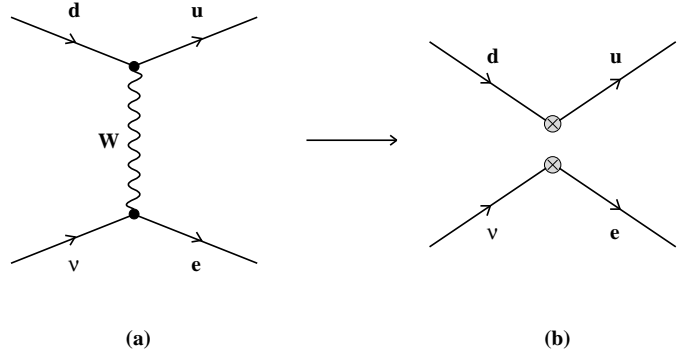}
}
\vspace{0.08in}
\caption[]{
$\beta$-decay at the quark level in the full (a) and effective (b)
theory.
\label{L:1}}
\end{figure}

In the simplest case of the $\beta$-decay, ${\cal H}_{eff}$ takes 
the familiar form
\be\label{beta}
{\cal H}^{(\beta)}_{eff}=\frac{G_F}{\sqrt{2}}
\cos\theta_c[\bar u\gamma_\mu(1-\gamma_5)d \otimes
\bar e \gamma^\mu (1-\gamma_5)\nu_e]~,
\ee
where $V_{ud}$ has been expressed in terms of the Cabibbo angle. In this
particular case the Wilson Coefficient is equal to unity and the local
operator, the object between the square brackets, is given by a product 
of two $V-A$ currents. This local operator is represented by the
diagram (b) in Fig. \ref{L:1}.
Equation (\ref{beta}) represents the Fermi theory for $\beta$-decays 
as formulated by Sudarshan and
Marshak \cite{SUMA} and Feynman and Gell-Mann \cite{GF} almost fifty 
years ago, 
except that in (\ref{beta})
the quark language has been used and following Cabibbo a small departure of
$V_{ud}$ from unity has been incorporated. In this context the basic 
formula (\ref{b1})
can be regarded as a generalization of the Fermi Theory to include all known
quarks and leptons as well as their strong and electroweak interactions as
summarized by the SM. It should be stressed that the formulation
of weak decays in terms of effective Hamiltonians is very suitable for the
inclusion of new physics effects. We will discuss this issue later on.

Now, I am aware of the fact that the formal operator language used here is
hated by experimentalists and frequently disliked by more phenomenologically
minded theorists. Consequently the literature on weak decays, in particular
on $B$-meson decays, is governed by Feynman diagram drawings with 
$W$-, $Z^0$- and top
quark exchanges, rather than by the operators in (\ref{b1}). 
In the case of the $\beta$-decay we have the diagram (a) in Fig.~\ref{L:1}.
Yet such Feynman
diagrams with full $W$-propagators, $Z^0$-propagators and top-quark propagators
really represent the situation at very short distance scales 
$\ord ({\rm M_{W,Z}, m_t})$, whereas the
true picture of a decaying hadron with masses 
$\ord(\mb,\mc,m_K)$ is more properly described by
effective point-like vertices which are represented by the local operators
$Q_i$. The Wilson coefficients $C_i$ can then be regarded as coupling constants
associated with these effective vertices.

Thus ${\cal H}_{eff}$ in (\ref{b1}) is simply a series of effective 
vertices multiplied 
by effective coupling constants $C_i$. This series is known under the name 
of the operator product expansion (OPE) \cite{OPE,ZIMM,WIT}. 
Due to the interplay of electroweak 
and strong interactions the structure of the local operators (vertices) is 
much richer than in the case of the $\beta$-decay. They can be classified 
with respect to the Dirac structure, colour structure and the type of quarks 
and leptons relevant for a given decay. Of particular interest are the 
operators involving quarks only. They govern the non-leptonic decays.

As an example we give a list of operators that play the role 
in weak $B$ decays.
Typical diagrams in
the full theory from which these operators originate are 
 shown in Fig.~\ref{fig:fdia}. The cross in Fig.~5d indicates  
that magnetic penguins originate from the mass-term on the external
line in the usual QCD or QED penguin diagrams.
The six classes of operators 
are given as follows ($\alpha$ and $\beta$ are colour
indices):

{\bf Current--Current (Fig.~5a):}
\begin{equation}\label{O1} 
Q_1 = (\bar c_{\alpha} b_{\beta})_{V-A}\;(\bar s_{\beta} c_{\alpha})_{V-A}
~~~~~~Q_2 = (\bar c b)_{V-A}\;(\bar s c)_{V-A} 
\end{equation}

{\bf QCD--Penguins (Fig.~5b):}
\begin{equation}\label{O2}
Q_3 = (\bar s b)_{V-A}\sum_{q=u,d,s,c,b}(\bar qq)_{V-A}~~~~~~   
 Q_4 = (\bar s_{\alpha} b_{\beta})_{V-A}\sum_{q=u,d,s,c,b}(\bar q_{\beta} 
       q_{\alpha})_{V-A} 
\end{equation}
\begin{equation}\label{O3}
 Q_5 = (\bar s b)_{V-A} \sum_{q=u,d,s,c,b}(\bar qq)_{V+A}~~~~~  
 Q_6 = (\bar s_{\alpha} b_{\beta})_{V-A}\sum_{q=u,d,s,c,b}
       (\bar q_{\beta} q_{\alpha})_{V+A} 
\end{equation}

{\bf Electroweak Penguins (Fig.~5c):}
\begin{equation}\label{O4} 
Q_7 = {3\over 2}\;(\bar s b)_{V-A}\sum_{q=u,d,s,c,b}e_q\;(\bar qq)_{V+A} 
~~~~~ Q_8 = {3\over2}\;(\bar s_{\alpha} b_{\beta})_{V-A}\sum_{q=u,d,s,c,b}e_q
        (\bar q_{\beta} q_{\alpha})_{V+A}
\end{equation}
\begin{equation}\label{O5} 
 Q_9 = {3\over 2}\;(\bar s b)_{V-A}\sum_{q=u,d,s,c,b}e_q(\bar q q)_{V-A}
~~~~~Q_{10} ={3\over 2}\;
(\bar s_{\alpha} b_{\beta})_{V-A}\sum_{q=u,d,s,c,b}e_q\;
       (\bar q_{\beta}q_{\alpha})_{V-A} 
\end{equation}

{\bf Magnetic Penguins (Fig.~5d):}
\begin{equation}\label{O6}
Q_{7\gamma}  =  \frac{e}{8\pi^2} m_b \bar{s}_\alpha \sigma^{\mu\nu}
          (1+\gamma_5) b_\alpha F_{\mu\nu}\qquad            
Q_{8G}     =  \frac{g}{8\pi^2} m_b \bar{s}_\alpha \sigma^{\mu\nu}
   (1+\gamma_5)T^a_{\alpha\beta} b_\beta G^a_{\mu\nu}  
\end{equation}

{\bf \boldmath{$\Delta S = 2 $} and \boldmath{$ \Delta B = 2 $}
 Operators (Fig.~5e):}
\begin{equation}\label{O7}
Q(\Delta S = 2)  = (\bar s d)_{V-A} (\bar s d)_{V-A}~~~~~
 Q(\Delta B = 2)  = (\bar b d)_{V-A} (\bar b d)_{V-A} 
\end{equation}

{\bf Semi--Leptonic Operators (Fig.~5f):}
\begin{equation}\label{9V}
Q_{9V}  = (\bar s b  )_{V-A} (\bar \mu\mu)_{V}~~~~~
Q_{10A}  = (\bar s b )_{V-A} (\bar \mu\mu)_{A}
\end{equation}

\begin{equation}\label{10V}
Q_{\nu\bar\nu}  = (\bar s b  )_{V-A} (\bar \nu\nu)_{V-A}~~~~~
Q_{\mu\bar\mu}  = (\bar s b )_{V-A} (\bar \mu\mu)_{V-A}~.
\end{equation}

Now what about the couplings $C_i(\mu)$ and the scale $\mu$? The 
important point is that $C_i(\mu)$
summarize the physics contributions from scales higher than $\mu$ and due to
asymptotic freedom of QCD they can be calculated in perturbation theory as
long as $\mu$ is not too small. $C_i$ include the top quark contributions and
contributions from other heavy particles such as $W$, $Z$-bosons, charged
Higgs particles, supersymmetric particles in the supersymmetric extensions
of the SM, Kaluza--Klein modes in models with extra dimensions and 
the heavy top quark $T$ in the Little Higgs models. 
At higher orders in the electroweak coupling the
neutral Higgs may also contribute. Consequently $C_i(\mu)$ depend generally 
on $m_t$ and also on the masses of new particles if extensions of the 
Standard Model are considered. This dependence can be found by evaluating 
so-called {\it box} and {\it penguin} diagrams with full $W$-, $Z$-, top- and 
new particles exchanges (see Fig.~\ref{fig:fdia}) 
and {\it properly} including short distance QCD 
effects. The latter govern the $\mu$-dependence of the couplings $C_i(\mu)$.

The value of $\mu$ can be chosen arbitrarily. It serves 
to separate the physics contributions to a given decay amplitude into
short-distance contributions at scales {\it higher}
 than $\mu$ and long-distance
contributions corresponding to scales {\it lower} than $\mu$. It is customary 
to choose
$\mu$ to be of the order of the mass of the decaying hadron. 
This is $\ord (\mb)$ and $\ord(\mc)$ for $B$-decays and
$D$-decays respectively. In the case of $K$-decays the typical choice is
 $\mu=\ord(1-2\gev)$
instead of $\ord(m_K)$, which is much too low for any perturbative 
calculation of the couplings $C_i$.

Now due to the fact that $\mu\ll  M_{W,Z},~ m_t$, large logarithms 
$\ln\mw/\mu$ compensate in the evaluation of
$C_i(\mu)$ the smallness of the QCD coupling constant $\alpha_s$ and 
terms $\alpha^n_s (\ln\mw/\mu)^n$, $\alpha^n_s (\ln\mw/\mu)^{n-1}$ 
etc. have to be resummed to all orders in $\alpha_s$ before a reliable 
result for $C_i$ can be obtained.
This can be done very efficiently by means of the renormalization group
methods that are discussed in detail in \cite{AJBLH}.
The resulting {\it renormalization group improved} perturbative
expansion for $C_i(\mu)$ in terms of the effective coupling constant 
$\alpha_s(\mu)$ does not involve large logarithms and is more reliable.

It should be stressed at this point that the construction of the effective
Hamiltonian ${\cal H}_{eff}$ by means of the operator product expansion and 
the
renormalization group methods can be done fully in the perturbative framework.
The fact that the decaying hadrons are bound states of quarks is irrelevant
for this construction. Consequently the coefficients $C_i(\mu)$ are 
independent of the
particular decay considered in the same manner in which the usual gauge
couplings are universal and process independent.

So far so good. Having constructed the effective Hamiltonian we can proceed
to evaluate the decay amplitudes. An amplitude for a decay of a given meson 
$M= K, B,..$ into a final state $F=\pi\nu\bar\nu,~\pi\pi,~DK$ is simply 
given by
\be\label{amp5}
A(M\to F)=\langle F|{\cal H}_{eff}|M\rangle
=\frac{G_F}{\sqrt{2}}\sum_i V^i_{CKM}C_i(\mu)\langle F|Q_i(\mu)|M\rangle,
\ee
where $\langle F|Q_i(\mu)|M\rangle$ 
are the hadronic matrix elements of $Q_i$ between $M$ and $F$. As indicated
in (\ref{amp5}) these matrix elements depend similarly to $C_i(\mu)$ 
on $\mu$. They summarize the physics contributions to the amplitude 
$A(M\to F)$ from scales lower than $\mu$.

We realize now the essential virtue of OPE: it allows to separate the problem
of calculating the amplitude
$A(M\to F)$ into two distinct parts: the {\it short distance}
(perturbative) calculation of the couplings $C_i(\mu)$ and 
the {\it long-distance} (generally non-perturbative) calculation of 
the matrix elements $\langle Q_i(\mu)\rangle$. The scale $\mu$, as
advertised above, separates then the physics contributions into short
distance contributions contained in $C_i(\mu)$ and the long distance 
contributions
contained in $\langle Q_i(\mu)\rangle$. By evolving this scale from 
$\mu=\ord(\mw)$ down to lower values one
simply transforms the physics contributions at scales higher than $\mu$ 
from the hadronic matrix elements into $C_i(\mu)$. Since no information 
is lost this way the full amplitude cannot depend on $\mu$. Therefore 
the $\mu$-dependence of the couplings $C_i(\mu)$ has to cancel the 
$\mu$-dependence of $\langle Q_i(\mu)\rangle$. In other words it is a
matter of choice what exactly belongs to $C_i(\mu)$ and what to 
$\langle Q_i(\mu)\rangle$. This cancellation
of $\mu$-dependence involves generally several terms in the expansion 
in (\ref{amp5}).

Clearly, in order to calculate the amplitude $A(M\to F)$, the matrix 
elements $\langle Q_i(\mu)\rangle$ have to be evaluated. 
Since they involve long distance contributions one is forced in
this case to use non-perturbative methods such as lattice calculations, the
$1/N$ expansion ($N$ is the number of colours), QCD sum rules, hadronic sum rules,
chiral perturbation theory and so on. In the case of certain $B$-meson decays,
the {\it Heavy Quark Effective Theory} (HQET) and recent approaches 
to non-leptonic decays like QCDF \cite{BBNS1}, PQCD \cite{PQCD} and 
SCET \cite{SCET} also turn out to be    
useful tools.
Needless to say, all these non-perturbative methods have some limitations.
Consequently the dominant theoretical uncertainties in the decay amplitudes
reside in the matrix elements $\langle Q_i(\mu)\rangle$.

The fact that in most cases the matrix elements $\langle Q_i(\mu)\rangle$
 cannot be reliably
calculated at present is very unfortunate. One of the main goals of the
experimental studies of weak decays is the determination of the CKM factors 
$V^i_{\rm CKM}$
and the search for the physics beyond the SM. Without a reliable
estimate of $\langle Q_i(\mu)\rangle$ this goal cannot be achieved unless 
these matrix elements can be determined experimentally or removed from the 
final measurable quantities
by taking the ratios or suitable combinations of amplitudes or branching
ratios. However, this can be achieved only in a handful of decays and
generally one has to face directly the calculation of 
$\langle Q_i(\mu)\rangle$.

Now in the case of semi-leptonic decays, in which there is at most one hadron
in the final state, the chiral perturbation theory in the case of $K$-decays
and HQET in the case of $B$-decays have already provided useful estimates of
the relevant matrix elements. This way it was possible to achieve
satisfactory determinations of the CKM elements $V_{us}$ and $V_{cb}$ in 
$K\to\pi e\nu$ and $B\to D^*e\nu$ respectively. 
We will also see that some rare decays like $K\to\pi\nu\bar\nu$ and
$B\to\mu\bar\mu$ can be calculated very reliably.

The case of non-leptonic decays in which the final state consists exclusively
of hadrons is a completely different story. Here even the matrix
elements entering the simplest decays, the two-body decays like 
$K\to\pi\pi$, $B\to K\pi$ or $B\to \pi\pi$ cannot be
calculated in QCD satisfactorly at present. 
For this reason approximative schemes 
for these decays like QCDF, PQCD and SCET can be found in the literature. 
In the case of $B$ decays they use the fact that the mass $m_b$ is much
larger than the typical hadronic scale.
We will also see in later sections that purely phenomenological approaches
supplemented by isospin symmetry, the
approximate SU(3) flavour symmetry and various plausible dynamical
assumptions can provide useful results.

Returning to the Wilson coefficients $C_i(\mu)$ it should be stressed that 
similar
to the effective coupling constants they do not depend only on the scale $\mu$
but also on the renormalization scheme used: this time on the 
scheme for the renormalization of local operators. That the local operators 
undergo renormalization is not surprising. After all they represent effective
vertices and as the usual vertices in a field theory they have to be
renormalized when quantum corrections like QCD or QED corrections are taken
into account. As a consequence of this, the hadronic matrix elements 
$\langle Q_i(\mu)\rangle$
are
renormalization scheme dependent and this scheme dependence must be cancelled
by the one of $C_i(\mu)$ so that the physical amplitudes are 
renormalization scheme
independent. Again, as in the case of the $\mu$-dependence, the 
cancellation of
the renormalization scheme dependence involves generally several 
terms in the
expansion (\ref{amp5}).

Now the $\mu$ and the renormalization scheme dependences of the couplings 
$C_i(\mu)$ can
be evaluated efficiently in the renormalization group improved perturbation
theory. Unfortunately the incorporation of these dependences in the
non-perturbative evaluation of the matrix elements  
$\langle Q_i(\mu)\rangle$
remains as an important
challenge for non-perturbative methods
but during the last years some progress 
has been done also here.

So far I have discussed only  {\it exclusive} decays. It turns out that
in the case of {\it inclusive} decays of heavy mesons, like $B$-mesons,
things turn out to be easier. In an inclusive decay one sums over all 
(or over
a special class) of accessible final states and eventually one can show 
 that 
 the resulting branching ratio can be calculated
in the expansion in inverse powers of $\mb$ with the leading term 
described by the spectator model
in which the $B$-meson decay is modelled by the decay of the $b$-quark:
\be\label{hqe}
{\rm Br}(B\to X)={\rm Br}(b\to q) +\ord(\frac{1}{\mb^2})~. 
\ee
This formula is known under the name of the Heavy Quark Expansion (HQE)
\cite{HQE1}-\cite{HQE3}.
Since the leading term in this expansion represents the decay of the quark,
it can be calculated in perturbation theory or more correctly in the
renormalization group improved perturbation theory. It should be realized
that also here the basic starting point is the effective Hamiltonian 
 (\ref{b1})
and that the knowledge of the couplings $C_i(\mu)$ is essential for 
the evaluation of
the leading term in (\ref{hqe}). But there is an important difference 
relative to the
exclusive case: the matrix elements of the operators $Q_i$ can be 
"effectively"
evaluated in perturbation theory. 
This means, in particular, that their $\mu$ and renormalization scheme
dependences can be evaluated and the cancellation of these dependences by
those present in $C_i(\mu)$ can be investigated.

Clearly in order to complete the evaluation of $Br(B\to X)$ also the 
remaining terms in
(\ref{hqe}) have to be considered. These terms are of a non-perturbative 
origin, but
fortunately they are suppressed often by  two powers of $m_b$. 
They have been
studied by several authors in the literature with the result that they affect
various branching ratios by less than $10\%$ and often by only a few percent.
Consequently the inclusive decays give generally more precise theoretical
predictions at present than the exclusive decays. On the other hand their
measurements are harder. There are of course some important theoretical
issues related to the validity of HQE in (\ref{hqe}) which appear in the 
literature under the name of quark-hadron duality but  
I will not discuss them here.

We have learned now that the matrix elements of $Q_i$ are easier to handle in
inclusive decays than in the exclusive ones. On the other hand the evaluation
of the couplings $C_i(\mu)$ is equally  difficult in both cases although 
as stated
above it can be done in a perturbative framework. Still in order to achieve
sufficient precision for the theoretical predictions it is desirable to have
accurate values of these couplings. Indeed it has been realized at the end of
the 1980's
that the leading term (LO) in the renormalization group improved perturbation
theory, in which the terms $\alpha^n_s (\ln\mw/\mu)^n$ are summed, is 
generally insufficient and the
inclusion of next-to-leading corrections  (NLO) which correspond to summing
the terms $\alpha^n_s (\ln\mw/\mu)^{n-1}$ is necessary. 
In particular, unphysical left-over $\mu$-dependences
in the decay amplitudes and branching ratios resulting from the truncation of
the perturbative series are considerably reduced by including NLO
corrections. These corrections are known by now for the most important and
interesting decays. Reviews can be found in \cite{BBL,AJBLH,Erice,REV4}. 

\subsection{Penguin Box Expansion}
\begin{figure}[hbt]
\centerline{
\epsfysize=4.3in
\epsffile{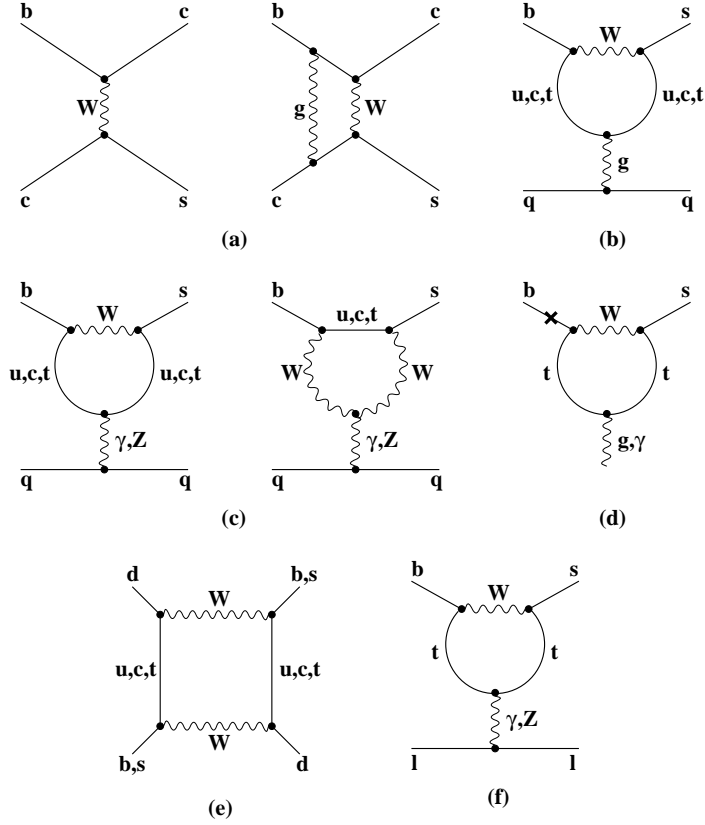}
}
\caption{Typical Penguin and Box Diagrams in the SM.}
\label{fig:fdia}
\end{figure}

Formula (\ref{amp5}) can be cast into 
 a more intuitive master formula for weak decay amplitudes in the SM
\cite{PBE} 
\be\label{smmaster}
{\rm A(Decay)}= \sum_i B_i \eta^i_{\rm QCD}V^i_{\rm CKM}  F_i(x_t),
\ee 
where $x_t=m^2_t/\mw^2$ and we suppressed $G_{\rm F}$.
Here non-perturbative parameters $B_i$ represent the matrix elements of local 
operators present in the SM. For instance in the case of 
$K^0-\bar K^0$ mixing, the matrix element of the operator
$\bar s \gamma_\mu(1-\gamma_5) d \otimes \bar s \gamma^\mu(1-\gamma_5) d $
is represented by the parameter $\hat B_K$.
There are other non-perturbative parameters in the SM that represent 
matrix elements of operators $Q_i$ with different colour and Dirac 
structures. The objects $\eta^i_{\rm QCD}$ are the QCD factors resulting 
from RG-analysis of the corresponding operators and $F^i_{\rm SM}$ stand for 
the so-called Inami-Lim functions \cite{IL} that result from the calculations 
of various
box and penguin diagrams in the SM shown in Fig.~\ref{fig:fdia}. 
They depend on the top-quark mass. 
Analogous diagrams are present in 
the extensions of the SM. 

In order to find the  functions $F_i(x_t)$, one first looks
at various functions resulting from penguin diagrams:
$C$ ($Z^0$ penguin), $D$ ($\gamma$ penguin), $E$ (gluon penguin), 
$D'$ ($\gamma$-magnetic penguin) and $E'$ (chromomagnetic 
penguin). Subsequently box diagrams have to be considered.
Here we have  the box function $S$ ($\Delta F=2$ transitions), 
as well as $\Delta F=1$ box functions
$B^{\nu\bar\nu}$ and $B^{\mu\bar\mu}$ relevant for decays with 
${\nu\bar\nu}$  and ${\mu\bar\mu}$ in the final state, respectively.

While the $\Delta F=2$ box function $S$ and the penguin functions 
$E$, $D'$ and $E'$ are gauge independent, this is not the case for 
$C$, $D$ and the $\Delta F=1$ box diagram functions 
$B^{\nu\bar\nu}$ and $B^{\mu\bar\mu}$.
In the phenomenological applications it is more 
convenient to work with gauge independent functions \cite{PBE}
\begin{equation}\label{XYZ} 
X(x_t)=C(x_t)+B^{\nu\bar\nu}(x_t),\qquad  
Y(x_t)  =C(x_t)+B^{\mu\bar\mu}(x_t), \qquad
Z(x_t)  =C(x_t)+\frac{1}{4}D(x_t).
\end{equation}
Indeed, the box diagrams have the Dirac structure $(V-A)\otimes (V-A)$, 
the $Z^0$ penguin diagram has the $(V-A)\otimes(V-A)$ and 
$(V-A)\otimes V$ components and the $\gamma$ penguin is pure
$(V-A)\otimes V$.
The $X$ and $Y$ correspond then to linear combinations of the 
$(V-A)\otimes(V-A)$ component 
of the $Z^0$ penguin diagram and box diagrams with final quarks and leptons 
having weak isospin $T_3=1/2$ and $T_3=-1/2$, respectively. $Z$ corresponds 
to the linear combination of  the $(V-A)\otimes V$ component 
of the $Z^0$ penguin diagram and the $\gamma$ penguin.

Then the set of seven gauge independent functions which govern
the FCNC processes in the SM models is given by
\be\label{masterfsm}
S(x_t),~X(x_t),~Y(x_t),~Z(x_t),~E(x_t),~ D'(x_t),~ E'(x_t)~.
\ee

In the SM we have  to a very good
approximation :
\begin{equation}\label{S0}
 S_0(x_t)=2.46~\left(\frac{\mt}{170\gev}\right)^{1.52},
\ee
\be\label{XA0}
X_0(x_t)=1.57~\left(\frac{\mt}{170\gev}\right)^{1.15},
\quad\quad
Y_0(x_t)=1.02~\left(\frac{\mt}{170\gev}\right)^{1.56},
\end{equation}
\begin{equation}
 Z_0(x_t)=0.71~\left(\frac{\mt}{170\gev}\right)^{1.86},\quad\quad
   E_0(x_t)= 0.26~\left(\frac{\mt}{170\gev}\right)^{-1.02},
\end{equation}
\begin{equation}\label{DP0}
 D'_0(x_t)=0.38~\left(\frac{\mt}{170\gev}\right)^{0.60}, \quad\quad 
E'_0(x_t)=0.19~\left(\frac{\mt}{170\gev}\right)^{0.38}.
\end{equation}
The subscript ``$0$'' indicates that 
these functions
do not include QCD corrections to the relevant penguin and box diagrams.
Exact expressions for all functions can be found in \cite{AJBLH}.

Generally, several master functions contribute to a given decay,
although decays exist which depend only on a single function. 
This will be discussed in the context of models with minimal flavour
violation below.

\subsection{Master Formula Beyond the SM}

Formula (\ref{smmaster}) can be generalized to 
 a master formula for weak decay amplitudes that 
 goes beyond the SM \cite{Pisa}. 
It reads 
\be\label{master}
{\rm A(Decay)} = \sum_i B_i \eta^i_{\rm QCD}V^i_{\rm CKM} 
\lbrack F^i_{\rm SM}+F^i_{\rm New}\rbrack   +
\sum_k B_k^{\rm New} \lbrack\eta^k_{\rm QCD}\rbrack^{\rm New} V^k_{\rm New} 
\lbrack G^k_{\rm New}\rbrack\, ,
\ee
where the first terms represent the SM contributions with 
$F^i_{\rm SM}=F^i(x_t)$ given explicitly in (\ref{S0})--(\ref{DP0}).

New physics can contribute to our master formula in two ways. It can 
modify the importance of a given operator, present already in the SM, 
through the new short distance functions $F^i_{\rm New}$ that depend on 
the new 
parameters in the extensions of the SM like the masses of charginos, 
squarks, charged Higgs particles and $\tan\beta=v_2/v_1$ in the MSSM. 
These new 
particles enter the new box and penguin diagrams. In more complicated 
extensions of the SM new operators (Dirac structures) that are either 
absent or very strongly suppressed in the SM, can become important. 
Their contributions are described by the second sum in 
(\ref{master}) with 
$B_k^{\rm New}, \lbrack\eta^k_{\rm QCD}\rbrack^{\rm New}, V^k_{\rm New}, 
G^k_{\rm New}$
being analogs of the corresponding objects in the first sum of the master 
formula. The $V^k_{\rm New}$ show explicitly that the second sum describes 
generally new sources of flavour and CP violation beyond the CKM matrix. 
This sum may, however, also include contributions governed by the CKM 
matrix that are strongly suppressed in the SM but become important in 
some extensions of the SM. In this case $V^k_{\rm New}=V^k_{\rm CKM}$. 
Clearly the new functions $F^i_{\rm New}$ and $G^k_{\rm New}$ as well as the 
factors $V^k_{\rm New}$ may depend on new CP violating phases complicating 
considerably phenomenological analyses.

\subsection{Classification of New Physics}
Classification of new physics (NP) contributions can be done in various ways. 
Having at hand the formula (\ref{master})  let us classify
these contributions from the point
of view of the operator structure of the effective weak Hamiltonian,
the complex phases present in the Wilson coefficients of the
relevant operators and the distinction whether the flavour changing
transitions are governed by the CKM matrix or by new sources of
flavour violation \cite{Erice,Dubro}.
For the first four classes below we assume that there are only three
generations of quarks and leptons. The last class allows for
more generations.

{\bf Class A}

This is the simplest class to which also the SM belongs.
In this class there are no new complex phases and flavour changing 
transitions are  governed  by the CKM matrix.
Moreover, the only relevant operators are those that are relevant in the SM.
Consequently NP enters only through the Wilson coefficients of the SM 
operators that can receive new contributions
  through diagrams involving new internal particles.

The models with these properties will be called  
Minimal Flavour Violation (MFV) models, as defined in 
\cite{UUT}. Other definitions can be found in \cite{AMGIISST,BOEWKRUR}. 
In this case  our master formula simplifies to 
\be\label{mmaster}
{\rm A(Decay)}= \sum_i B_i \eta^i_{\rm QCD}V^i_{\rm CKM} 
 F_i(v),
\qquad
 F_i=F^i_{\rm SM}+F^i_{\rm New}~~{\rm (real),}
\ee 
where $F_i(v)$ are the {\it  master functions} of MFV models \cite{Zakopane}
\be\label{masterf}
S(v),~X(v),~Y(v),~Z(v),~E(v),~ D'(v),~ E'(v)
\ee
with $v$ denoting collectively the parameters of a
given MFV model. A very detailed account of the MFV can be found in
\cite{Zakopane}. In Section~\ref{MFVM}
 some of its main features will be recalled.
Examples of models in this class are the Two Higgs Doublet Model II and 
the  MSSM without new sources of flavour violation and for
 $\tan\beta$ not too large. Also models with one extra universal 
dimension and the simplest little Higgs models are of MFV type.

We have the following correspondence between the most interesting FCNC
processes and the master functions in the MFV models \cite{Zakopane,BH92}:
\begin{center}
\begin{tabular}{lcl}
$K^0-\bar K^0$-mixing ($\varepsilon_K$) 
&\qquad\qquad& $S(v)$ \\
$B_{d,s}^0-\bar B_{d,s}^0$-mixing ($\Delta M_{s,d}$) 
&\qquad\qquad& $S(v)$ \\
$K \to \pi \nu \bar\nu$, $B \to X_{d,s} \nu \bar\nu$ 
&\qquad\qquad& $X(v)$ \\
$K_{\rm L}\to \mu \bar\mu$, $B_{d,s} \to l\bar l$ &\qquad\qquad& $Y(v)$ \\
$K_{\rm L} \to \pi^0 e^+ e^-$ &\qquad\qquad& $Y(v)$, $Z(v)$, 
$E(v)$ \\
$\varepsilon'$, Nonleptonic $\Delta S=1$ &\qquad\qquad& $X(v)$,
$Y(v)$, $Z(v)$,
$E(v)$ \\
Nonleptonic $\Delta B=1$ &\qquad\qquad& $X(v)$,
$Y(v)$, $Z(v)$,
$E(v)$,$E'(v)$ \\
$B \to X_s \gamma$ &\qquad\qquad& $D'(v)$, $E'(v)$ \\
$B \to X_s~{\rm gluon}$ &\qquad\qquad& $E'(v)$ \\
$B \to X_s l^+ l^-$ &\qquad\qquad&
$Y(v)$, $Z(v)$, $E(v)$, $D'(v)$, $E'(v)$
\end{tabular}
\end{center}

This table means that the observables like branching ratios, mass differences
$\Delta M_{d,s}$ in $B_{d,s}^0-\bar B_{d,s}^0$-mixing and the CP violation 
parameters $\varepsilon$ und $\varepsilon'$, all can be to a very good 
approximation  expressed in
terms of the corresponding master functions and the relevant CKM factors. The
remaining entries in the relevant formulae for these observables are low 
energy parameters such as the parameters $B_i$ that can be calculated within 
the SM and the QCD factors $\eta^i_{\rm QCD}$ describing the renormalization
group evolution of operators for scales $\mu\le M_W$. These factors 
being universal can be calculated, similarly to $B_i$, in the SM. The 
remaining, model specific QCD corrections can be absorbed in the functions 
$F_i$. Further simplifications are discussed in \cite{Zakopane}.

The formulae for the processes listed above in the SM, given in terms of 
the master 
functions and CKM factors can be found in many papers.  The full list using 
the same notation is given in \cite{BBL}. An update of these formulae 
with additional references is given in two papers on universal extra
dimensions \cite{BSW02,BPSW}, where one has to replace $F_i(x_t,1/R)$ by   
$F_i(v)$ to obtain the formulae in a general MFV model.
The supersymmetric contributions to the functions $S$, $X$,
$Y$, $Z$ and $E$ within the  MSSM with minimal
flavour violation are compiled
in \cite{BRMSSM}. See also \cite{MFV0,BERTOL,MW96}, where the remaining 
functions can be found. The QCD corrections to these functions can be found 
in \cite{BJW90,BB1,BB2,BB98,MU98,BGH,QCDMSSM,Bobeth,BSGAMMA,BBE04}. 
The full set of $F_i(v)$ in the SM with one extra universal dimension
is given in \cite{BSW02,BPSW}. The functions $S$, $X$ and $Y$ in the 
littlest Higgs Model can be found in \cite{BPU04,DEC05,BPU05}.

{\bf Class B}

This class of models differs from class A through the contributions
of new operators not present in the SM. It is assumed, however,
that no new complex phases
beyond the CKM phase are present. We have then
\be\label{masterB}
{\rm A(Decay)} = \sum_i B_i \eta^i_{\rm QCD}V^i_{\rm CKM} 
\lbrack F^i_{\rm SM}+F^i_{\rm New}\rbrack   +
\sum_k B_k^{\rm New} \lbrack\eta^k_{\rm QCD}\rbrack^{\rm New} V^k_{\rm CKM} 
\lbrack G^k_{\rm New}\rbrack\,
\ee
with all the functions $F^i_{\rm SM}$, $F^i_{\rm New}$ and $G^k_{\rm New}$ 
being {\rm real}. 
Typical examples of new Dirac structures are the 
operators $(V-A)\otimes(V+A)$, 
$(S-P)\otimes (S\pm P)$ and 
$\sigma_{\mu\nu} (S-P) \otimes \sigma^{\mu\nu} (S-P)$ contributing to 
$B_{d,s}^0-\bar B^0_{d,s}$ mixings that become relevant in the MSSM 
with a large $\tan\beta $. 
A subset of relevant references can be found in 
\cite{NLONEWOPS,BabuKolda,BCRS,Dedes,Foster,Kolda04}. 

{\bf Class C}

This class of models differs from class A through the presence of
new complex phases in the Wilson coefficients of the usual SM
operators. Contributions of new operators can
be, however, neglected.  An example is the
MSSM with  a small $\tan\beta$  and with non-diagonal elements
in the squark mass matrices.
This class can be summarized by
\be\label{mmaster3}
{\rm A(Decay)}= \sum_i B_i \eta^i_{\rm QCD}V^i_{\rm CKM} 
 F_i(v)~,
\qquad
 F_i(v)~~{\rm (complex).}
\ee
A simple example of this class will be discussed in the context of $B\to\pi
K$ decays in Section~\ref{NPHASE}.

 {\bf Class D}

 Here we group models with new complex phases, new operators
 and new flavour changing contributions which are not governed
 by the CKM matrix. As now the amplitudes are given by the most general
 expression (\ref{master}), the phenomenology in this class of models
 is more involved than in the classes A--C \cite{MPR,GGMS}.
  Examples of models in class D are multi-Higgs models 
 with complex phases in the  Higgs sector, general SUSY models, 
 models with spontaneous
 CP violation and left-right symmetric  models.

 {\bf Class E}

 Here we group  models in which
the unitarity of the three generation CKM matrix does not
hold. 
Examples are four generation models and models with tree
level FCNC transitions. If this type of physics is present,
the unitarity triangle does not close. 

We observe that the structure of weak decays beyond the SM could be very
rich. However, until today, there are no fully convincing signs in the
existing data for any contributions beyond the SM. Exceptions could be some
hints for new physics seen in the $B\to\pi K$ data with possible 
spectacular implications for rare $K$ and $B$ decays. They will be discussed 
in Section~\ref{NPHASE}. Also the decays $B\to\phi K_S$, $B\to\omega K_S$ 
 and $B\to \eta' K$
give some signs of new physics but we should wait until the data improve.

\section{Particle-Antiparticle Mixing and Various Types\\ of CP
Violation}
        \label{PPCP}
\setcounter{equation}{0}
\subsection{Preliminaries}
Let us next discuss the formalism of particle--antiparticle mixing
and CP violation. Much more elaborate discussion can be found in
two books \cite{Branco,Bigi}. We will concentrate here on
$K^0-\bar K^0$ mixing, $B_{d,s}^0-\bar B^0_{d,s}$ mixings and
CP violation in $K$-meson and $B$-meson decays. 
Due to GIM mechanism \cite{GIM}
 the phenomena discussed in this section
appear first at the one--loop level and as such they are
sensitive measures of the top quark couplings $V_{ti}(i=d,s,b)$ and 
in particular of the phase $\delta=\gamma$.
They allow then to construct the unitarity triangle as explicitly
demonstrated in Section~\ref{UT-Det}.

\begin{figure}[hbt]
\vspace{0.10in}
\centerline{
\epsfysize=1.5in
\epsffile{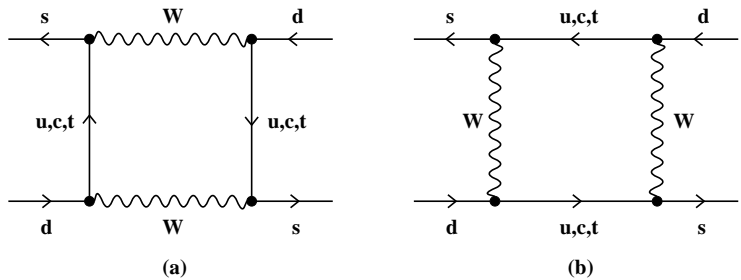}
}
\vspace{0.08in}
\caption[]{Box diagrams contributing to $K^0-\bar K^0$ mixing
in the SM.
\label{L:9}}
\end{figure}

\subsection{Express Review of $K^0-\bar K^0$ Mixing}
$K^0=(\bar s d)$ and $\bar K^0=(s\bar d)$ are flavour eigenstates which 
in the SM
may mix via weak interactions through the box diagrams in Fig.
\ref{L:9}.
We will choose the phase conventions so that 
\be
CP|K^0\rangle=-|\bar K^0\rangle, \qquad   CP|\bar K^0\rangle=-|K^0\rangle.
\ee

In the absence of mixing the time evolution of $|K^0(t)\rangle$ is
given by
\be
|K^0(t)\rangle=|K^0(0)\rangle \exp(-iHt)~, 
\qquad H=M-i\frac{\Gamma}{2}~,
\ee
where $M$ is the mass and $\Gamma$ the width of $K^0$. A similar formula
exists for $\bar K^0$.

On the other hand, in the presence of flavour mixing the time evolution 
of the $K^0-\bar K^0$ system is described by
\be\label{SCH}
i\frac{d\psi(t)}{dt}=\hat H \psi(t) \qquad  
\psi(t)=
\left(\begin{array}{c}
|K^0(t)\rangle\\
|\bar K^0(t)\rangle
\end{array}\right)
\ee
where
\be
\hat H=\hat M-i\frac{\hat\Gamma}{2}
= \left(\begin{array}{cc} 
M_{11}-i\frac{\Gamma_{11}}{2} & M_{12}-i\frac{\Gamma_{12}}{2} \\
M_{21}-i\frac{\Gamma_{21}}{2}  & M_{22}-i\frac{\Gamma_{22}}{2}
    \end{array}\right)
\ee
with $\hat M$ and $\hat\Gamma$ being hermitian matrices having positive
(real) eigenvalues in analogy with $M$ and $\Gamma$. $M_{ij}$ and
$\Gamma_{ij}$ are the transition matrix elements from virtual and physical
intermediate states respectively.
Using
\be
M_{21}=M^*_{12}~, \qquad 
\Gamma_{21}=\Gamma_{12}^*~,\quad\quad {\rm (hermiticity)}
\ee
\be
M_{11}=M_{22}\equiv M~, \qquad \Gamma_{11}=\Gamma_{22}\equiv\Gamma~,
\quad {\rm (CPT)}
\ee
we have
\be\label{MM12}
\hat H=
 \left(\begin{array}{cc} 
M-i\frac{\Gamma}{2} & M_{12}-i\frac{\Gamma_{12}}{2} \\
M^*_{12}-i\frac{\Gamma^*_{12}}{2}  & M-i\frac{\Gamma}{2}
    \end{array}\right)~.
\ee

Diagonalizing (\ref{SCH}) we find:

{\bf Eigenstates:}
\be\label{KLS}
K_{L,S}=\frac{(1+\bar\varepsilon)K^0\pm (1-\bar\varepsilon)\bar K^0}
        {\sqrt{2(1+\mid\bar\varepsilon\mid^2)}}
\ee
where $\bar\varepsilon$ is a small complex parameter given by
\be\label{bare3}
\frac{1-\bar\varepsilon}{1+\bar\varepsilon}=
\sqrt{\frac{M^*_{12}-i\frac{1}{2}\Gamma^*_{12}}
{M_{12}-i\frac{1}{2}\Gamma_{12}}}=
\frac{2 M^*_{12}-i\Gamma^*_{12}}{\Delta M-i\frac{1}{2}\Delta\Gamma}
\equiv r\exp(i\kappa)~.
\ee
with $\Delta\Gamma$ and $\Delta M$ given below.

{\bf Eigenvalues:}
\be
M_{L,S}=M\pm \RE Q~,  \qquad \Gamma_{L,S}=\Gamma\mp 2 \IM Q
\ee
where
\be
Q=\sqrt{(M_{12}-i\frac{1}{2}\Gamma_{12})(M^*_{12}-i\frac{1}{2}\Gamma^*_{12})}.
\ee
Consequently we have
\be\label{deltak}
\Delta M= M_L-M_S = 2\RE Q~,
\quad\quad
\Delta\Gamma=\Gamma_L-\Gamma_S=-4 \IM Q.
\ee

It should be noted that the mass eigenstates $K_S$ and $K_L$ differ from 
the CP eigenstates
\begin{equation}
K_1={1\over{\sqrt 2}}(K^0-\bar K^0),
  \qquad\qquad CP|K_1\rangle=|K_1\rangle~,
\end{equation}
\begin{equation}
K_2={1\over{\sqrt 2}}(K^0+\bar K^0),
  \qquad\qquad CP|K_2\rangle=-|K_2\rangle~,
\end{equation}
by 
a small admixture of the
other CP eigenstate:
\begin{equation}
K_{\rm S}={{K_1+\bar\varepsilon K_2}
\over{\sqrt{1+\mid\bar\varepsilon\mid^2}}},
\qquad
K_{\rm L}={{K_2+\bar\varepsilon K_1}
\over{\sqrt{1+\mid\bar\varepsilon\mid^2}}}\,.
\end{equation}

Since $\bar\varepsilon$ is $\ord(10^{-3})$, one has
 to a very good approximation:
\be\label{deltak1}
\Delta M_K = 2 \RE M_{12}, \qquad \Delta\Gamma_K=2 \RE \Gamma_{12}~,
\ee
where we have introduced the subscript $K$ to stress that these formulae apply
only to the $K^0-\bar K^0$ system.

The 
$K_{\rm L}-K_{\rm S}$
mass difference is experimentally measured to be \cite{PDG}
\begin{equation}\label{DMEXP}
\Delta M_K=M(K_{\rm L})-M(K_{\rm S}) = 
(3.490\pm 0.006) \cdot 10^{-15} \gev\,.
\end{equation}
In the SM roughly $80\%$ of the measured $\Delta M_K$
is described by the real parts of the box diagrams with charm quark
and top quark exchanges, whereby the contribution of the charm exchanges
is by far dominant. 
The remaining $20 \%$ of the measured $\Delta M_K$ is attributed to long 
distance contributions which are difficult to estimate \cite{GERAR}.
Further information with the relevant references can be found in 
\cite{HNa}.
The situation with $\Delta \Gamma_K$ is rather different.
It is fully dominated by long distance effects. Experimentally
one has $\Delta\Gamma_K\approx-2 \Delta M_K$.

Generally to observe CP violation one needs an interference between
various amplitudes that carry complex phases. As these phases are
obviously convention dependent, the CP-violating effects depend only
on the differences of these phases. 
In particular
the parameter $\bar\varepsilon$  depends on the 
phase convention
chosen for $K^0$ and $\bar K^0$. Therefore it may not 
be taken as a physical measure of CP violation.
On the other hand $\RE~\bar\varepsilon$ and $r$, defined in
(\ref{bare3})  are independent of
phase conventions. In fact the departure of $r$ from 1
measures CP violation in the $K^0-\bar K^0$ mixing:
\be
r=1+\frac{2 |\Gamma_{12}|^2}{4 |M_{12}|^2+|\Gamma_{12}|^2}
    \IM\left(\frac{M_{12}}{\Gamma_{12}}\right)
\approx 1-\IM\left(\frac{\Gamma_{12}}{M_{12}}\right)~.
\ee

This type of CP violation can be best isolated in semi-leptonic
decays of the $ K_L$ meson. The non-vanishing
asymmetry $a_{\rm SL}(K_L)$:
\be\label{ASLK}
\frac{\Gamma(K_L\to \pi^-e^+\nu_e )-
                 \Gamma( K_L\to \pi^+e^-\bar\nu_e )}
{\Gamma(K_L\to \pi^-e^+\nu_e )+
                 \Gamma( K_L\to \pi^+e^-\bar\nu_e )}
          = \left(\IM\frac{\Gamma_{12}}{M_{12}}\right)_K
          = 2 \RE \bar\varepsilon
\ee
signals this type of CP violation.
Note that $a_{\rm SL}(K_L)$ is determined purely by the quantities
related to  $K^0-\bar K^0$ mixing. 
Specifically, it measures
the difference between the phases of $\Gamma_{12}$ and
$M_{12}$.

That a non--vanishing $a_{\rm SL}(K_L)$ is indeed a signal
of CP violation can also be understood in the following
manner. $K_L$, that should be a CP eigenstate $K_2$ in the case
of CP conservation, decays into CP conjugate final states with
different rates. As $\RE \bar\varepsilon>0$, $K_L$ prefers slightly
to decay into $\pi^-e^+\nu_e$ than $\pi^+e^-\bar\nu_e$.
This would not be possible in a CP-conserving world.

\subsection{The First Look at $\varepsilon$ and $\varepsilon'$}
Since two pion final states, $\pi^+\pi^-$ and $\pi^0\pi^0$, are CP even 
while the three pion final state $3\pi^0$ is CP
odd, $K_{\rm S}$ and $K_{\rm L}$ decay to $2\pi$ and $3\pi^0$, 
respectively
via the following CP-conserving decay modes:
\begin{equation}
K_{\rm L}\to 3\pi^0 {\rm ~~(via~K_2),}\qquad K_{\rm S}\to 2 
\pi {\rm ~~(via~K_1).}
\end{equation}
Moreover, $K_{\rm L}\to \pi^+\pi^-\pi^0$ is also CP conserving provided 
the orbital angular momentum of $\pi^+\pi^-$ is even. 
This difference between $K_L$ and $K_S$ decays 
is responsible for the large disparity in their
life-times. A factor of 579. 
However, $K_{\rm L}$ and $K_{\rm S}$ are not CP eigenstates and 
may decay with small branching fractions as follows:
\begin{equation}
K_{\rm L}\to 2\pi {\rm ~~(via~K_1),}\qquad K_{\rm S}\to 3 
\pi^0 {\rm ~~(via~K_2).}
\end{equation}
This violation of CP is called {\it indirect} as it
proceeds not via explicit breaking of the CP symmetry in 
the decay itself but via the admixture of the CP state with opposite 
CP parity to the dominant one.
 The measure for this
indirect CP violation is defined as (I=isospin)
\begin{equation}\label{ek}
\varepsilon
\equiv {{A(K_{\rm L}\rightarrow(\pi\pi)_{I=0}})\over{A(K_{\rm 
S}\rightarrow(\pi\pi)_{I=0})}}.
\end{equation}
Note that the decay $K_{\rm S}\to \pi^+\pi^-\pi^0$ is  
CP violating (conserving) if
the orbital angular momentum of $\pi^+\pi^-$ is even (odd). 

Following the derivation in \cite{CHAU83} one finds
\begin{equation}
\eps = \bar\varepsilon+i\xi= \frac{\exp(i \pi/4)}{\sqrt{2} \Delta M_K} \,
\left( \IM M_{12} + 2 \xi \RE M_{12} \right),
\quad\quad
\xi = \frac{\IM A_0}{\RE A_0}.
\label{eq:epsdef}
\end{equation}
The phase convention dependence of $\xi$ cancels
the one of $\bar\varepsilon$ so that $\varepsilon$
is free from this dependence. The isospin amplitude $A_0$ is defined
below.

The important point in the definition (\ref{ek}) is that only the
transition to $(\pi\pi)_{I=0}$ enters. The transition to
$(\pi\pi)_{I=2}$ is absent. This allows to remove a certain type
of CP violation that originates in decays only. Yet as 
$\varepsilon\not=\bar\varepsilon$ and only $\RE\varepsilon=
\RE\bar\varepsilon$, it is clear that $\varepsilon$ includes
a type of CP violation represented by $\IM\varepsilon$ which is
absent in the semileptonic asymmetry (\ref{ASLK}). We will
identify this type of CP violation in Section~\ref{CDECAYS}, where a more
systematic classification of different types of CP violation
will be given.

While {\it indirect} CP violation reflects the fact that the mass
eigenstates are not CP eigenstates, so-called {\it direct}
CP violation is realized via a 
direct transition of a CP odd to a CP even state: $K_2\to \pi\pi$.
A measure of such a direct CP violation in $K_L\to \pi\pi$ is characterized
by a complex parameter $\varepsilon'$  defined as
\be\label{eprime0}
\varepsilon'\equiv\frac{1}{\sqrt{2}}\left(\frac{A_{2,L}}{A_{0,S}}-
\frac{A_{2,S}}{A_{0,S}}\frac{A_{0,L}}{A_{0,S}}\right)
\ee
where $A_{I,L}\equiv A(K_L\to (\pi\pi)_I)$
and $A_{I,S}\equiv A(K_S\to (\pi\pi)_I)$.

This time the transitions to $(\pi\pi)_{I=0}$ and
$(\pi\pi)_{I=2}$ are included which allows to study CP violation in
the decay itself. We will discuss this issue in general terms
in Section~\ref{CDECAYS}. It is useful to cast (\ref{eprime0})
into 
\be\label{eprime}
\varepsilon'=\frac{1}{\sqrt{2}}\IM\left(\frac{A_2}{A_0}\right)
              \exp(i\Phi_{\varepsilon'}), 
   \qquad \Phi_{\varepsilon'}=\frac{\pi}{2}+\delta_2-\delta_0, 
\ee
where
the isospin amplitudes $A_I$ in $K\to\pi\pi$
decays are introduced through
\begin{equation}\label{ISO1} 
A(K^+\rightarrow\pi^+\pi^0)=\sqrt{3\over 2} A_2 e^{i\delta_2}~,
\end{equation}
\begin{equation}\label{ISO2}
A(K^0\rightarrow\pi^+\pi^-)=\sqrt{2\over 3} A_0 e^{i\delta_0}+ \sqrt{1\over
3} A_2 e^{i\delta_2}~,
\end{equation}
\begin{equation}\label{ISO3}
A(K^0\rightarrow\pi^0\pi^0)=\sqrt{2\over 3} A_0 e^{i\delta_0}-2\sqrt{1\over
3} A_2 e^{i\delta_2}\,.
\end{equation} 
Here the subscript $I=0,2$ denotes states with isospin $0,2$
equivalent to $\Delta I=1/2$ and $\Delta I = 3/2$ transitions,
respectively, and $\delta_{0,2}$ are the corresponding strong phases. 
The weak CKM phases are contained in $A_0$ and $A_2$.
The isospin amplitudes $A_I$ are complex quantities which depend on
phase conventions. On the other hand, $\varepsilon'$ measures the 
difference between the phases of $A_2$ and $A_0$ and is a physical
quantity.
The strong phases $\delta_{0,2}$ can be extracted from $\pi\pi$ scattering. 
Then $\Phi_{\varepsilon'}\approx \pi/4$. See \cite{Meissner} for more details.

Experimentally $\varepsilon$ and $\varepsilon'$
can be found by measuring the ratios
\begin{equation}
\eta_{00}={{A(K_{\rm L}\to\pi^0\pi^0)}\over{A(K_{\rm S}\to\pi^0\pi^0)}},
            \qquad
  \eta_{+-}={{A(K_{\rm L}\to\pi^+\pi^-)}\over{A(K_{\rm S}\to\pi^+\pi^-)}}.
\end{equation}

Indeed, assuming $\varepsilon$ and $\varepsilon'$ to be small numbers one
finds
\be
\eta_{00}=\varepsilon-{{2\varepsilon'}\over{1-\sqrt{2}\omega}}
            ,~~~~
  \eta_{+-}=\varepsilon+{{\varepsilon'}\over{1+\omega/\sqrt{2}}},
\end{equation}
where $\omega=\RE A_2/\RE A_0=0.045$.
In the absence of direct CP violation $\eta_{00}=\eta_{+-}$.
The ratio ${\varepsilon'}/{\varepsilon}$  can then be measured through
\begin{equation}\label{BASE}
\RE(\epe)=\frac{1}{6(1+\omega/\sqrt{2})}
\left(1-\left|{{\eta_{00}}\over{\eta_{+-}}}\right|^2\right)~.
\end{equation}

\subsection{Basic Formula for $\eps$}
            \label{subsec:epsformula}
With all this information at hand one can derive a formula for $\varepsilon$
which can be efficiently used in phenomenological applications.
As this derivation has been presented in detail in \cite{Erice}, we will be 
very brief here.

Calculating the box diagrams of Fig.~\ref{L:9} and including
leading and next-to-leading QCD corrections one finds
\begin{equation}
M_{12} = \frac{G_{\rm F}^2}{12 \pi^2} F_K^2 \hat B_K m_K \mw^2
\left[ {\lambda_c^*}^2 \eta_1 S_0(x_c) + {\lambda_t^*}^2 \eta_2 S_0(x_t) +
2 {\lambda_c^*} {\lambda_t^*} \eta_3 S_0(x_c, x_t) \right],
\label{eq:M12K}
\end{equation}
where $F_K=160~\mev$ is the $K$-meson decay constant and $m_K$
the $K$-meson mass. 
Next, the renormalization group 
invariant parameter $\hat B_K$ is defined by
\cite{BJW90,BSS}
\begin{equation}
\hat B_K = B_K(\mu) \left[ \alpha_s^{(3)}(\mu) \right]^{-2/9} \,
\left[ 1 + \frac{\alpha_s^{(3)}(\mu)}{4\pi} J_3 \right]~,
\label{eq:BKrenorm}
\end{equation}
\begin{equation}
\langle \bar K^0| (\bar s d)_{V-A} (\bar s d)_{V-A} |K^0\rangle
\equiv \frac{8}{3} B_K(\mu) F_K^2 m_K^2,
\label{eq:KbarK}
\end{equation}
where $\alpha_s^{(3)}$ is the strong coupling constant
in an effective three flavour theory and $J_3=1.895$ in the NDR scheme 
\cite{BJW90}. The CKM factors are given by $\lambda_i = V_{is}^* V_{id}^{}$ 
and the functions $S_0$  by ($x_i=m^2_i/\mw^2$) \cite{BSS}
\begin{equation}\label{S0SM}
 S_0(x_t)=2.46~\left(\frac{\mt}{170\gev}\right)^{1.52},
\quad\quad S_0(x_c)=x_c,
\ee
\begin{equation}\label{BFF}
S_0(x_c, x_t)=x_c\left[\ln\frac{x_t}{x_c}-\frac{3x_t}{4(1-x_t)}-
 \frac{3 x^2_t\ln x_t}{4(1-x_t)^2}\right].
\end{equation}

Short-distance NLO QCD effects are described through the correction
factors $\eta_1$, $\eta_2$, $\eta_3$ \cite{HNa,BJW90,HNb,Nierste}:
\begin{equation}
\eta_1=(1.32\pm 0.32) \left[\frac{1.30\gev}{m_c(m_c)}\right]^{1.1},\quad
\eta_2=0.57\pm 0.01,\quad
  \eta_3=0.47\pm0.05~.
\end{equation}

To proceed further we neglect the last term in (\eqn{eq:epsdef}) as it
 constitutes at most a 5\,\% correction to $\eps$. A recent discussion 
can be found in \cite{Vysotsky}. 
This is justified
in view of other uncertainties, in particular those connected with
$\hat B_K$.
Inserting (\eqn{eq:M12K}) into (\eqn{eq:epsdef}) we find
\begin{equation}
\eps=C_{\eps} \hat B_K \IM\lambda_t \left\{
\RE\lambda_c \left[ \eta_1 S_0(x_c) - \eta_3 S_0(x_c, x_t) \right] -
\RE\lambda_t \eta_2 S_0(x_t) \right\} e^{i \pi/4}\,,
\label{eq:epsformula}
\end{equation}
where the numerical constant $C_\eps$ is given by
\begin{equation}
C_\eps = \frac{G_{\rm F}^2 F_K^2 m_K \mw^2}{6 \sqrt{2} \pi^2 \Delta M_K}
       = 3.837 \cdot 10^4 \, .
\label{eq:Ceps}
\end{equation}
Comparing (\eqn{eq:epsformula}) with the
experimental value for $\eps$ \cite{PDG}
\begin{equation}\label{eexp}
\varepsilon_{exp}
=(2.280\pm0.013)\cdot10^{-3}\;\exp{i\Phi_{\varepsilon}},
\qquad \Phi_{\varepsilon}={\pi\over 4},
\end{equation}
one obtains a constraint on the unitarity triangle in 
Fig.~\ref{fig:utriangle}. 
See Section~\ref{UT-Det}.

\subsection{Express Review of $B_{d,s}^0$-$\bar B_{d,s}^0$ Mixing}
The flavour eigenstates in this case are
\be\label{fl}
B^0_d=(\bar bd),\qquad
\bar B^0_d=(b \bar d),\qquad
B^0_s=(\bar bs),\qquad
\bar B^0_s=( b \bar s)~.
\ee
They mix via the box diagrams in Fig.~\ref{L:9} with $s$ replaced
by $b$ in the case of $B_{d}^0$-$\bar B_{d}^0$ mixing.
In the case of $B_{s}^0$-$\bar B_{s}^0$ mixing also $d$ has to be replaced
by $s$.

Dropping the subscripts $(d,s)$ for a moment, it is customary to
denote the mass eigenstates by
\be\label{HL}
B_H=p B^0+q \bar B^0, \qquad B_L=p B^0-q \bar B^0,
\ee
\be\label{pq}
p=\frac{1+\bar\varepsilon_B}{\sqrt{2(1+|\bar\varepsilon_B|^2)}},
\qquad
q=\frac{1-\bar\varepsilon_B}{\sqrt{2(1+|\bar\varepsilon_B|^2)}},
\ee
with $\bar\varepsilon_B$ corresponding to $\bar\varepsilon$ in
the $K^0-\bar K^0$ system. Here ``H'' and ``L'' denote 
{\it Heavy} and {\it Light} respectively. As in the $B^0-\bar B^0$ 
system one has $\Delta\Gamma\ll\Delta M$, 
 it is more suitable to distinguish the mass eigenstates by their 
masses than by the corresponding life-times.

The strength of the $B^0_{d,s}-\bar B^0_{d,s}$ mixings
is described by the mass differences
\begin{equation}\label{DHL}
\Delta M_{d,s}= M_H^{d,s}-M_L^{d,s}~.
\end{equation}
In contrast to $\Delta M_K$, in this case the long distance contributions
are estimated to be very small and $\Delta M_{d,s}$ is very well
approximated by the relevant box diagrams. 
Moreover, due to $m_{u,c}\ll m_t$ 
only the top sector is relevant.

 $\Delta M_{d,s}$ can be expressed
in terms of the off-diagonal element in the neutral $B$-meson mass matrix
by using the formulae developed previously for the $K$-meson system.
One finds \cite{BSS}
\begin{equation}
\Delta M_q= 2 |M_{12}^{(q)}|, \qquad
\Delta \Gamma_q=2 \frac{\RE(M_{12}\Gamma_{12}^*)}{|M_{12}|} \ll\Delta M_q, \qquad
 q=d,s.
\label{eq:xdsdef}
\end{equation}
These formulae differ from (\ref{deltak1}) because in the
B-system $\bar\varepsilon_B$ cannot be neglected and $\Gamma_{12}\ll M_{12}$.

We also have
\be\label{q/p}
\frac{q}{p}=\sqrt{\frac{M^*_{12}-i\frac{1}{2}\Gamma^*_{12}}
{M_{12}-i\frac{1}{2}\Gamma_{12}}}=
\frac{2 M^*_{12}-i\Gamma^*_{12}}{\Delta M-i\frac{1}{2}\Delta\Gamma}
=\frac{M_{12}^*}{|M_{12}|}
\left[1-\frac{1}{2}\IM\left(\frac{\Gamma_{12}}{M_{12}}\right)\right]
\ee
where higher order terms in the small quantity $\Gamma_{12}/M_{12}$
have been neglected.

As $\IM(\Gamma_{12}/M_{12})< \ord(10^{-3})$,
\bi
\item
The semileptonic asymmetry $a_{\rm SL}(B)$ discussed a few
pages below is even smaller than $a_{\rm SL}(K_L)$. Typically 
$\ord(10^{-4})$. These are bad news.
\item
The ratio $q/p$ is a pure phase to an excellent approximation.
These are very good news as we will see below.
\ei

Calculating the relevant box diagrams we find
\begin{equation}\label{M12Q}
(M_{12})_q = \frac{G_F^2}{12\pi^2}F_{B_q}^2
\hat B_{B_q} m_{B_q} M_W^2 (V_{tq}^*V_{tb})^2 S_0(x_t)
\eta_B,
\end{equation}
where $F_{B_q}$ is the $B_q$-meson decay constant, $\hat B_q$
renormalization group invariant parameters defined
in analogy to (\ref{eq:BKrenorm}) and (\ref{eq:KbarK}) and $\eta_B$ stands 
for short distance QCD corrections \cite{BJW90,UKJS}
\begin{equation}
\eta_B=0.55\pm0.01.
\end{equation}
Consequently,
\be
(M_{12}^*)_d \propto (V_{td}V_{tb}^*)^2~,
\qquad
(M_{12}^*)_s \propto (V_{ts}V_{tb}^*)^2~.
\ee
Now, from Section~\ref{CKMM} we know that
\be
V_{td}=\vtd e^{-i\beta}, \qquad
V_{ts}=-\vts e^{-i\beta_s}
\ee
with $\beta_s=\ord(10^{-2})$. Consequently to an excellent approximation 
\be\label{pureph}
\left(\frac{q}{p}\right)_{d,s}= e^{i2\phi_M^{d,s}},
\qquad
\phi^d_M=-\beta, \qquad \phi^s_M=-\beta_s,
\ee
with $\phi_M^{d,s}$ given entirely by the weak phases in the
CKM matrix.

\subsection{Basic Formulae for $\Delta M_{d,s}$}
            \label{subsec:BBformula}
From (\ref{eq:xdsdef}) and (\ref{M12Q}) we have with $V_{tb}=1$
\begin{equation}
\Delta M_q = \frac{G_{\rm F}^2}{6 \pi^2} \eta_B m_{B_q} 
(\hat B_{B_q} F_{B_q}^2 ) \mw^2 S_0(x_t) |V_{tq}|^2,
\label{eq:xds}
\end{equation}
Using (\ref{S0SM}) and  (\ref{eq:xds}) we obtain two useful formulae
\begin{equation}\label{DMD}
\Delta M_d=
0.50/{\rm ps}\cdot\left[ 
\frac{\sqrt{\hat B_{B_d}}F_{B_d}}{230\mev}\right]^2
\left[\frac{\mtb(\mt)}{167\gev}\right]^{1.52} 
\left[\frac{\vtd}{7.8\cdot10^{-3}} \right]^2 
\left[\frac{\eta_B}{0.55}\right]  
\end{equation}
and
\begin{equation}\label{DMS}
\Delta M_{s}=
17.2/{\rm ps}\cdot\left[ 
\frac{\sqrt{\hat B_{B_s}}F_{B_s}}{260\mev}\right]^2
\left[\frac{\mtb(\mt)}{167\gev}\right]^{1.52} 
\left[\frac{\vts}{0.040} \right]^2
\left[\frac{\eta_B}{0.55}\right] \,.
\end{equation}

\subsection{Classification of CP Violation}\label{CDECAYS}
\subsubsection{Preliminaries}
We have mentioned in Section~\ref{THF} that due to the presence of hadronic
matrix elements, various decay amplitudes contain large theoretical
uncertainties. It is of interest to investigate which measurements
of CP-violating effects do not suffer from hadronic uncertainties.
To this end it is useful to make a classification of CP-violating
effects that is more transparent than the division into the
{\it indirect} and {\it direct} CP violation considered so far.
A nice detailed presentation has been given by Nir \cite{REV2}.

Generally complex phases may enter particle--antiparticle mixing
and the decay process itself. It is then natural to consider
three types of CP violation:
\bi
\item
CP Violation in Mixing
\item
CP Violation in Decay
\item
CP Violation in the Interference of Mixing and Decay
\ei

As the phases in mixing and decay are convention dependent,
the CP-violating effects depend only
on the differences of these phases. This is clearly seen in
the classification given below.

\subsubsection{CP Violation in Mixing}
This type of CP violation can be best isolated in semi-leptonic
decays of neutral $B$ and $K$ mesons. We have discussed the asymmetry
$a_{SL}(K_L)$ before. In the case of $B$ decays the non-vanishing
asymmetry $a_{SL}(B)$ (we suppress the indices $(d,s)$),
\be\label{ASLB}
\frac{\Gamma(\bar B^0(t)\to l^+\nu X)-
                 \Gamma( B^0(t)\to l^-\bar\nu X)}
                {\Gamma(\bar B^0(t)\to l^+\nu X)+
                 \Gamma( B^0(t)\to l^-\bar\nu X)}
            =\frac{1-|q/p|^4}{1+|q/p|^4}
          = \left(\IM\frac{\Gamma_{12}}{M_{12}}\right)_B
\ee
signals this type of CP violation. Here $\bar B^0(0)=\bar B^0$, 
$B^0(0)= B^0$. For $t\not=0$ the formulae analogous to 
(\ref{SCH}) should be used.
Note that the final states in (\ref{ASLB}) contain ``wrong charge''
leptons and can only be reached in the presence of $B^0-\bar B^0$
mixing. That is one studies effectively the difference between the
rates for $\bar B^0\to B^0\to l^+\nu X$
and $ B^0 \to \bar B^0 \to l^-\bar\nu X$. 
As the phases in
the transitions $B^0 \to \bar B^0$ and $\bar B^0 \to B^0$ 
differ from each other, a non-vanishing CP asymmetry follows.
Specifically $a_{\rm SL}(B)$ measures
the difference between the phases of $\Gamma_{12}$ and
$M_{12}$.

As $M_{12}$ and 
in particular $\Gamma_{12}$ suffer from large hadronic uncertainties,
no precise extraction of CP-violating phases from this type of CP
violation can be expected.  
Moreover as $q/p$ is almost a pure phase, see (\ref{q/p}) and 
(\ref{pureph}), the
asymmetry is very small and very difficult to measure.

\subsubsection{CP Violation in Decay}
This type of CP violation is best isolated in charged $B$ and charged $K$
decays as mixing effects do not enter here. However, it can also
be measured in neutral $B$ and $K$ decays. The relevant asymmetry is
given by
\be\label{ADECAY}
\caladir(B^\pm\to f^\pm)=\frac{\Gamma(B^+\to f^+)-\Gamma(B^-\to f^-)}
                          {\Gamma(B^+\to f^+)+\Gamma(B^-\to f^-)}
=\frac{1-|\bar A_{f^-}/A_{f^+}|^2}{1+| \bar A_{f^-}/A_{f^+}|^2}
\ee
where
\be\label{AH}
A_{f^+}=\langle f^+|{\cal H}^{\rm weak}| B^+\rangle,
\qquad
\bar A_{f^-}=\langle f^-|{\cal H}^{\rm weak}| B^-\rangle~.
\ee
For this asymmetry to be non-zero one needs at least two different 
contributions with different {\it weak} ($\phi_i$) and {\it strong}
($\delta_i$) phases. These could be for instance two tree diagrams,
two penguin diagrams or one tree and one penguin. Indeed writing the
decay amplitude $A_{f^+}$ and its CP conjugate $\bar A_{f^-}$ as
\be\label{AMPL}
A_{f^+}=\sum_{i=1,2} A_i e^{i(\delta_i+\phi_i)},
\qquad
\bar A_{f^-}=\sum_{i=1,2} A_i e^{i(\delta_i-\phi_i)},
\ee
with $A_i$ being real, one finds 
\be\label{BDECAY}
\caladir(B^\pm\to f^\pm)=\frac{ -2 A_1 A_2 \sin(\delta_1-\delta_2)
\sin(\phi_1-\phi_2)}{A_1^2+A_2^2+2 A_1 A_2 \cos(\delta_1-\delta_2)
\cos(\phi_1-\phi_2)}~.
\ee
The sign of the strong phases $\delta_i$ is the same for $A_{f^+}$
and $\bar A_{f^-}$ because CP is conserved by strong interactions.
The weak phases have opposite signs. 

The presence of hadronic uncertainties in $A_i$ and 
of strong phases $\delta_i$ complicates the extraction of the
phases $\phi_i$ from data. An example of this type of
CP violation in $K$ decays is $\varepsilon'$. We will demonstrate
this below.
\subsubsection{CP Violation in the Interference of Mixing and Decay}
This type of CP violation is only possible in neutral $B$ and $K$
decays. We will use $B$ decays for illustration suppressing the
subscripts $d$ and $s$. Moreover, we set $\Delta\Gamma=0$. Formulae
with $\Delta\Gamma\not =0$ can be found in \cite{BF97,REV1}.

 Most interesting are the decays into final states which
are CP-eigenstates. Then a time dependent asymmetry defined by
\be\label{TASY}
\calaCP(t,f)=\frac{\Gamma(B^0(t)\to f)-
                 \Gamma( \bar B^0(t)\to f)}
                {\Gamma(B^0(t)\to f)+
                 \Gamma( \bar B^0(t)\to f)}
\ee
is given by
\begin{equation}\label{e8}
\calaCP(t,f)=
\caladir(f)\cos(\Delta M
t)+\calamix(f)\sin(\Delta M t)
\end{equation}
where we have separated the {\it decay} CP-violating contributions 
($\caladir$)
from those describing CP violation in the interference of
mixing and decay $(\calamix$). The latter type of CP violation is usually 
called the
{\it mixing-induced} CP violation. One has
\begin{equation}\label{e9}
\caladir(f)=\frac{1-\left\vert\xi_f\right\vert^2}
{1+\left\vert\xi_f\right\vert^2}\equiv C_f,
\quad
\calamix(f)=\frac{2\mbox{Im}\xi_f}{1+
\left\vert\xi_f\right\vert^2}\equiv -S_f~,
\end{equation}
where $C_f$ and $S_f$ are popular notations found in the literature.
Unfortunately, the signs in the literature differ from paper to 
paper and some papers interchange $B^0$ and $\bar B^0$ with respect 
to the one used by us in the definition of the asymmetry in (\ref{TASY}). 
 Therefore in Table~\ref{TABRECK}, provided by Stefan Recksiegel, 
we give the relations between various definitions.

\begin{table}[hbt]
\begin{center}
\begin{tabular}{|c||c|c|}
\hline
  These Lectures  &   $\caladir(f)$  &  $\calamix(f)$
 \\ \hline
  BaBar Book \cite{BABAR}  &   $ C_f$  & $S_f$ \\\hline
  BaBar  &  $C_f$ &  $-S_f$ \\\hline
Belle &   $-A_f$  &  $-S_f$ \\
\hline
\end{tabular}
\caption{Comparison between various definitions present in the literature.
}
\label{TABRECK}
\end{center}
\end{table}

The quantity $\xi_f$ containing  all the information
needed to evaluate the asymmetries (\ref{e9}) 
is given by
\begin{equation}\label{e11}
\xi_f=\frac{q}{p}\frac{A(\bar B^0\to f)}{A(B^0 \to f)}=
\exp(i2\phi_M)\frac{A(\bar B^0\to f)}{A(B^0 \to f)}
\end{equation}
with $\phi_M$, introduced in (\ref{pureph}), 
denoting the weak phase in the $B^0-\bar B^0$ mixing.
 $A(B^0 \to f)$ and $A(\bar B^0 \to f)$ are  decay amplitudes. 
The time dependence of $\calaCP(t,f)$ allows to extract
$\caladir$ and $\calamix$ as
coefficients of $\cos(\Delta M t)$ and $\sin(\Delta M t)$,
respectively.

Generally several decay mechanisms with different weak and
strong phases can contribute to $A(B^0 \to f)$. These are
tree diagram (current-current) contributions, QCD penguin
contributions and electroweak penguin contributions. If they
contribute with similar strength to a given decay amplitude
the resulting CP asymmetries suffer from hadronic uncertainties
related to matrix elements of the relevant operators $Q_i$.
The situation is then analogous to the class just discussed.
Indeed
\be\label{ratiocp}
\frac{A(\bar B^0\to f)}{A(B^0 \to f)}=-\eta_f
\left[\frac{A_T e^{i(\delta_T-\phi_T)}+A_P e^{i(\delta_P-\phi_P)}}
{A_T e^{i(\delta_T+\phi_T)}+A_P e^{i(\delta_P+\phi_P)}}\right]
\ee
with $\eta_f=\pm 1$ being the CP-parity of the final state,
depends on the strong phases $\delta_{T,P}$ and the hadronic matrix
elements present in $A_{T,P}$. Thus the measurement of the
asymmetry does not allow a clean determination of the weak
phases $\phi_{T,P}$. The minus sign in (\ref{ratiocp}) follows
from our CP phase convention $ CP |B^0\rangle= -|\bar B^0\rangle$,
 that has also been used in writing the phase factor in (\ref{e11}).
Only $\xi$ is phase convention independent. See Section 8.4.1 of 
\cite{BF97} for details.

An interesting case arises when a single mechanism dominates the 
decay amplitude or the contributing mechanisms have the same weak 
phases. Then the hadronic matrix elements and strong phases drop out and
\be\label{cp}
\frac{A(\bar B^0\to f)}{A(B^0 \to f)}=-\eta_f e^{-i2\phi_D}
\ee
is a pure phase with $\phi_D$ being the weak phase in 
$A(B^0 \to f)$.
Consequently
\begin{equation}\label{e111}
\xi_f=-\eta_f\exp(i2\phi_M) \exp(-i 2 \phi_D),
\qquad
\mid \xi_f \mid^2=1~.
\end{equation}
In this particular case 
$\caladir(f)=C_f$
vanishes and the CP asymmetry is given entirely
in terms of the weak phases $\phi_M$ and $\phi_D$:
\begin{equation}\label{simple}
\calaCP(t,f)= \calamix(f) \sin(\Delta Mt) \qquad
\calamix(f)=\IM\xi_f=\eta_f \sin(2\phi_D-2\phi_M)=-S_f~.
\end{equation}
Thus the corresponding measurement of weak phases is free from
hadronic uncertainties. A well known example is the decay
$B_d\to \psi K_S$. Here $\phi_M=-\beta$ and $\phi_D=0$. As
in this case $\eta_f=-1$,  we find
\begin{equation}
\calaCP(t,f)= -\sin(2\beta) \sin(\Delta Mt), \qquad S_f=\sin(2\beta)
\end{equation}
which allows a very clean measurement of the angle $\beta$ in the
unitarity triangle. We will discuss other examples in
Section~\ref{STRATEGIES}. 

We observe that the asymmetry $\calamix(f)$ measures directly the
difference between the phases of the $B^0-\bar B^0$-mixing $(2\phi_M)$
and of the decay amplitude $(2\phi_D)$. This tells us immediately 
that we are dealing with the interference of mixing and decay.
As $\phi_M$ and $\phi_D$
are phase convention dependent quantities, only their
difference is physical, it is
impossible to state on the basis of a single asymmetry 
whether CP violation takes place in the decay or in the mixing.
To this end at least two asymmetries for $B^0 (\bar B^0)$
decays to different final states $f_i$ have to be measured.
As $\phi_M$ does not depend on the final state, 
$\IM\xi_{f_1}\not=\IM\xi_{f_2}$ is a signal of CP violation
in the decay. 

We will see in Section~\ref{STRATEGIES} 
that the ideal situation presented above 
does not always take place and two or more different mechanisms with 
different weak and strong phases contribute to the CP asymmetry. One
finds then
\begin{equation}\label{e8a}
\calaCP(t,f)=C_f\cos(\Delta Mt)-S_f\sin(\Delta M t),
\end{equation}
\be\label{Cf}
C_f=-2 r \sin(\phi_1-\phi_2)\sin(\delta_1-\delta_2)~,
\ee
\be\label{Sf}
S_f=-\eta_f\left[\sin 2(\phi_1-\phi_M)+
2r \cos 2(\phi_1-\phi_M) \sin(\phi_1-\phi_2)\cos(\delta_1-\delta_2)\right]
\ee
where 
$r=A_2/A_1\ll 1$ has been assumed
and $\phi_i$ and $\delta_i$ are weak and strong phases, respectively.
For $r=0$ the previous formulae are obtained.

In the case of $K$ decays, this type of CP violation can be
cleanly measured in the rare decay $K_L\to\pi^0\nu\bar\nu$.
Here the difference between the weak phase in the $K^0-\bar K^0$
mixing and in the decay $\bar s \to \bar d \nu\bar\nu$ matters.
We will discuss this decay in Section~\ref{KPNN}.

We can now compare the two classifications of different types
of CP violation. CP violation in mixing is a manifestation
of indirect CP violation. CP violation in decay is a manifestation
of direct CP violation. The third type contains elements of both 
the indirect and direct CP
violation.

It is clear from this discussion that only in the case of the
third type of CP violation there are possibilities to measure directly weak
phases without hadronic uncertainties and moreover without invoking sophisticated 
methods. This takes place provided
a single mechanism (diagram) is responsible for the decay or the
contributing decay mechanisms have the same weak phases.
However, we will see in Section~\ref{STRATEGIES}  
that there are other strategies, involving 
also decays to CP non-eigenstates, that provide clean measurements of the 
weak phases.

\subsubsection{Another Look at $\varepsilon$ and $\varepsilon'$}
Let us finally investigate what type of CP violation is
represented by $\varepsilon$ and $\varepsilon'$.
Here instead of different mechanisms it is sufficient to talk
about different isospin amplitudes.

In the case of $\varepsilon$, CP violation in decay is not
possible as only the isospin amplitude $A_0$ is involved.
See (\ref{ek}). We also know that only $\RE~\varepsilon=
\RE~\bar\varepsilon$ is related to CP violation in mixing.
Consequently:
\bi
\item 
$\RE\varepsilon$ represents CP violation in mixing,
\item
$\IM\varepsilon$ represents CP violation in the
interference of mixing and decay.
\ei

In order to analyze the case of $\varepsilon'$ we use the formula
(\ref{eprime}) to find
\be\label{ree}
\RE\,\varepsilon'=-\frac{1}{\sqrt{2}}\left\vert\frac{A_2}{A_0}\right\vert
\sin(\phi_2-\phi_0)\sin(\delta_2-\delta_0)
\ee
\be\label{iee}
\IM\,\varepsilon'=\frac{1}{\sqrt{2}}\left\vert\frac{A_2}{A_0}\right\vert
\sin(\phi_2-\phi_0)\cos(\delta_2-\delta_0)~.
\ee  
Consequently:
\bi
\item 
$\RE~\varepsilon'$ represents CP violation in decay as it is only
non zero provided simultaneously $\phi_2\not=\phi_0$ and 
$\delta_2\not=\delta_0$.
\item
$\IM~\varepsilon'$ exists even for $\delta_2=\delta_0$ but as
it requires $\phi_2\not=\phi_0$ it represents CP violation in 
decay as well.
\ei
Experimentally  $\delta_2\not=\delta_0$.
Within the SM, $\phi_2$ and $\phi_0$ are connected with
electroweak penguins and QCD penguins, respectively.
We will briefly discuss the ratio $\epe$ at the end of Section~\ref{UT-Det}.

\section{Standard Analysis of the Unitarity Triangle (UT)}\label{UT-Det}
\setcounter{equation}{0}
\subsection{General Procedure}
After this general discussion of basic concepts let us concentrate on
the standard analysis of the Unitarity Triangle (see 
Fig.~\ref{fig:utriangle}) within the SM. 
A very detailed description of this analysis with the participation of 
the leading experimentalists and theorists in this field 
can be found in  \cite{CERNCKM}.
 
Setting $\lambda=\vus=0.224$, the analysis
proceeds in the following five steps:

{\bf Step 1:}

{}From  the $b\to c$ transition in inclusive and exclusive 
leading $B$-meson decays
one finds $\vcb$ and consequently the scale of the UT:
\begin{equation}
\vcb\quad \Longrightarrow\quad\lambda \vcb= \lambda^3 A~.
\end{equation}

{\bf Step 2:}

{}From  the $b\to u$ transition in inclusive and exclusive $B$ meson decays
one finds $\vub$ and consequently using (\ref{2.94}) 
the side $CA=R_b$ of the UT:
\begin{equation}\label{rb}
\left| \frac{V_{ub}}{V_{cb}} \right|
 \quad\Longrightarrow \quad R_b=\sqrt{\bar\varrho^2+\bar\eta^2}=
4.35 \cdot \left| \frac{V_{ub}}{V_{cb}} \right|~.
\end{equation}

{\bf Step 3:}

{}From the experimental value of $\varepsilon_K$ in (\ref{eexp})   
and the formula (\ref{eq:epsformula}) rewritten in terms of Wolfenstein 
parameters
one derives  
the constraint on $(\bar \varrho, \bar\eta)$ \cite{WARN}
\begin{equation}\label{100}
\bar\eta \left[(1-\bar\varrho) A^2 \eta_2 S_0(x_t)
+ P_c(\varepsilon) \right] A^2 \hat B_K = 0.187,
\end{equation}
where
\begin{equation}\label{102}
P_c(\varepsilon) = 
\left[ \eta_3 S_0(x_c,x_t) - \eta_1 x_c \right] \frac{1}{\lambda^4},
\qquad
x_i=\frac{m^2_i}{\mw^2}
\end{equation}
with all symbols defined in the previous Section and
$P_c(\varepsilon)=0.29\pm0.07$ \cite{Nierste} summarizing the contributions
of box diagrams with two charm quark exchanges and the mixed 
charm-top exchanges.  

As seen in Fig.~\ref{L:10}, equation (\ref{100}) specifies 
a hyperbola in the $(\bar \varrho, \bar\eta)$
plane.
The position of the hyperbola depends on $\mt$, $|V_{cb}|=A \lambda^2$
and $\hat B_K$. With decreasing $\mt$, $|V_{cb}|$ and $\hat B_K$ it
moves away from the origin of the
$(\bar\varrho,\bar\eta)$ plane. 

\begin{figure}[hbt]
  \vspace{0.10in} \centerline{
\begin{turn}{-90}
  \mbox{\epsfig{file=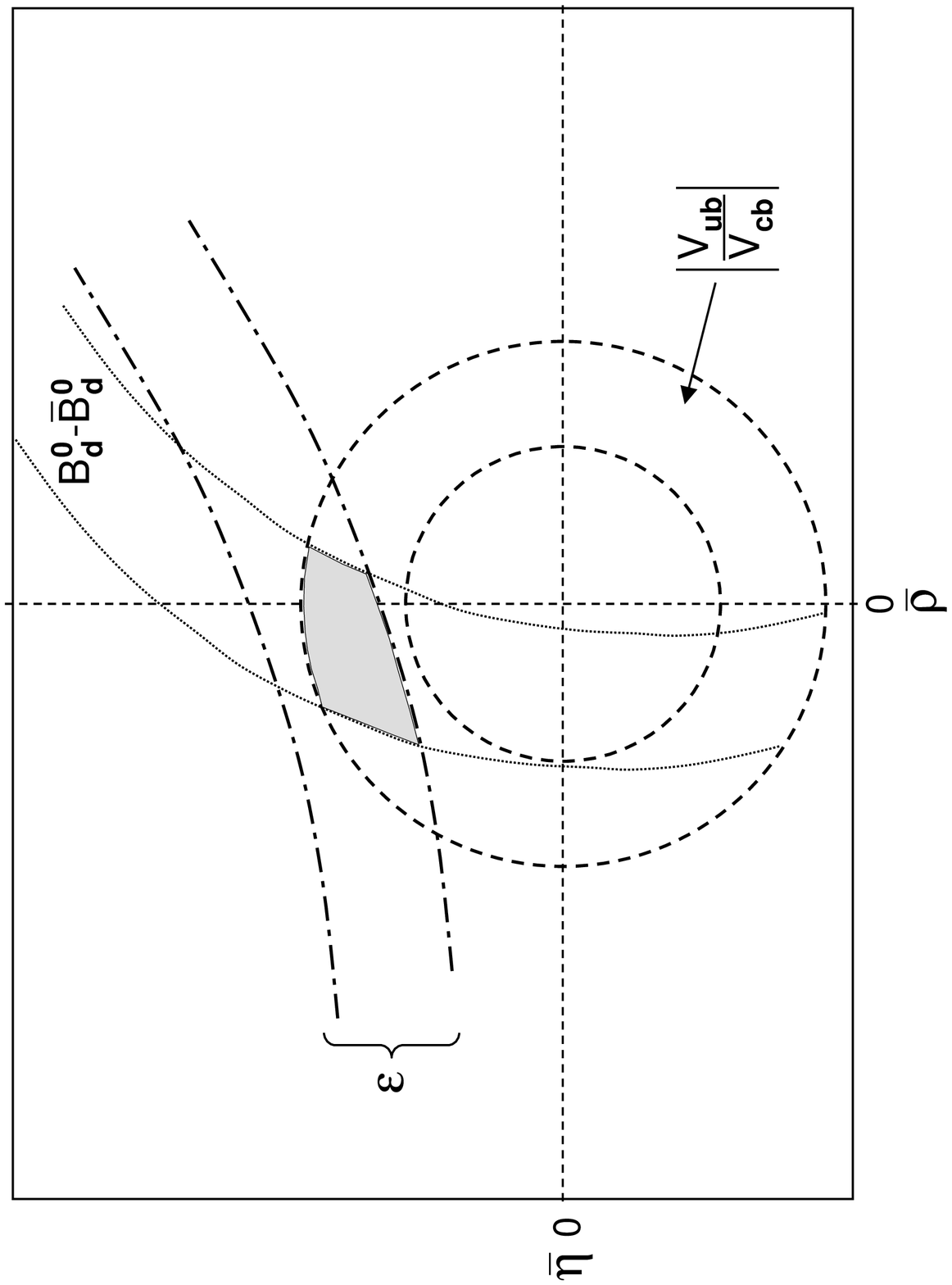,width=0.5\linewidth}}
\end{turn}
} \vspace{-0.18in}
\caption[]{Schematic determination of the Unitarity Triangle.
\label{L:10}}
 \end{figure}

{\bf Step 4:}

{}From the measured $\Delta M_d$ 
and the formula (\ref{DMD}),
the side $AB=R_t$ of the UT can be determined:
\begin{equation}\label{106}
 R_t= \frac{1}{\lambda}\frac{|V_{td}|}{\vcb} = 0.85 \cdot
\left[\frac{|V_{td}|}{7.8\cdot 10^{-3}} \right] 
\left[ \frac{0.041}{\vcb} \right],
\end{equation}
\begin{equation}\label{VT}
\vtd=
7.8\cdot 10^{-3}\left[ 
\frac{230\mev}{\sqrt{\hat B_{B_d}}F_{B_d}}\right]
\left[\frac{167\gev}{\mtb(\mt)} \right]^{0.76} 
\left[\frac{\Delta M_d}{0.50/{\rm ps}} \right ]^{0.5} 
\sqrt{\frac{0.55}{\eta_B}}
\end{equation}
with all symbols defined in the previous Section.
$\mtb(\mt)=(168\pm 4)$ GeV.
Note that $R_t$ suffers from the additional uncertainty in $\vcb$,
which is absent in the determination of $\vtd$ this way. 
The constraint in the $(\bar\varrho,\bar\eta)$ plane coming from
this step is illustrated in Fig.~\ref{L:10}.

{\bf Step 5:}

{}The measurement of  $\Delta M_s$
together with $\Delta M_d$  allows to determine $R_t$ in a different
manner:
\be\label{Rt}
R_t=0.90~\left[\frac{\xi}{1.24}\right] \sqrt{\frac{18.4/{\rm ps}}{\Delta M_s}} 
\sqrt{\frac{\Delta M_d}{0.50/{\rm ps}}},
\qquad
\xi = 
\frac{\sqrt{\hat B_{B_s}}F_{B_s} }{ \sqrt{\hat B_{B_d}}F_{B_d}}.
\ee
One should 
note that $\mt$ and $|V_{cb}|$ dependences have been eliminated this way
 and that $\xi$ should in principle 
contain much smaller theoretical
uncertainties than the hadronic matrix elements in $\Delta M_d$ and 
$\Delta M_s$ separately.  

The main uncertainties in these steps originate in the theoretical 
uncertainties in  $\hat B_K$ and 
$\sqrt{\hat B_d}F_{B_d}$ and to a lesser extent in $\xi$ 
\cite{CERNCKM}: 
\be
\hat B_K=0.86\pm0.15, \quad  \sqrt{\hat B_d}F_{B_d}=(235^{+33}_{-41})\mev,
\quad \xi=1.24\pm 0.08~.
\ee
It should also be emphasized here that the most recent approach is 
to use $\sqrt{\hat B_s}F_{B_s}$ instead of $\sqrt{\hat B_d}F_{B_d}$ as 
it is subject to smaller theoretical uncertainties in lattice calculations 
that the latter quantity. The most recent value used by the UTfit 
collaboration \cite{UTfit} 
is $\sqrt{\hat B_d}F_{B_d}=276\pm 38\mev$ with $\xi$ 
being unchanged. This results in a slightly lower value for $\sqrt{\hat
B_d}F_{B_d}$ that given above but well within the quoted errors.

Also the uncertainties due to $\vub$ in step 2  
are substantial. The QCD sum rules results for  
the parameters in question are similar and can be found in 
\cite{CERNCKM}. 
Finally \cite{CERNCKM}
\be
\Delta M_d=(0.503\pm0.006)/{\rm ps}, \qquad 
\Delta M_s>14.4/{\rm ps}~~ {\rm at }~~ 95\%~{\rm C.L.}
\ee 
\boldmath
\subsection{The Angle $\beta$ from $B_d\to \psi K_S$}
\unboldmath
One of the highlights of the last two years were the considerably improved 
measurements of 
$\sin2\beta$ by means of the time-dependent CP asymmetry
\be\label{asy}
\calaCP(\psi K_S,t)\equiv \calamix(\psi K_S)\sin(\Delta M_d t)=
-\sin 2 \beta \sin(\Delta M_d t)~.
\ee
The BaBar \cite{BaBar} and Belle \cite{Belle} collaborations find
\begin{displaymath}\label{sinb}
(\sin 2\beta)_{\psi K_S}=\left\{
\begin{array}{ll}
0.741\pm 0.067 \, \mbox{(stat)} \pm0.033 \, \mbox{(syst)} & \mbox{(BaBar)}\\
0.719\pm 0.074 \, \mbox{(stat)} \pm0.035  \, \mbox{(syst)} & \mbox{(Belle).}
\end{array}
\right.
\end{displaymath}
Combining these results with earlier measurements by CDF 
$(0.79^{+0.41}_{-0.44})$, ALEPH $(0.84^{+0.82}_{-1.04}\pm 0.16)$ and OPAL 
gives the grand average 
\be
(\sin 2\beta)_{\psi K_S}=0.726\pm0.037~.
\label{ga}
\ee
This is a milestone in the field of CP violation and in the tests of the
SM as we will see in a moment. Not only violation of this symmetry has been 
confidently established 
in the $B$ system, but also its size has been measured very accurately.
Moreover in contrast to the five constraints listed above, the determination 
of the angle $\beta$ in this manner is theoretically very clean.
\boldmath
\subsection{The Sign of $\Delta M_d$ and $\sin 2\beta$}
\unboldmath

The result in (\ref{ga}) leads to a two fold ambiguity in the value of $\beta$ 
\be\label{bCKM}
\beta_{\rm CKM}=23.3\pm 1.6^\circ, \qquad 
\tilde\beta=\frac{\pi}{2}-\beta_{\rm CKM}
\ee
with the second possibility inconsistent with the SM expectations as 
discussed below. Measuring
$\cos 2\beta$ will tell us which of these two solutions is chosen by nature.
The first direct experimental result of BaBar \cite{cos} for $\cos 2\beta$
and other analyses \cite{REF1,BFRS-BIG,CKMfit} disfavour  
in fact the second solution in (\ref{bCKM}).

In extracting the value given in (\ref{bCKM}) it is usually  assumed that the
mass difference $\Delta M_d=M_1-M_2 > 0$ with $M_1$ and $M_2$ denoting the
masses of the neutral $B$ meson eigenstates. 
See (\ref{DHL}) and (\ref{eq:xdsdef}).
As the
sign of $\Delta M_d$ is not known experimentally by itself, 
it is instructive to see 
whether the sign of  $\Delta M_d$ matters at all here.  
If it mattered one would conclude
that with $\Delta M_d<0$, the BaBar and Belle data imply 
$\sin 2\beta=-0.726\pm 0.037$. This would mean that instead 
of only two solutions in (\ref{bCKM}) two 
additional solutions for the angle $\beta$ exist. These findings, 
if correct, would weaken significantly the present believe that the BaBar and
Belle data combined with the standard analysis of the unitarity triangle
imply that the CKM matrix 
is very likely the dominant source of CP violation
observed in low energy experiments.

Fortunately, it can be straightforwardly shown that
the sign of $\Delta M_d$ is {\it irrelevant} for the determination of $\sin
2\beta$ \cite{MBB04} and
the  only relevant quantity for this determination is the weak phase of the
mixing amplitude $M_{12}$. 

Let us recall this derivation here.
Using the expressions of the previous Section but not assuming 
$\Delta M_d$ to be positive we have  
\be\label{E6}
\Delta M_d = M_1 - M_2 = 2 \mathrm{Re} \left( \frac{q}{p} M_{12} \right) = \pm
2|M_{12}|, 
\ee
\be\label{E7}
 B_1=pB^0+q\bar B^0, \qquad B_2 = pB^0 - q \bar B^0~, 
\ee
where $B_1$ and $B_2$ denote the mass eigenstates
and
\begin{equation}\label{q/pnew}
\frac{q}{p} = \pm \sqrt{\frac{M^*_{12}}{M_{12}}} = \pm \frac{M_{12}^*}{|M_{12}|}
=  2 \frac{M_{12}^*}{\Delta M_d}
\end{equation}
with $\pm$ corresponding to $\pm$ in (\ref{E6}). (\ref{q/pnew}) is the
generalization of (\ref{q/p}) to include both signs except
that we have used the fact that the width difference $\Delta \Gamma$ and 
$\Gamma_{12}$ can be neglected.
Consequently,
\begin{equation}\label{e110}
\xi_{\psi K_S}=\frac{q}{p}\frac{A(\bar B^0\to f)}{A(B^0 \to f)}=
\frac{q}{p}=2\frac{M_{12}^*}{\Delta M_d},
\end{equation}
where we have used (\ref{cp}) with $\eta_{\psi K_S}=-1$ and $\phi_D=0$.

Inserting (\ref{e110}) into (\ref{simple}) we find
\begin{equation}\label{bas}
\calaCP(\psi K_S,t) =  2 \mathrm{Im}\left(\frac{M^*_{12}}{\Delta M_d}\right)
\sin(\Delta M_d t).
\end{equation}

This formula demonstrates explicitly that 
the sign of $\Delta M_d$ is irrelevant and only the phase of 
$M_{12}$ matters. Therefore let us have a look at $M_{12}$ including
already some new physics effects. 
Assuming that $M_{12}$ is governed by the usual $(V-A)\otimes(V-A)$ operator, 
we have
quite generally (compare with (\ref{M12Q}))
\begin{equation}\label{M12}
M_{12} = \frac{G_F^2}{12\pi^2}F_B^2
\hat B_B m_B M_W^2 (V_{td}^*V_{tb})^2 S_0(x_t)
\eta_B^{\mathrm{QCD}} r e^{i2\theta_d},
\end{equation}
with all symbols defined in Section 5. 
The last factor in (\ref{M12}) 
describes possible new physics contributions to the
Wilson coefficient of the $(V-A)\otimes (V-A)$ operator that have been 
discussed at
various occasions in the literature 
\cite{WPDELTAF2,Laplace,FLISMA,AI01,Berg,BSU}.
Without loss of generality we take $r>0$.
$\theta_d$ is a new weak phase.

Using
\begin{equation}
V_{tb}=1, \qquad V_{td}= |V_{td}|e^{-i\beta}
\end{equation}
and inserting (\ref{M12}) into (\ref{bas}) we find
\begin{equation}\label{mast}
\calaCP(\psi K_S,t) =- \mathrm{sign}(\hat B_B)~
\sin 2(\beta+\theta_d)~\mathrm{sign}(\Delta M_B) 
\sin(\Delta M_B t)~.
\end{equation}
This formula generalizes and summarizes various discussions of 
$\calaCP(\psi K_S,t)$ in
the SM and its simplest extensions that appeared in the 
literature. 
In particular in \cite{GKN}
 the relevance of the sign of $\hat B_B$ has been 
discussed. 
In these extensions only the
usual $(V-A)\otimes (V-A)$ operator is present and as new physics has no 
impact on its
matrix element between $B^0$ and $\bar B^0$ states, $\hat B_B>0$ 
\cite{CERNCKM}. 
 With
$\theta_d=0$ formula (\ref{mast}) reduces to the usual formula used by 
BaBar and
Belle, except that sign$(\Delta M_d)$ in front of $\sin(\Delta M_d t)$ 
demonstrates
that the sign of $\Delta M_d$ is immaterial.

With $\theta_d=90^\circ$ one recovers a particular minimal flavour violation 
scenario of 
\cite{BF01} in
which the sign of $S_0(x_t)$ is reversed. In this case indeed the BaBar and 
Belle
measurement implies $\sin 2 \beta = -0.726\pm0.037$, but this has nothing to do
with the sign of $\Delta M_d$.

\subsection{Unitarity Triangle 2004}
We are now in the position to combine all these constraints in order to 
construct the unitarity triangle and determine various quantities of interest.
In this context the important issue is the error analysis of these formulae, 
in particular the treatment of theoretical uncertainties. In the 
literature the most popular are the 
Bayesian approach \cite{C00} and the frequentist approach \cite{FREQ}. 
For the PDG analysis see \cite{PDG}.
A critical comparison of these and other methods can be found in 
\cite{CERNCKM}. I can recommend this reading. Most recent analyses of the 
UT by the bayesians (``UTfitters") and frequentists (``CKMfitters") can 
be found in \cite{UTfit} and \cite{CKMfit}, respectively. 

In Fig.~\ref{fig:figmfv} we show the result of  
an analysis in collaboration 
with Felix Schwab and Selma Uhlig \cite{BSU}. 
The allowed region for $(\bar\varrho,\bar\eta)$ 
is the area inside the  ellipse.
We observe that the region
$\bar\varrho<0$ is disfavoured by the lower bound on
$\Delta M_s$.
It is clear
from this figure that the measurement of $\Delta M_s$
giving $R_t$ through (\ref{Rt}) will have a large impact
on the plot in Fig.~\ref{fig:figmfv}.  
Most importantly there is 
an excellent agreement  
between the direct measurement in (\ref{ga})
and the standard analysis of
the UT within the SM.
This gives a strong indication that the CKM matrix is very likely 
the dominant source of CP violation in flavour violating decays.
In order to be sure whether this is indeed the case other theoretically
clean quantities have to be measured. In particular the angle $\gamma$. 
We will discuss other processes that are 
useful for the determination of the UT in the next two sections.

\begin{figure}[htb!]
\begin{center}
{\epsfig{figure=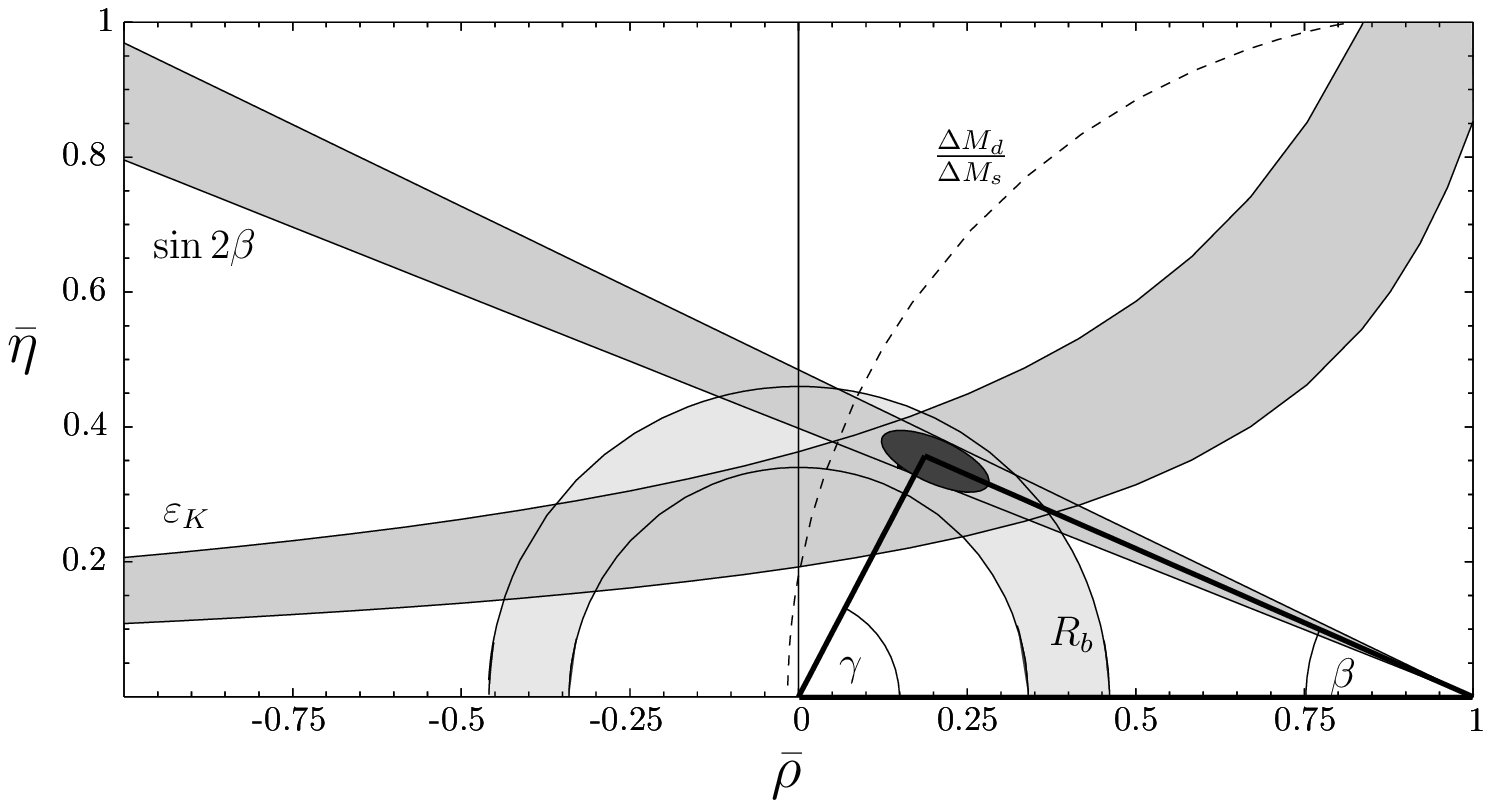,height=6cm}}
\caption[]{The allowed  region in the 
$(\bar\varrho,\bar\eta)$ plane in the SM \cite{BSU}. 
}
\label{fig:figmfv}
\end{center}
\end{figure}

Finally,
the ranges for various quantities that result from this analysis 
are given in the SM column of table~\ref{mfv}. The UUT column 
will be discussed in Section~9.

\begin{table}[htb!]
\begin{center}
\caption[]{ \small { Values for different quantities from 
the UT fit \cite{BSU}.
$\lambda_t=V_{ts}^*V_{td}$.}}
\begin{tabular}{|c|c|c|}
\hline
  Strategy    &            SM     &               UUT                  \\
  $\bar {\eta}$ &   ~~0.354 $\pm$ 0.027~~  &       ~ ~ 0.360 $\pm$ 0.031~~ \\ 
  $\bar {\varrho}$     &   ~~ 0.187 $\pm$ 0.059~~ &     ~~0.174 $\pm$ 0.068~~     \\
  $\sin 2\beta$  & ~~0.732 $\pm0.049$~~ &  ~~0.735 $\pm$0.049~~  \\
  $\beta$  & ~~ ~~$ (23.5\pm 2.1)^\circ$~~  &     ~~$ (23.7 \pm 2.1)^\circ$~~            \\
  $\gamma$     &   ~~$ (62.2 \pm 8.2)^\circ $~~   &   ~~$ (64.2 \pm 9.6)^\circ$~~   \\
  $R_b$      & ~~0.400 $\pm$ 0.039~~      &       ~~0.400 $\pm$ 0.044~~       \\
  $R_t$     &    ~~$0.887 \pm 0.056 $~~     &       ~~$0.901\pm 0.064$~~         \\
$\vtd~(10^{-3})$  &      ~~8.24 $\pm$ 0.54~~  &        ~~8.38 $\pm$ 0.62~~            \\
~~${\rm Im} \lambda_t$ ($10^{-4}$)~~   &    ~~1.40 $\pm$ 0.12~~  & ~~1.43 $\pm$ 0.14~~   ~    \\
%
~~${\rm Re} \lambda_t$ ($10^{-4}$)~~ & $- (3.06\pm 0.25)$~   & ~~$- (3.11\pm 0.28) $~~    \\
\hline
\end{tabular}
\label{mfv}
\end{center}
\end{table}

\subsection{$\epe$ in the Standard Model}\label{EpsilonPrime}
The ratio $\epe$ that parametrizes the size of direct CP violation with
respect to the indirect CP violation in $K_L\to \pi\pi$ decays has been the 
subject of very intensive experimental and theoretical studies in the last
three decades. After tremendous efforts, on the experimental side the world
average based on the results from NA48 \cite{NA48E} and KTeV
\cite{KTeVE}, and previous results from NA31 \cite{NA31} and E731
\cite{E731}, reads
\begin{equation}
  \label{eps}
  \epe=(16.6\pm 1.6) \cdot 10^{-4} \qquad\qquad (2003)~.
\end{equation}
On the other hand, the theoretical estimates of this ratio are subject to
very large hadronic uncertainties. While several analyzes of recent years
within the Standard Model (SM) find results that are compatible with 
(\ref{eps}) \cite{EP99,BRMSSM,ROME,DORT,Trieste,PP,Lund,MARS}), 
it is fair to say
that the chapter on the theoretical calculations of
$\epe$ is far from being closed. A full historical account of the theoretical
efforts before 1998 can for example be found in \cite{AJBLH,BFE00}.
Most recent reviews can be found in
\cite{BJ03,PICH04}.

\boldmath
\section{ The Angles {$\alpha$}, {$\beta$} and 
{$\gamma$} from B Decays}\label{STRATEGIES}
\unboldmath
\setcounter{equation}{0}
\subsection{Preliminaries}

CP violation in $B$ decays is certainly one of the most important 
targets of $B$-factories and of dedicated $B$-experiments at hadron 
facilities. It is well known that CP-violating effects are expected
to occur in a large number of channels at a level attainable 
experimentally and in fact as we have seen above and we will see below, 
clear signals of CP violation in $B$ decays have already been observed, 
Moreover there exist channels which
offer the determination of CKM phases essentially without any hadronic
uncertainties. 

The  results on
$\sin 2\beta$ from BaBar and Belle are 
very encouraging. These results should be further improved over the coming 
years 
through  new measurements of $\calaCP(t,\psi K_S)$ by both collaborations 
and by CDF and D0 at Fermilab. Moreover measurements of CP 
asymmetries in other $B$ decays and the measurements of the angles 
$\alpha,~\beta$ and $\gamma$ by means of various strategies using
two-body $B$ decays should contribute substantially to our understanding
of CP violation and will test the KM picture of CP violation.
In fact some interesting results are already available. They will be
discussed later on.

The various types of CP violation have been already classified in 
Section~\ref{CDECAYS}.
It turned out that CP violation in the interference 
of mixing and decay, in a $B$ meson decay into a CP eigenstate, is very
suitable for a theoretically clean determination of the angles of the 
unitarity triangle provided a single CKM phase governs  the decay. 
However as we will see below several useful strategies 
for the determination of the angles $\alpha,~\beta$ and $\gamma$
have been developed that are effective also in the presence of competing 
CKM phases and when the final state is not a CP eigenstate. 
The notes below should only be considered as an introduction to this 
rich field. For more details the references in Section 1 should be 
contacted.

\begin{figure}[t]
\vspace*{0.2truecm}
{\small
\hspace*{3.2truecm}\begin{picture}(80,50)(80,20)
\Line(10,45)(80,45)\Photon(40,45)(80,5){2}{10}
\Line(80,5)(105,20)\Line(80,5)(105,-10)
\Text(5,45)[r]{$b$}\Text(85,45)[l]{$q_1$}
\Text(110,20)[l]{$\overline{q_2}$}
\Text(110,-10)[l]{$d\,(s)$}
\Text(45,22)[tr]{$W$}
\end{picture}}
\hspace*{-0.6truecm}
{\small
\begin{picture}(140,60)(0,20)
\Line(10,50)(130,50)\Text(5,50)[r]{$b$}\Text(140,50)[l]{$d\,(s)$}
\PhotonArc(70,50)(30,0,180){3}{15}
\Text(69,56)[b]{$u,c,t$}\Text(109,75)[b]{$W$}
\Gluon(70,50)(120,10){2}{10}
\Line(120,10)(135,23)\Line(120,10)(135,-3)
\Text(85,22)[tr]{$G$, $Z$, $\gamma$}\Text(140,-3)[l]{$q$}
\Text(140,23)[l]{$\overline{q}$}
\end{picture}}
\vspace*{1.2truecm}
\caption{Tree and penguin diagrams.}\label{TP}
\end{figure}
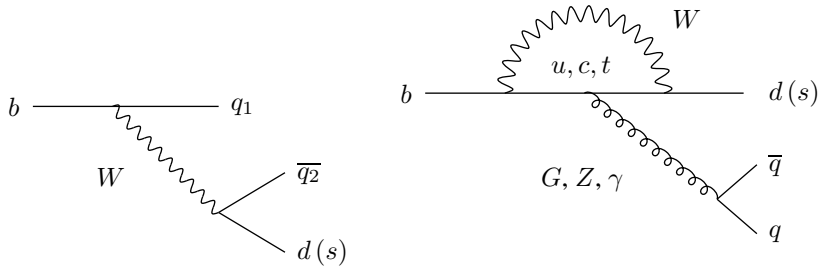

\subsection{Classification of Elementary Processes}
Non-leptonic $B$ decays are caused by elementary decays of $b$ quarks that
are represented by tree and penguin diagrams in Fig.~\ref{TP}. Generally we 
have
\be
b\to q_1 \bar q_2 d(s), \qquad  b \to q\bar q d(s)
\ee
for tree and penguin diagrams, respectively.

There are twelve basic transitions that can be divided into three classes:

{\bf Class I}: both tree and penguin diagrams contribute. Here 
$q_1=q_2=q=u,c$ and consequently the basic transitions are
\be
b\to c\bar c s, \qquad b \to c\bar c d, \qquad
b\to u\bar u s, \qquad b \to u\bar u d.
\ee

{\bf Class II}: only tree diagrams contribute. Here 
$q_1\not=q_2\in\{u,c\}$ and 
\be
b\to c\bar u s, \qquad b \to c\bar u d, \qquad
b\to u\bar c s, \qquad b \to u\bar c d. 
\ee

{\bf Class III}: only penguin diagrams contribute. Here 
$q=d,s$ and 
\be
b\to s\bar s s, \qquad b \to s\bar s d, \qquad
b\to d\bar d s, \qquad b \to d\bar d d. 
\ee

Now in presenting various decays below, we did not show the corresponding 
diagrams on purpose. After all these are lectures and the exercise for the 
students is to draw these diagrams by embedding the elementary diagrams 
of Fig.~\ref{TP}  into a given $B$ meson decay.  
In case of difficulties the student 
should look at \cite{LHCB,REV1} where these diagrams can be found.

\boldmath
\subsection{Neutral B Decays into CP eigenstates}
\subsubsection{{$B^0_d\to \psi K_S$} and 
{$\beta$}}
\unboldmath
The amplitude for this decay can be written as follows
\be\label{FA1}
A(B^0_d\to \psi K_S)=V_{cs}V_{cb}^*(A_T+P_c)+ V_{us}V_{ub}^*P_u+
V_{ts}V_{tb}^*P_t
\ee
where $A_T$ denotes tree diagram contributions and $P_i$ with $i=u,c,t$ 
stand for penguin diagram contributions with internal $u$, $c$ and $t$ 
quarks. Now
\be
V_{cs}V_{cb}^*\approx A\lambda^2, \quad
V_{us}V_{ub}^*\approx A \lambda^4 R_b e^{i\gamma}, 
\quad V_{ts}V_{tb}^*=- V_{us}V_{ub}^*- V_{cs}V_{cb}^*
\ee
with the last relation following from the unitarity of the CKM matrix.
Thus
\be\label{FA1a}
A(B^0_d\to \psi K_S)=V_{cs}V_{cb}^*(A_T+P_c-P_t)+ V_{us}V_{ub}^*(P_u-P_t)~.
\ee
We next note that
\be
\left|\frac{V_{us}V_{ub}^*}{V_{cs}V_{cb}^*}\right|\le 0.02, \qquad
\frac{P_u-P_t}{A_T+P_c-P_t}\ll 1
\ee
where the last inequality is very plausible as the Wilson coefficients 
of the current--current operators responsible for $A_T$ are much larger 
than the ones of the penguin operators \cite{AJBLH,BBL}. 
Consequently this decay is dominated 
by a single CKM factor and as discussed in 
Section~\ref{CDECAYS}, a clean determination 
of the relevant CKM phase is possible. Indeed in this decay $\phi_D=0$  
and $\phi_M=-\beta$. Using (\ref{simple})
we find once more ($\eta_{\psi K_S}=-1$)
\begin{equation}\label{psiKS}
\calamix(\psi K_S)=
\eta_{\psi K_S}\sin(2\phi_D-2\phi_M)=-\sin 2\beta,
\end{equation}
\be
C_{\psi K_S}=0, \qquad S_{\psi K_S}=\sin 2\beta
\ee
that is confirmed by experiment as discussed in the previous Section.
\boldmath
\subsubsection{{$B^0_s\to \psi \phi$} and 
{$\beta_s$}}
\unboldmath
This decay differs from the previous one by the spectator quark, with 
$d\to s$  so that the formulae above remain unchanged except that 
now $\phi_M=-\beta_s=-\lambda^2\bar\eta$. A complication arises as 
the final state is a mixture of $CP=+$ and $CP=-$ states. This issue 
can be resolved experimentally \cite{LHCB}. 
Choosing $\eta_{\psi \phi}=1$ we then find
\begin{equation}\label{psiphi}
\calamix(\psi \phi)=
\sin(2\phi_D-2\phi_M)=2\beta_s=2\lambda^2\bar\eta\approx 0.03,
 \qquad C_{\psi \phi}=0~.
\end{equation}
Thus this asymmetry measures the phase of $V_{ts}$ that is predicted to 
be very small from the unitarity of the CKM matrix alone. Because of this
there is a lot of room for 
new physics contributions to $\calamix(\psi \phi)$. 
\boldmath
\subsubsection{{$B^0_d\to \phi K_S$} and 
{$\beta$}}
\unboldmath
This decay proceeds entirely through penguin diagrams and consequently 
should be much more sensitive to new physics contributions than the decay
$B^0_d\to \psi K_S$. Assuming $\phi=(s\bar s)$, the decay amplitude 
is given by (\ref{FA1a}) 
with $A_T$ removed:
\be\label{FA1b}
A(B^0_d\to \phi K_S)=V_{cs}V_{cb}^*(P_c-P_t)+ V_{us}V_{ub}^*(P_u-P_t)~.
\ee
With 
\be
\left|\frac{V_{us}V_{ub}^*}{V_{cs}V_{cb}^*}\right|\le 0.02, \qquad
\frac{P_u-P_t}{P_c-P_t}=\ord( 1)
\ee
also in this decay a single CKM phase dominates and as 
$\phi_D$ and $\phi_M$ are the same as in $B^0_d\to  \psi K_S$ we 
find 
\be\label{phi K_S}
C_{\phi K_S}=0, \qquad S_{\phi K_S}= S_{\psi K_S}=\sin 2\beta~.
\ee
The equality of these two asymmetries need not be perfect as the $\phi$ 
meson is not entirely a $s\bar s$ state and the approximation of neglecting 
the second amplitude in (\ref{FA1b}) could be only true within a few percent.
However, a detailed analysis shows \cite{Worah} that these two 
asymmetries should 
be very close to each other within the SM:  
$|S_{\phi K_S}-S_{J/\psi K_S}|\le 0.04$~. Any strong violation of this 
bound would be a signal for new physics.

In view of this prediction, the first results on this asymmetry from 
BaBar \cite{Bab} and Belle \cite{Bela} were truly exciting:
\begin{displaymath}\label{sinbphi}
(\sin 2\beta)_{\phi K_S}=\left\{
\begin{array}{ll}
-0.19 \pm 0.51 \, \mbox{(stat)} \pm0.09 \, \mbox{(syst)} & \mbox{(BaBar)}\\
-0.73\pm 0.64 \, \mbox{(stat)} \pm0.18  \, \mbox{(syst)} & \mbox{(Belle),}
\end{array}
\right.
\end{displaymath}
implying
\be\label{gaphi}
S_{\phi K_s}=-0.39\pm 0.41, \qquad C_{\phi K_s}=0.56\pm 0.43, 
\ee
\be
|S_{\phi K_S}-S_{J/\psi K_S}|=1.12\pm 0.41
\ee
and the violation of the bound 
$|S_{\phi K_S}-S_{J/\psi K_S}|\le 0.04$  by $2.7\sigma$. 
These results invited 
a number of theorists to speculate what kind of new physics could be 
responsible for this difference. 
Some references are given in \cite{PHIKS}.
Enhanced QCD penguins, enhanced $Z^0$ 
penguins, rather involved supersymmetric scenarios have been suggested as 
possible origins of the departure from the SM prediction. 

Unfortunately the new data presented at the 2004 summer conferences  
by both collaborations
look much closer to the SM predictions 
\begin{displaymath}\label{sinbphinew}
(\sin 2\beta)_{\phi K_S}=\left\{
\begin{array}{ll}
0.50 \pm 0.25 \, \mbox{(stat)} \pm0.06 \, \mbox{(syst)} & \mbox{(BaBar)}\\
0.06\pm 0.33 \, \mbox{(stat)} \pm0.09  \, \mbox{(syst)} & \mbox{(Belle),}
\end{array}
\right.
\end{displaymath}
implying
\be\label{gaphinew}
S_{\phi K_s}=0.34\pm 0.20, \qquad C_{\phi K_s}=-0.04\pm 0.17. 
\ee
In particular the BaBar result is in full agreement with the SM.
Still some room for new physics contributions is left and it will be
interesting to follow the development in the values of
$S_{\phi K_s}$ and $C_{\phi K_s}$ and similar values in other channels such
as $B\to\eta^\prime K_S$.
Some recent discussions can be found in \cite{PhiKSNEW}.

\boldmath
\subsubsection{{$B^0_d\to \pi^+\pi^-$} and 
{$\alpha$}}
\unboldmath
This decay receives the contributions from both tree and penguin diagrams.
The amplitude  can be written as follows
\be\label{FALPHA}
A(B^0_d\to \pi^+\pi^-)=V_{ud}V_{ub}^*(A_T+P_u)+ V_{cd}V_{cb}^*P_c+
V_{td}V_{tb}^*P_t
\ee
where 
\be
V_{cd}V_{cb}^*\approx A\lambda^3, \quad
V_{ud}V_{ub}^*\approx A \lambda^3 R_b e^{i\gamma}, 
\quad V_{td}V_{tb}^*=- V_{ud}V_{ub}^*- V_{cd}V_{cb}^*~.
\ee
Consequently
\be\label{FALPHAa}
A(B^0_d\to \pi^+\pi^-)=V_{ud}V_{ub}^*(A_T+P_u-P_t)+ V_{cd}V_{cb}^*(P_c-P_t).
\ee
We next note that
\be
\left|\frac{V_{cd}V_{cb}^*}{V_{ud}V_{ub}^*}\right|=\frac{1}{R_b}\approx 2.5,
 \qquad
\frac{P_c-P_t}{A_T+P_u-P_t}\equiv \frac{P_{\pi\pi}}{T_{\pi\pi}}.
\ee
Now the dominance of a single CKM amplitude in contrast to the cases 
considered until now is very uncertain and takes only place provided
$P_{\pi\pi}\ll T_{\pi\pi}$. Let us assume first that this is indeed the 
case. Then this decay is dominated 
by a single CKM factor and a clean determination 
of the relevant CKM phase is possible. Indeed in this decay $\phi_D=\gamma$  
and $\phi_M=-\beta$. Using (\ref{simple})
we find then ($\eta_{\pi\pi}=1$)
\begin{equation}\label{pipi}
\calamix(\pi^+\pi^-)=
\eta_{\pi\pi}\sin(2\phi_D-2\phi_M)=\sin 2(\gamma+\beta)=-\sin 2\alpha
\end{equation}
and 
\be
C_{\pi\pi}=0, \qquad S_{\pi\pi}=\sin 2\alpha~.
\ee
This should be compared with the  2004 results from 
BaBar \cite{BaBar-CP-Bpipi} and Belle \cite{Belle-CP-Bpipi}:
\begin{displaymath}\label{sinaC}
C_{\pi\pi}=\left\{
\begin{array}{ll}
-0.09 \pm 0.15 \, \mbox{(stat)} \pm0.04 \, \mbox{(syst)} & \mbox{(BaBar)}\\
-0.58\pm 0.15 \, \mbox{(stat)} \pm0.07  \, \mbox{(syst)} & \mbox{(Belle)}
\end{array}
\right.
\end{displaymath}

\begin{displaymath}\label{sinaS}
S_{\pi\pi}=\left\{
\begin{array}{ll}
-0.30 \pm 0.17 \, \mbox{(stat)} \pm0.03 \, \mbox{(syst)} & \mbox{(BaBar)}\\
-1.00\pm 0.21 \, \mbox{(stat)} \pm0.07 \, \mbox{(syst)} & \mbox{(Belle).}
\end{array}
\right.
\end{displaymath}
giving the averages
\be\label{ave}
C_{\pi\pi}=-0.37\pm0.11, \qquad S_{\pi\pi}=-0.61\pm0.14~.
\ee

The results from BaBar are consistent with earlier expectations that 
the direct CP violation is very small. After all 
$\alpha$ from the UT fit is in the ballpark of $90^\circ$.
On the other hand Belle results indicate a non-zero asymmetry and a 
sizable contribution of the penguin diagrams invalidating our assumption
$P_{\pi\pi}\ll T_{\pi\pi}$. While the results from BaBar and Belle are
not fully compatible with each other, the average in (\ref{ave}) did not 
change by much over the last two years.

The ``QCD penguin pollution" discussed above  has to be
taken care of in order to extract $\alpha$. 
The well known strategy to deal with this "penguin problem''
is the isospin analysis of Gronau and London \cite{CPASYM}. It
requires however the detailed measurements  of all $B\to \pi\pi$ 
modes.
For this reason several other strategies  for extraction of $\alpha$
have been proposed.
They are reviewed in 
\cite{BF97,BABAR,SUPERB,LHCB,REV1}. 
Some recent analyses of $B\to \pi\pi$ data can be found in 
\cite{PIPICOL,CKMfit}. Others will be discussed in Section~\ref{NPHASE}.

While it is not clear that a precise value of $\alpha$ will 
follow in a foreseeable future from this enterprise,
one should  stress \cite{BB4,BB96,BBNS2,BUPAST} that only a 
moderately precise measurement of $\sin 2\alpha$ can be as useful for 
the UT as a precise measurement of the angle $\beta$.

From my point of view a more promising approach is to use the full 
$B\to\pi\pi$ system in conjunction with the known value of $\beta$ in order 
to determine the angle $\gamma$ and subsequently $(\bar\varrho,\bar\eta)$. 
Some analyses of this type have already been presented  in
\cite{CERNCKM,ALPHA,FLISMA},
but recently in view of the new $B\to\pi\pi$ data they have been generalized 
and improved. 
One of such analyses done in collaboration with Robert
Fleischer, Stefan Recksiegel and Felix Schwab will be presented in 
Section~\ref{NPHASE}.

\subsection{Decays to CP Non-Eigenstates}
\subsubsection{Preliminaries}
The strategies discussed below have the following general properties:
\begin{itemize}
\item
$B^0_d(B^0_s)$ and their antiparticles $\bar B^0_d(\bar B^0_s)$ can decay to 
the same final state,
\item
Only tree diagrams contribute to the decay amplitudes,
\item
A full time dependent analysis of the four processes is required:
\be
B^0_{d,s}(t)\to f, \quad \bar B^0_{d,s}(t)\to f, \quad
 B^0_{d,s}(t)\to \bar f, \quad \bar B^0_{d,s}(t)\to \bar f~.
\ee
\end{itemize}
The latter analysis allows to measure
\begin{equation}\label{4rates}
\xi_f=\exp(i2\phi_M)\frac{A(\bar B^0\to f)}{A(B^0 \to f)}, \qquad
\xi_{\bar f}=\exp(i2\phi_M)\frac{A(\bar B^0\to \bar f)}{A(B^0 \to \bar f)}.
\end{equation}
It turns out then that
\be
\xi_f\cdot \xi_{\bar f}=F(\gamma,\phi_M)
\ee
without any hadronic uncertainties, so that determining $\phi_M$ from 
other decays as discussed above, allows the determination of $\gamma$.
Let us show this and find an explicit expression for $F$.
\boldmath
\subsubsection{{$B^0_d\to D^\pm\pi^\mp$}, 
{$\bar B^0_d\to D^\pm\pi^\mp$} and {$\gamma$} }
\unboldmath
With $f=D^+\pi^-$ the four decay amplitudes are given by
\be\label{amp12}
A(B^0_d\to D^+\pi^-)=M_f A \lambda^4 R_b e^{i\gamma}, \qquad
A(\bar B^0_d\to D^+\pi^-)= \bar M_f A \lambda^2
\ee
\be\label{amp34}
A(\bar B^0_d\to D^-\pi^+)=\bar M_{\bar f} A \lambda^4 R_b e^{-i\gamma}, \qquad
A(B^0_d\to D^-\pi^+)=  M_{\bar f} A \lambda^2
\ee
where we have factored out the CKM parameters, $A$ is a Wolfenstein 
paramater and $M_i$ stand for the rest of the amplitudes that generally are 
subject to large hadronic uncertainties. The important point is that each 
of these transitions receives the contribution from a single phase so that
\begin{equation}\label{2xiD}
\xi_f^{(d)}=
e^{-i(2\beta+\gamma)}\frac{1}{\lambda^2 R_b}\frac{\bar M_f}{M_f}, \qquad
\xi_{\bar f}^{(d)}=
e^{-i(2\beta+\gamma)}\lambda^2 R_b\frac{\bar M_{\bar f}}{M_{\bar f}}~.
\end{equation}
Now, as CP is conserved in QCD we simply have
\be
M_f=\bar M_{\bar f}, \qquad \bar M_f= M_{\bar f} 
\ee
and consequently \cite{DPI}
\be\label{dxi}
\xi_f^{(d)}\cdot \xi_{\bar f}^{(d)}=e^{-i2(2\beta+\gamma)}
\ee
as promised. The phase $\beta$ is already known with high precision 
and consequently $\gamma$ can be determined. Unfortunately as seen in
(\ref{amp12}) and (\ref{amp34}), the relevant interefences are 
$\ord(\lambda^2)$ and the execution of this strategy is a very difficult 
experimental task. See \cite{Silva} for an interesting discussion.


\boldmath
\subsubsection{{$B^0_s\to D_s^\pm K^\mp$}, 
{$\bar B^0_s\to D_s^\pm K^\mp$} and {$\gamma$}}
\unboldmath
Replacing the $d$-quark by the $s$-quark in the strategy just discussed allows 
to enhance the intereference between various contributions.
With $f=D^+_s K^-$ equations 
(\ref{amp12}) and (\ref{amp34}) are replaced by
\be\label{samp12}
A(B^0_s\to D_s^+K^-)=M_f A \lambda^3 R_b e^{i\gamma}, \qquad
A(\bar B^0_s\to D_s^+K^-)= \bar M_f A \lambda^3
\ee
\be\label{samp34}
A(\bar B^0_s\to D_s^- K^+)=\bar M_{\bar f} A \lambda^3 R_b e^{-i\gamma}, \qquad
A(B^0_s\to D_s^- K^+)=  M_{\bar f} A \lambda^3~.
\ee
Proceeding as in the previous strategy one finds \cite{adk}
\be\label{sxi}
\xi_f^{(s)}\cdot \xi_{\bar f}^{(s)}=e^{-i2(2\beta_s+\gamma)}
\ee
with $\xi_f^{(s)}$ and $\xi_{\bar f}^{(s)}$ being the analogs of 
$\xi_f^{(d)}$ and $\xi_{\bar f}^{(d)}$, respectively. Now, all interferring 
amplitudes are of a similar size. With $\beta_s$ extracted 
one day from the asymmetry in $B_s^0(\bar B_s^0)\to \psi\phi$, the angle 
$\gamma$ can be determined.
\boldmath
\subsubsection{{$B^\pm\to D^0 K^\pm$}, 
{$B^\pm\to \bar D^0 K^\pm$} and {$\gamma$} }
\unboldmath
By replacing the spectator $s$-quark in the last strategy through a
$u$-quark one arrives at decays of $B^\pm$ that can be used to extract 
$\gamma$. Also this strategy is unaffected by penguin contributions. 
Moreover, as particle-antiparticle mixing is absent here, $\gamma$ can 
be measured directly without any need for phases in the mixing. Both 
these features make it plausible that this strategy, not involving 
to first approximation any loop diagrams,  is particularly suited 
for the determination of $\gamma$ without any new physics pollution.

By considering six decay rates $B^{\pm}\to D^0_{CP} K^{\pm}$,
$B^+ \to D^0 K^+,~ \bar D^0 K^+$ and  $B^- \to D^0 K^-,~ \bar D^0 K^-$
where $D^0_{CP}=(D^0+\bar D^0)/\sqrt{2}$ is a CP eigenstate, and 
noting that
\be\label{CONST}
A(B^+ \to \bar D^0 K^+)= A(B^- \to D^0 K^-),
\ee
\be\label{CONSTa}
A(B^+ \to D^0 K^+)= A(B^- \to \bar D^0 K^-) e^{2 i\gamma}
\ee
the well known triangle construction due to Gronau and Wyler 
\cite{Wyler} allows to determine $\gamma$. However, the method is 
not without problems. The detection of $D^0_{CP}$, that is necessary 
for this determination because $K^+\bar D^0\not=K^+ D^0$, is experimentally 
challenging.
Moreover, the small
branching ratios of the colour supressed channels in 
(\ref{CONSTa}) and the absence of this suppression in the two
remaining channels in (\ref{CONST}) imply a rather 
squashed triangle 
thereby making the extraction of $\gamma$ very difficult.
Variants of this method
that could be more promising are discussed in \cite{DUN2,V97}.

\boldmath
\subsubsection{Other Clean Strategies for {$\gamma$} and 
{$\beta$}}
\unboldmath
The three strategies discussed above can be generalized to other decays. 
In particular \cite{DUN2,FLEISCHER}
\begin{itemize}
\item
$2\beta+\gamma$ and $\gamma$ can be measured in
\be
B^0_d\to K_S D^0,~K_S \bar D^0, \qquad B^0_d\to \pi^0 D^0,~\pi^0 \bar D^0 
\ee
and the corresponding CP conjugated channels,
\item
$2\beta_s+\gamma$ and $\gamma$ can be measured in
\be
B^0_s\to \phi D^0,~\phi \bar D^0, \qquad B^0_s\to K_S D^0,~K_S \bar D^0 
\ee
and the corresponding CP conjugated channels,
\item
$\gamma$ can be measured by generalizing the Gronau--Wyler construction 
to $B^\pm\to D^0\pi^\pm, \bar D^0\pi^\pm$ and to $B_c$ decays \cite{FW01}:
\be
B_c^\pm\to D^0 D^\pm_s,~\bar D^0 D^\pm_s, 
\qquad B_c^\pm\to D^0 D^\pm,~\bar D^0 D^\pm~. 
\ee
\end{itemize}
In this context I can recommend the papers by Fleischer 
\cite{FLEISCHER} 
that while discussing these decays go far beyond the methods presented
here. It appears that the methods discussed in this subsection  may give 
useful results at later stages of CP-B investigations, in particular 
at LHC-B. 
\subsection{U--Spin Strategies}
\subsubsection{Preliminaries}
 Useful strategies for $\gamma$ using the U-spin symmetry have
been proposed by Robert Fleischer in \cite{RF99,RF991}. 
The first strategy involves
the decays $B^0_{d,s}\to \psi K_S$ and $B^0_{d,s}\to D^+_{d,s} D^-_{d,s}$.
The second strategy involves $B^0_s\to K^+ K^-$ and $B^0_d\to\pi^+\pi^-$.
They are unaffected by FSI and are only limited
by U-spin breaking effects. They are promising for
Run II at FNAL and in particular for LHC-B. 

A method of determining $\gamma$, using $B^+\to K^0\pi^+$ and the
U-spin related processes $B_d^0\to K^+\pi^-$ and $B^0_s\to \pi^+K^-$,
was presented in \cite{GRCW}. A general discussion of U-spin symmetry 
in charmless $B$ decays and more references to this topic can be
found in \cite{REV1,G00}. I will only briefly discuss the 
method in \cite{RF991}.
\boldmath
\subsubsection{$B^0_d\to \pi^+\pi^-$,  
$B^0_s\to K^+K^-$ and {$(\gamma,\beta)$}}
\unboldmath
Replacing in $B^0_d\to \pi^+\pi^-$ the $d$ quark by an $s$ quark we obtain 
the decay $B^0_s\to K^+K^-$. 
The amplitude  can be then written in analogy to (\ref{FALPHAa}) as follows
\be\label{Fgamma}
A(B^0_s\to K^+K^-)=V_{us}V_{ub}^*(A'_T+P'_u-P'_t)+ V_{cs}V_{cb}^*(P'_c-P'_t).
\ee
This formula differs from (\ref{FALPHAa}) only by $d\to s$ and the primes on 
the hadronic matrix elements that in principle are different in these two 
decays. As
\be
V_{cs}V_{cb}^*\approx A\lambda^2, \qquad
V_{us}V_{ub}^*\approx A \lambda^4 R_b e^{i\gamma}, 
\ee
the second term in (\ref{Fgamma}) is even more important than the 
corresponding term in 
the case of $B^0_d\to \pi^+\pi^-$. Consequently $B^0_d\to K^+K^-$
taken alone does not offer a useful method for the determination of the CKM 
phases. On the other hand, with the help of the U-spin symmetry of strong 
interations, it allows roughly speaking to determine the penguin contributions
in $B^0_d\to \pi^+\pi^-$ and consequently the extraction of $\beta$ and 
$\gamma$.

Indeed, from the U-spin symmetry we have
\be
\frac{P_{\pi\pi}}{T_{\pi\pi}}=\frac{P_c-P_t}{A_T+P_u-P_t}
=\frac{P'_c-P'_t}{A'_T+P'_u-P'_t}=\frac{P_{KK}}{T_{KK}}
\equiv d e^{i\delta}
\ee
where $d$ is a real non-perturbative parameter and $\delta$ a strong phase.
Measuring $S_f$ and $C_f$ for both decays and extracting $\beta_s$ from
$B_s^0\to \psi \phi$, we can determine four unknowns: $d$, $\delta$, $\beta$ 
and $\gamma$ subject mainly to U-spin breaking corrections. 
An analysis using these ideas can be found in \cite{FLISMA}.
\boldmath
\subsection{Constraints for $\gamma$ from $B\to\pi K$}
\unboldmath
The four modes $B^\pm\to \pi^\mp K^0$, $B^\pm\to \pi^0 K^\pm$,  
$B^0_d\to \pi^\mp K^\pm$ and $B^0_d\to \pi^0 K^0$
have been observed 
by the CLEO, BaBar  and Belle collaborations and 
allow us already now to obtain some direct information on $\gamma$.
This information will improve 
when the errors on branching ratios and the CP asymmetries decrease. 
The progress on the accuracy of these measurements is 
slow but steady and they begin to give, in addition to $\gamma$,
 an interesting insight 
into the flavour and QCD dynamics.
In particular, one of the highlights of 2004 was the observation of 
the direct CP violation in $B_d\to\pi^\mp K^\pm$
decays \cite{BaBar-CP-dir-obs,Belle-CP-dir-obs}.
Other experimental results for  $B\to\pi K$ decays will be discussed in 
Section~\ref{NPHASE}.

There has been a large theoretical activity in this field during 
the last seven years.
The main issues here are the final state interactions (FSI), 
SU(3) symmetry
breaking effects and the importance of electroweak penguin
contributions. Several interesting ideas have been put forward
to extract the angle $\gamma$ in spite of large hadronic
uncertainties in $B\to \pi K$ decays 
\cite{FM,GRRO,GPAR1,GPAR3,GPAR2,NRBOUND}.

Three strategies for bounding and determining $\gamma$ have been 
proposed. The ``mixed" strategy \cite{FM} uses 
$B^0_d\to \pi^0 K^\pm$ and $B^\pm\to\pi^\pm K$. The ``charged" strategy
\cite{NRBOUND} involves $B^\pm\to\pi^0 K^\pm,~\pi^\pm K$ and
the ``neutral" strategy \cite{GPAR3} the modes 
$B_d^0\to \pi^\mp K^\pm,~\pi^0K^0$. 
General parametrizations for the 
study of the FSI, SU(3) symmetry
breaking effects and of the electroweak penguin
contributions in these channels have been presented 
in \cite{GPAR1,GPAR3,GPAR2}.
Moreover, general parametrizations by means
of Wick contractions \cite{IWICK,BSWICK} have been proposed. 
They can be used for all two-body B-decays.
These parametrizations should
turn out to be useful when the data improve.
Finally, recently a graphical approach using SCET ideas 
has been proposed in \cite{GSCET}.

Parallel to these efforts an important progress has been made 
by developing approaches for
the calculation of the hadronic matrix elements of local operators in 
QCD beyond the standard factorization method. These are in particular the
QCD factorization approach \cite{BBNS1}, the perturbative QCD approach
\cite{PQCD} and the soft-collinear effective theory \cite{SCET}.
Moreover new methods to calculate exclusive hadronic matrix
elements from QCD light-cone sum rules have been developed 
in \cite{KOD,PBALL}. 
While,
in my opinion, an important progress in evaluating non-leptonic 
amplitudes has been made in these papers, the usefulness of this 
recent progress
at the quantitative level has still to be demonstrated when the
data improve. In fact as discussed in Section~\ref{NPHASE} the most recent
data indicate that the present theoretical frameworks have real problems 
in certain channels.

As demonstrated in a  number of papers 
\cite{FM,GPAR1,GPAR3,GPAR2,NRBOUND} in the past, 
these strategies imply within the SM interesting bounds on $\gamma$ 
that do not necessarily agree with the values extracted from the UT analysis 
of Section~\ref{UT-Det}.  In particular already in 2000 combining the 
neutral and
charged strategies \cite{BF-neutral2} we have  found that the 2000 data 
on $B\to \pi K$ favoured $\gamma$  in the second quadrant, which was in 
conflict with the standard analysis of the unitarity triangle that 
implied $\gamma= (62\pm 7)^\circ$. Other arguments for $\cos\gamma<0$ using
$B\to PP,~PV$ and $VV$ decays were given also in 
\cite{CLEO99,CERNCKM,Neubert02}.

On the other hand it has been emphasized by various authors that in view of
sizable 
theoretical uncertainties in the analyses of
$B\to\pi K$ and of still significant experimental errors in the
corresponding branching ratios it is not yet clear whether the
discrepancy in question is serious. For instance \cite{CIFRMAPISI}
sizable contributions of the so-called charming penguins \cite{PAP0,c-pen}
 to the
$B\to\pi K$ amplitudes could in principle shift $\gamma$ extracted from these
decays below $90^\circ$ but at present these contributions cannot be
calculated reliably. A similar role could be played by annihilation
contributions \cite{PQCD} and large non-factorizable 
SU(3) breaking effects
\cite{BF-neutral2}.  

However, a much more attractive solution to all these problems appears to me
the proposal made by Robert Fleischer and myself already in 2000 that 
the puzzling features of the $B\to\pi K$ data indicate
 new physics contributions in the electroweak
penguin sector \cite{BF-neutral2}.  
In order to address this issue in the presence of improved data, we have
recently developed a strategy in collaboration with Stefan Recksiegel and
Felix Schwab that allows to analyze not only the $B\to \pi K$ system but also
the $B\to\pi\pi$ system and subsequently study possible implications of the
modified electroweak penguin sectors on rare $K$ and $B$ decays. We will
discuss this strategy in Section~\ref{NPHASE}. 
In order to be prepared for this
discussion we turn now to the analysis of $\kpn$ and $\klpn$ decays.

\boldmath
\section{$\kpn$ and $\klpn$ }\label{KPNN} 
\unboldmath
\setcounter{equation}{0}

\subsection{Preliminaries}
The rare decays $\kpn$ and $\klpn$ are very promising probes of 
flavour physics within the SM and possible extensions, since they are
governed by short distance interactions. They proceed through $Z^0$-penguin
 and box diagrams in Fig.~\ref{fig:ZP}. 
As the required hadronic matrix elements can be extracted 
from the leading semileptonic decays and other long distance contributions 
turn out to be small \cite{RS,LW,GHL,EPdR,FalkLP,GBGI,IMS05,CSMITH}, 
the relevant branching ratios can be 
computed
to an exceptionally high degree of precision \cite{BB3,BB98,MU98}. 
The main theoretical
uncertainty in the CP conserving decay $\kpn$ originates in the 
value of $\mu_c$ in
$\mc(\mu_c)$. It has been reduced through NLO corrections down to 
$\pm 8\%$ 
\cite{BB3,BB98}
at the level of the branching ratio. The dominantly CP-violating decay 
$\klpn$ \cite{littenberg:89} is even cleaner as only the internal top 
contributions matter. The 
theoretical error for $Br(\klpn)$ amounts to $\pm 2\%$ and is safely 
negligible.

\begin{figure}[hbt]
\vspace{0.10in}
\centerline{
\epsfysize=10.0cm
\epsffile{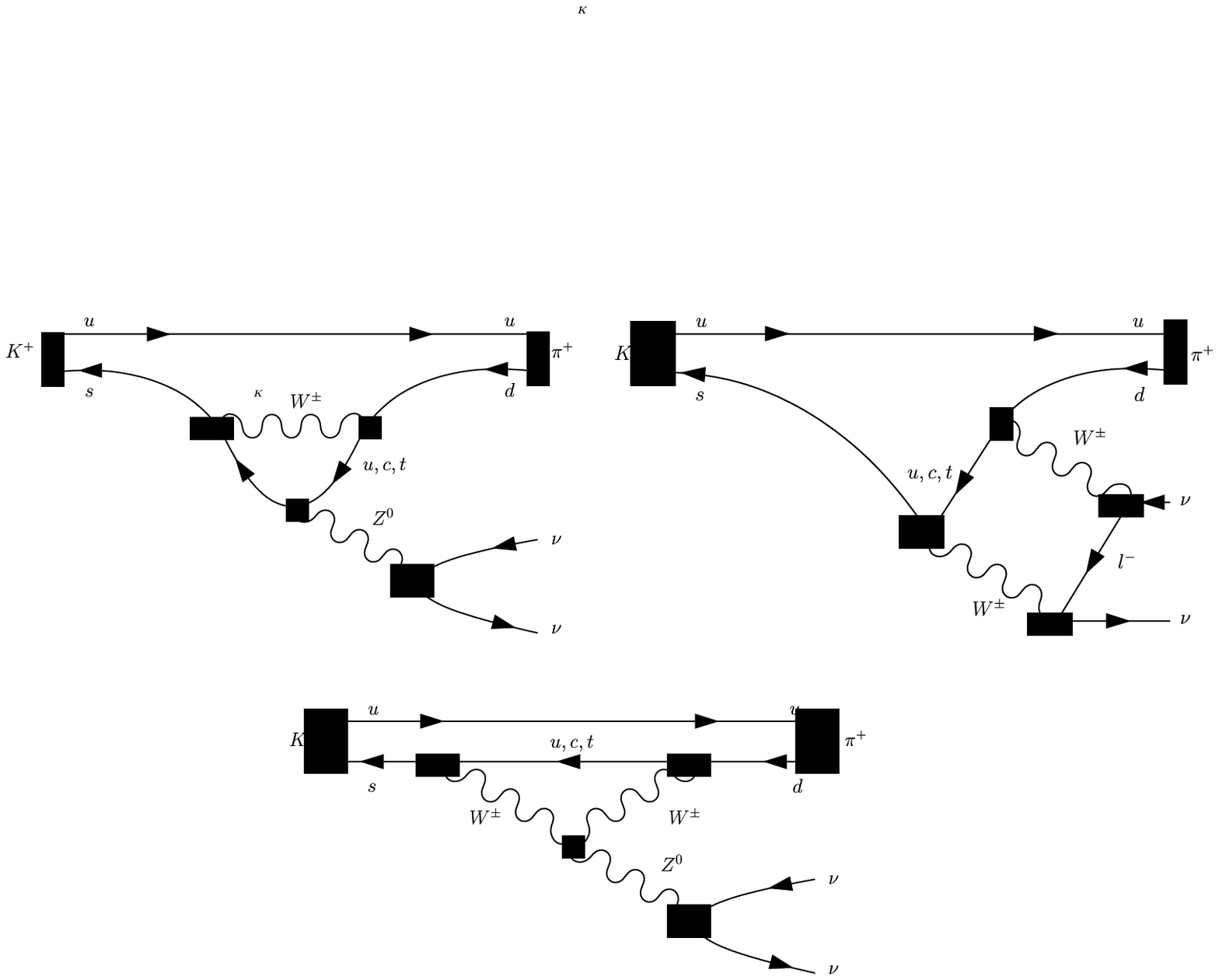}
}
\vspace{0.08in}
\caption{Penguin and box diagrams contributing to $\kpn$.
In $\klpn$ the spectator quark is changed from $u$ to $d$.
}\label{fig:ZP}
\end{figure}

On the experimental side the AGS E787 collaboration at Brookhaven was the 
first to observe the decay $\kpn$ \cite{Adler970}. 
The resulting branching ratio based on two 
events and published in 2002 was \cite{Adler02}
\be\label{EXP0}
Br(\kpn)=
(15.7^{+17.5}_{-8.2})\cdot 10^{-11} \qquad (2002).
\ee
In 2004 a new $\kpn$ experiment, AGS E949 \cite{E949}, 
released its first results 
 that are based on the 2002 running. One additional event has been 
observed. Including the result of AGS E787 the present branching ratio reads
\be\label{EXP1}
Br(\kpn)=
(14.7^{+13.0}_{-8.9})\cdot 10^{-11} \qquad (2004).
\ee
It is not clear, at present, how this result will be improved in the coming 
years but AGS E949 should be able to collect in total 10 SM events.
One should also hope that the efforts at Fermilab around the CKM 
experiment \cite{CKMEXP}, the corresponding efforts at CERN around the NA48
collaboration \cite{NA48EXP} and at J-PARC in Japan 
\cite{JPAR} will provide additional 50-100 events in the next five years.

The situation is different for $\klpn$. 
While the present upper bound on its branching ratio from KTeV \cite{KTeV00X}, 
\be\label{EXP2}
Br(\klpn)<5.9 \cdot 10^{-7},
\end{equation}
is about four orders of magnitude above the SM expectation, the prospects for 
an accurate measurement of $\klpn$ appear almost better than for 
$\kpn$ from the present perspective.

Indeed, a $\klpn$ experiment at KEK, E391a \cite{E391}, which 
just started taking data, 
should in its first stage improve the bound in (\ref{EXP2}) by  three orders 
of magnitude. While this 
is insufficient to reach the SM level, a few events could be observed if 
$Br(\klpn)$ turned out to be by one order of magnitude larger due to 
new physics contributions. 

Next, a very interesting experiment at Brookhaven, KOPIO \cite{KOPIO}, 
should in due time 
provide 40-60 events of $\klpn$ at the SM level. Finally, the second stage of 
the E391 experiment could, using the high intensity 50 GeV/c proton beam from
J-PARC \cite{JPAR}, provide more than 100 SM events of 
$\klpn$, which would be truly fantastic! Perspectives of a search for $\klpn$
at a $\Phi$-factory have been discussed in \cite{Bossi}.
Further reviews on experimental prospects for $K\to\pi\nu\bar\nu$ 
can be found in \cite{KOPIO,REVK}.

In view of these prospects a new very detailed review of $K\to\pi\nu\bar\nu$
decays in collaboration with Felix Schwab and Selma Uhlig has been presented 
in \cite{BSU}. Essentially everything that is known about these decays on the
theoretical side can be found there. Therefore, we will only 
summarize the main virtues of 
$K\to\pi \nu\bar\nu$ decays here. Other reviews can be found in 
\cite{Gino03}. For a very recent summary of the exceptional virtues of 
$\klpn$ in probing new physics see \cite{BBIL}.

\subsection{Branching Ratios}
The basic formulae for the branching ratios are given as follows
\cite{BB98}  
\begin{equation}\label{bkpnn}
Br(K^+\to\pi^+\nu\bar\nu)=\kappa_+\cdot
\left[\left(\frac{{\rm Im}\lambda_t}{\lambda^5}X(x_t)\right)^2+
\left(\frac{{\rm Re}\lambda_c}{\lambda}P_c(X)+
\frac{{\rm Re}\lambda_t}{\lambda^5}X(x_t)\right)^2\right],
\end{equation}
\begin{equation}\label{bklpn}
Br(K_L\to\pi^0\nu\bar\nu)=\kappa_L\cdot
\left(\frac{{\rm Im}\lambda_t}{\lambda^5}X(x_t)\right)^2.
\end{equation}
Here $x_t=m^2_t/M^2_W$, $\lambda_i=V^*_{is}V_{id}$
 and
\begin{equation}\label{kapp}
\kappa_+=(4.84\pm 0.06) \cdot 10^{-11}\,, \qquad
\kappa_{\rm L}=(2.12\pm0.03)\cdot 10^{-10}
\end{equation}
include isospin
breaking corrections in relating $\kpn$ and $\klpn$ to $K^+\to\pi^0e^+\nu$,  
respectively \cite{MP}.
Due to the update of input parameters made 
in \cite{BSU},
the numbers in (\ref{kapp}) differ from the ones in the
original papers \cite{BB96,BB98} 

Next
\be\label{XT}
X(x_t)=1.53\pm 0.04 
\ee
represents the internal top contribution and 
$P_c(X)$ results from the internal charm 
contribution \cite{BB3}. 
A recent analysis of the uncertainties in $P_c(X)$ resulted 
in \cite{BSU}
\be\label{PC}
P_c(X)=0.389\pm 0.033_{m_c} \pm 0.045_{\mu_c} \pm 0.010_{\alpha_s}=
0.39\pm 0.07,
\ee
where the errors correspond to $m_c(m_c)$, $\mu_c$ and 
$\alpha_s(M_Z)$, respectively. The latter parameters 
 have been varied as follows
\be
1.25\gev\le\mc(m_c)\le 1.35\gev,
 \qquad
1.0\gev\le\mu_c\le 3.0\gev,
\ee
\be
0.116\le \alpha_s(M_Z)\le 0.120~.
\ee

The result in (\ref{PC}) does not include the recently calculated 
\cite{IMS05,CSMITH}
contributions of dimension-eight four fermion operators generated at the
charm scale and genuine long distance contributions which can be described
within the framework of chiral perturbation theory. Including these
contributions one finds
\be\label{PCNEW}
P_c(X)=0.43\pm 0.07.
\ee
We anticipate that all remaining long distance uncertainties, that are well 
below the
error in (\ref{PCNEW}), are already included in the error quoted above.

We observe that the error in $P_c(X)$ is dominated by the left-over scale 
uncertainty ($\mu_c$), implying that a  calculation of $P_c(X)$ at the 
NNLO level 
is certainly desirable. The uncertainty due to $m_c$ is smaller but still 
significant. On the other hand, the uncertainty due to $\alpha_s$ is small. 

We expect that a NNLO calculation, that is now in progress \cite{BGHN},
 will reduce  the error in $P_c(X)$
due to $\mu_c$ by a factor of 2-3 and the reduction of the error in
$\alpha_s(M_Z)$ to $\pm 0.001$ will decrease the corresponding error 
to $0.005$, making it negligible.  
Concerning the error due to $m_c(m_c)$, 
we have to a good 
approximation \cite{BSU}
\be\label{FS}
\sigma (P_c(X))_{m_c}= \left[\frac{0.67}{\rm GeV}\right]
 \sigma(\mc(\mc)).
\ee

This discussion shows that after a NNLO analysis has been performed, the 
main uncertainty in $P_c(X)$ will be due to $m_c$.
From the present perspective, 
unless some important advances in the determination of $\mc(m_c)$ will be made,
it will be very difficult to decrease the error on $P_c(X)$ below 
$\pm 0.03$, although $\pm 0.02$ cannot be fully excluded. 

Imposing all existing constraints on the CKM matrix one finds 
using (\ref{PC})
\cite{BSU}
\begin{equation}\label{kpnr}
Br(\kpn)=(7.77 \pm 0.82_{P_c}\pm 0.91)\cdot 10^{-11}=
(7.8 \pm 1.2)\cdot 10^{-11}, 
\end{equation}
\begin{equation}\label{klpnr4}
Br(\klpn)=
(3.0 \pm 0.6)\cdot 10^{-11} 
\end{equation}
Similar results are found in
 \cite{Gino03,kettel}.

On the other hand, using the most recent result on $P_c(X)$ in 
(\ref{PCNEW}) one finds
\begin{equation}\label{kpnrnew}
Br(\kpn)=(8.3 \pm 1.2)\cdot 10^{-11}, 
\end{equation}
without any change in (\ref{klpnr4}).

The central value of $Br(\kpn)$ in (\ref{kpnrnew}) is below the central 
experimental value
in (\ref{EXP1}), but within theoretical, parametric and experimental 
uncertainties, the SM result is fully consistent with the data.
We also observe that the error in $P_c(X)$ constitutes still a significant 
portion of the full error. 
The present upper bound on $Br(K_{\rm L}\to \pi^0\nu\bar\nu)$ from
the KTeV  experiment at Fermilab \cite{KTeV00X} and given in (\ref{EXP2})
is about four orders of magnitude above the SM expectation
(\ref{klpnr4}).
Moreover this bound is substantially weaker than the 
{\it model independent} bound \cite{GRNIR}
from isospin symmetry:
\begin{equation}\label{NRBOUND}
Br(\klpn) < 4.4 \cdot Br(\kpn)
\end{equation}
which through (\ref{EXP1})  gives
\begin{equation}\label{B108}
Br(\klpn) < 1.4 \cdot 10^{-9} ~(90\%~ C.L.)
\end{equation}
\begin{figure}[hbt]
\vspace{0.10in}
\centerline{
\epsfysize=2.0in
\epsffile{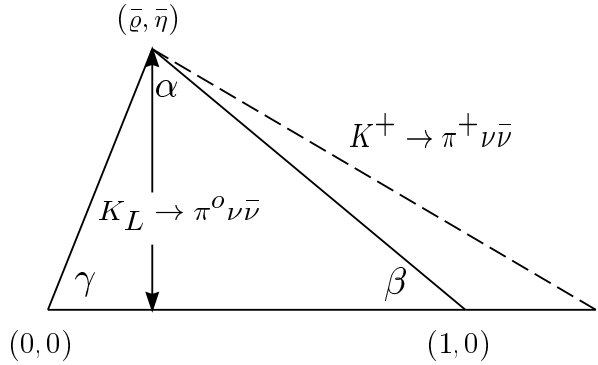}
}
\vspace{0.08in}
\caption{Unitarity triangle from $K\to\pi\nu\bar\nu$.}\label{fig:KPKL}
\end{figure}

\boldmath
\subsection{Unitarity Triangle, {$\sin 2\beta$} and $\gamma$ from 
{$K\to\pi\nu\bar\nu$}}
\unboldmath
\label{sec:Kpnn:Triangle}
The measurement of $Br(\kpn)$ and $Br(\klpn)$ can determine the
unitarity triangle completely, (see Fig.~\ref{fig:KPKL}) \cite{BB4,BB96}.
The explicit formulae can be found in \cite{BLO,Erice,BB4,BB96,BSU}. Most 
interesting in this context are very clean determinations of $\sin 2\beta$ 
and ${\rm Im}\lambda_t$ that are free not only from hadronic uncertainties 
but also parametric uncertainties like $\vcb$ and $m_c$. The determination 
of $\vtd$ is also theoretically clean but its precision depends on the 
accuracy with which $\vcb$ and $m_c$ are known. Also the scale uncertainties 
in $\vtd$ amount to $4.5\%$ at the NLO \cite{BB96,BSU}. They should be 
significantly reduced through
a calculation of NNLO corrections \cite{BGHN} to the charm contribution 
that is in progress and should be available in 2005. 

As emphasized in \cite{BSU} an interesting determination of the angle $\gamma$
can also be made by means of $K\to\pi\nu\bar\nu$.
Assuming that the branching ratios will be  known to within $\pm 10\%$
we expect the following accuracy \cite{BSU}
\be
\sigma(\sin 2\beta)=\pm 0.05, \quad
\sigma({\rm Im}\lambda_t)=\pm 5\%, \quad
\sigma(|V_{td}|)= \pm 7\%, \quad \sigma(\gamma)=\pm 11^\circ~. 
\ee
With the measurments of the branching ratios at the $\pm 5\%$ level 
these estimates change to
\be
\sigma(\sin 2\beta)=\pm 0.03, \quad
\sigma({\rm Im}\lambda_t)=\pm 3\%, \quad
\sigma(|V_{td}|)= \pm 4\%, \quad \sigma(\gamma)=\pm 6^\circ~.
\ee
Further details can be found in \cite{BSU}.

\begin{figure}[htb!]
\begin{center}
{\epsfig{figure=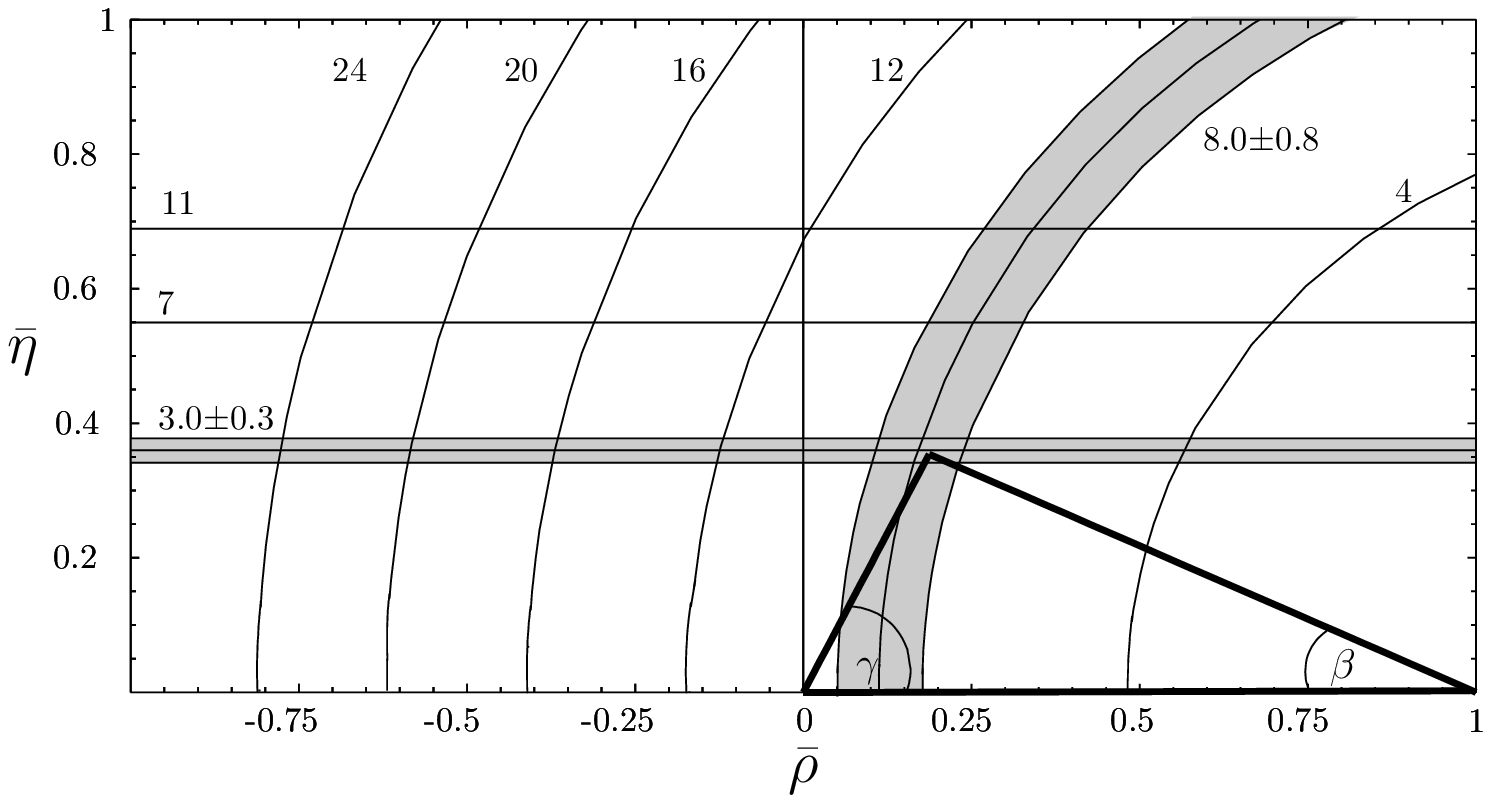,height=6cm}}
\caption[]{The UT from $\klpn$ and $\kpn$ \cite{BSU}. 
}
\label{fig:2012p}
\end{center}
\end{figure}

Clearly the UT resulted from $K\to\pi\nu\bar\nu$ decays could significantly 
differ from the one obtained in Fig.~\ref{fig:figmfv} by means of the
standard UT analysis. This we show in Fig.~\ref{fig:2012p}, where each
crossing point between the horizontal lines from $\klpn$ and the circles 
from $\kpn$ represents possible apex of the UT. The UT shown in this figure 
is the one of Fig.~\ref{fig:figmfv}.

\subsection{Golden Relations}
The comparison of these results with the corresponding determinations 
in $B$ decays
will offer a very good test of flavour dynamics and CP violation in the
SM and a powerful tool to probe the physics
beyond it. 
To this end a number of theoretically clean relations will play an important
role.  We list them here for completeness.

In \cite{BB98} an upper bound on $Br(K^+ \rightarrow \pi^+
\nu \bar{\nu})$ 
has been derived within the SM. This bound depends only on $\vcb$, $X$, 
$\xi$ and $\Delta M_d/\Delta M_s$. With the precise value for the angle 
$\beta$ now available this bound can be turned into a useful formula for 
$Br(K^+ \rightarrow \pi^+ \nu \bar{\nu})$ \cite{AI01}
that expresses 
this branching ratio in terms of theoretically clean observables. 
In the SM and any MFV model this formula reads:
\be \label{AIACD}
Br(K^+ \rightarrow \pi^+ \nu \bar{\nu})=
\bar\kappa_+~\vcb^4 X^2
\Bigg[ \sigma   R^2_t\sin^2\beta+
\frac{1}{\sigma}\left(R_t\cos\beta +
\frac{\lambda^4P_c(X)}{\vcb^2X}\right)^2\Bigg],
\ee
where
\be\label{kapbar}
\bar\kappa_+=\frac{\kappa_+}{\lambda^8}=(7.64\pm 0.09) \cdot 10^{-6},
\qquad
\sigma = \left( \frac{1}{1- \frac{\lambda^2}{2}} \right)^2\,.
\end{equation}
It can be considered as the fundamental formula for a correlation between 
$Br(\kpn)$, $\beta$ and any observable used to determine $R_t$.
This formula is theoretically very clean with
the uncertainties residing only 
in $\vcb$ and $P_c(X)$. However, when one relates $R_t$ to some observable 
new uncertainties could enter. 
In \cite{BB98} and \cite{AI01} it has been proposed to express
$R_t$ through $\Delta M_d/\Delta M_s$ by means of (\ref{Rt}).
This implies an additional uncertainty due to
the value of $\xi$.

In \cite{BSU} it has been pointed out that
if the strategy $(\beta,\gamma)$ of Section~\ref{CKMM} is used to determine 
$R_t$ by means of (\ref{S3}), the resulting formula that relates 
$Br(\kpn)$, $\beta$ and $\gamma$ is even cleaner than the one 
that relates $Br(\kpn)$, $\beta$ and $\Delta M_d/\Delta M_s$.
We have then \cite{BSU}
\be \label{AJBNEW}
Br(K^+ \rightarrow \pi^+ \nu \bar{\nu})=
\bar\kappa_+~\vcb^4 X^2
\Bigg[ \sigma   T_1^2+
\frac{1}{\sigma}\left(T_2 +
\frac{\lambda^4P_c(X)}{\vcb^2X}\right)^2\Bigg],
\ee
where
\be\label{T1T2}
T_1=\frac{\sin\beta\sin\gamma}{\sin(\beta+\gamma)}, \qquad
T_2=\frac{\cos\beta\sin\gamma}{\sin(\beta+\gamma)}.
\ee 

Next, the branching ratio for $\klpn$ allows to determine
 $\bar\eta$ 
\be\label{bareta}
\bar\eta=0.351 
\sqrt{\frac{3.34\cdot 10^{-5}}{\bar\kappa_L}}
\left[\frac{1.53}{X(x_t)}\right]
\left[{\frac{0.0415}{\vcb}}\right]^2
\sqrt{\frac{Br(\klpn)}{3\cdot 10^{-11}}},
\ee
where
\begin{equation}\label{BAKL}
\bar\kappa_L=\frac{\kappa_L}{\lambda^8}=(3.34\pm0.05)\cdot 10^{-5}.
\end{equation}
The determination of $\bar\eta$ in this manner requires the knowledge
of $\vcb$ and $\mt$. With the improved determination of these two
parameters a useful determination of $\bar\eta$ should be possible.

On the other hand, the uncertainty due to $\vcb$ is not present in the 
determination of $\imlt$ as \cite{BB96}:
\begin{equation}\label{imlta}
\IM\lambda_t=1.39\cdot 10^{-4} 
\left[\frac{\lambda}{0.224}\right]
\sqrt{\frac{3.34\cdot 10^{-5}}{\bar\kappa_L}}
\left[\frac{1.53}{X(x_t)}\right]
\sqrt{\frac{Br(\klpn)}{3\cdot 10^{-11}}}.
\end{equation}
This formula  offers
 the cleanest method to measure $\IM\lambda_t$ in the SM and all MFV models
in which the function $X$ takes generally different values than $X(x_t)$.
This determination is even better than the one with the help of 
the CP asymmetries
in $B$ decays that require the knowledge
of $\vcb$ to determine $\IM\lambda_t$. Measuring $Br(\klpn)$
with $10\%$ accuracy allows to determine $\IM\lambda_t$
with an error of $5\%$ \cite{BBL,BB96}.

Next, in the spirit of the analysis in \cite{AJB94} we can use 
the clean CP asymmetries in $B$ decays and determine $\bar\eta$ through 
the $(\beta,\gamma)$ strategy. Using (\ref{S1}) and (\ref{S3}) in 
(\ref{bareta})
we obtain a new ``golden relation" \cite{BSU} 
\be\label{newrel}
\frac{\sin\beta\sin\gamma}{\sin(\beta+\gamma)}
=0.351 
\sqrt{\frac{3.34\cdot 10^{-5}}{\bar\kappa_L}}
\left[\frac{1.53}{X(x_t)}\right]
\left[{\frac{0.0415}{\vcb}}\right]^2
\sqrt{\frac{Br(\klpn)}{3\cdot 10^{-11}}}.
\ee

This relation between $\beta$, $\gamma$ and $Br(\klpn)$, 
is very clean and offers an excellent test of the SM and of its 
extensions. Similarly to the ``golden relation" in (\ref{sin}) it connects 
the observables in $B$ decays with those in $K$ decays and has other 
important virtues that are discussed in \cite{BSU}.

Finally, defining reduced branching ratios
\begin{equation}\label{b1b2}
B_1={Br(\kpn)\over \kappa_+},\qquad
B_2={Br(\klpn)\over \kappa_L}\,.
\end{equation}
one has
 \cite{BB4}
\begin{equation}\label{sin}
\sin 2\beta=\frac{2 r_s}{1+r^2_s}, \qquad
r_s=\sqrt{\sigma}{\sqrt{\sigma(B_1-B_2)}-P_c(X)\over\sqrt{B_2}}
=\cot\beta.
\end{equation}
Thus, within the approximation of \cite{BLO}, $\sin 2\beta$ is
independent of $V_{cb}$ (or $A$) and $m_t$ and as shown in
 \cite{BSU} these dependences are fully negligible.

It should be stressed that $\sin 2\beta$ determined this way depends
only on two measurable branching ratios and on the parameter
$P_c(X)$ which is completely calculable in perturbation theory as discussed 
in the previous section.
Consequently this determination is free from any hadronic
uncertainties and its accuracy can be estimated with a high degree
of confidence. The calculation of  NNLO QCD corrections to $P_c(X)$ 
in \cite{BGHN} will 
certainly improve the accuracy of the determination of $\sin 2
\beta$ from the $K\to\pi\nu\bar\nu$ complex.

\subsection{Concluding Remarks}
As the theorists were able to calculate the branching ratios for 
these decays rather precisely, the future of this field is in the 
hands of experimentalists and depends on the financial support 
that is badly needed.

\section{Minimal Flavour Violation Models}\label{MFVM}
\setcounter{equation}{0}
\subsection{Preliminaries}
As discussed in Section~\ref{THF}, these are the simplest extensions of the
SM. A detailed review of these models can be found in \cite{Zakopane}.
 Here I would like first
to list five interesting properties of these models that are independent 
of particular parameters present in these models. Other relations can be
found in \cite{REL,Zakopane}. These are:
\begin{itemize}
\item
There exists a universal unitarity triangle (UUT) \cite{UUT} common to all 
these models and the SM that can be constructed by using measurable 
quantities that depend on the CKM parameters but are not polluted by the 
new parameters present in the extensions of the SM. 
The UUT can be constructed, for instance, by using $\sin 2\beta$ from 
$\calamix(\psi K_S)$ and the ratio $\Delta M_s/\Delta M_d$. 
The relevant formulae can be found in Section~\ref{UT-Det} and in
 \cite{UUT,BF01}, where also other 
quantities suitable for the determination of the UUT are discussed.
\item
The golden relation \cite{BB4,BF01}:
\be
(\sin 2\beta)_{\psi K_S}=(\sin 2\beta)_{\pi\nu\bar\nu}
\ee
\item
For given $\sin 2\beta $ and $Br(\kpn)$ only two values of 
$Br(\klpn)$ are possible 
in the full class of MFV models, independently of any new parameters 
present in these models \cite{BF01}. 
These two values correspond to two signs of the function $X(v)$. 
Consequently, measuring $Br(\klpn)$ will 
either select one of these two possible values or rule out all MFV models.
A very recent analysis shows that the case of $X(v)<0$ is very unlikely 
\cite{MFV05}, living basically only one value for $Br(\klpn)$ once 
$Br(\kpn)$ has been precisely measured.
\item
There exists a correlation between $Br(B_{d,s}\to\mu\bar\mu)$ and 
$\Delta M_{d,s}$ \cite{AJB03}:
\be
\frac{Br(B_{s}\to\mu\bar\mu)}{Br(B_{d}\to\mu\bar\mu)}
=\frac{\hat B_{d}}{\hat B_{s}}
\frac{\tau( B_{s})}{\tau( B_{d})} 
\frac{\Delta M_{s}}{\Delta M_{d}} \nonumber 
\ee
where $\tau(B_{q})$ are $B$-meson life-times  and $\hat B_{q}$ 
non-perturbative parameters in $\Delta M_{q}$ with 
$\hat B_{d}=1.34\pm0.12$ and $\hat B_{s}=1.34\pm0.12$
obtained in lattice simulations \cite{CERNCKM}.
This correlation is practically free of theoretical uncertainties as 
$\hat B_{s}/\hat B_{d}=1$ up to small  $SU(3)$ breaking corrections.
\item
Similar correlations between $Br(B_{d,s}\to\mu\bar\mu)$ and 
$\Delta M_{d,s}$, respectively, allow rather precise predictions for 
$Br(B_{d,s}\to\mu\bar\mu)$ within the MFV models once $\Delta M_{d,s}$ 
are known  \cite{AJB03}.
Indeed one finds
$(q=s,d)$
\be\label{R2}
\frac{Br(B_{q}\to\mu\bar\mu)}{\Delta M_{q}}
=4.4 \cdot 10^{-10} \frac{\tau(B_{q})}{\hat B_{q}} F(v),
 \qquad F(v)=\frac{Y^2(v)}{S(v)},
\ee
In the SM, 
$F_{\rm SM}=0.40$. 
\end{itemize}
\subsection{Universal Unitarity Triangle}
The presently available quantities that do not depend on the new physics 
parameters within the MFV models and therefore can be used to determine 
the UUT are $R_t$ from $\Delta M_d/\Delta M_s$ by means of (\ref{Rt}),
$R_b$ from $\vub$ by means of (\ref{2.94})   and $\sin 2\beta$ 
extracted from the CP asymmetry in $B^0_d\to \psi K_S$. 
Using only these three quantities, we show in the UUT column of
table~\ref{mfv}   
the results for various quantities of 
interest related to this UUT \cite{BSU}.
A similar analysis has been done in \cite{BUPAST,AMGIISST,UTfit}.
In particular \cite{UTfit} finds
\be\label{UUTfit}
\bar\eta=0.353\pm0.028, \qquad \bar\varrho=0.191\pm 0.046
\ee 
in a good agreement with the results of table~\ref{mfv}.

It should be stressed that any MFV model that is inconsistent with the 
 UUT column in table~\ref{mfv} is ruled out. 
We observe that there is little room for MFV models that in their predictions 
for UT differ significantly from the SM. It is also clear that to 
distinguish the SM from the MFV models on the 
basis of the analysis of the UT of Section 6, will require 
a considerable reduction of 
theoretical uncertainties.
\subsection{Models with Universal Extra Dimensions}
In view of the difficulty in distinguishing various MFV models on the 
basis of the standard analysis of UT from each other, it is essential to 
study other FCNC processes as rare $B$ and $K$ decays and radiative 
$B$ decays like
$B\to X_s\gamma$ and $B\to X_s\mu^+\mu^-$. In the case of MSSM at low 
$\tan \beta$ such  analyses can be found in \cite{BRMSSM,ALIHILL}. 
In 2003 a very extensive analysis
of all relevant FCNC processes in a SM with one universal extra dimension 
\cite{appelquist:01} has been presented in \cite{BSW02,BPSW}. In this model 
all standard model fields can propagate in the fifth dimension and the 
FCNC processes are affected by the exchange of the Kaluza-Klein particles 
in loop diagrams.  
The most interesting 
results of \cite{BSW02,BPSW} are the enhancements of $Br(\kpn)$ and 
$Br(B\to X_s\mu^+\mu^-)$, strong suppressions of $Br(B\to X_s\gamma)$
(see also \cite{Deshpande})
and $Br(B\to X_s~{\rm gluon})$ and a significant downward shift of the 
zero $\hat s_0$ in the forward-backward asymmetry in $Br(B\to X_s\mu^+\mu^-)$.

\begin{figure}[hbt]

  \centering 
 \psfragscanon
  \psfrag{bsgammabsgammabsgamma}{ \shortstack{\\ \\ $(Br(B\to
 X_s\gamma)\times 10^4)^\frac12$ ${}$}}
  \psfrag{hats0}[][]{ \shortstack{\\ $\hat{s}_0 $ }}
      \resizebox{.36\paperwidth}{!}{\includegraphics[]{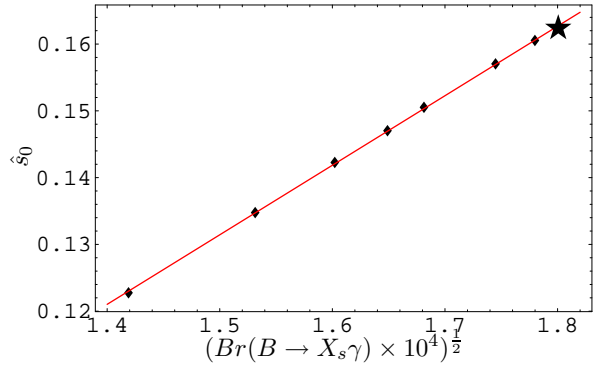}}
    
  \caption[]{\small\label{corrplot} Correlation between
    $\sqrt{Br(B\to X_s\gamma)}$  and $\hat s_0$ \cite{BPSW}. 
    The dots are the results in the ACD
    model (see below) with the compactification scale
    $200,250,300,350,400,600$ and $1000$ GeV  and the star
    denotes the SM value.
}
\end{figure}

 As pointed out in \cite{BPSW} 
 this downward shift of $\hat s_0$ is correlated with the suppression of 
$Br(B\to X_s\gamma)$. This is seen  in Fig.~\ref{corrplot}. This property 
is characteristic for all MFV models and is verified in a MSSM with MFV 
even after the inclusion of NNLO QCD corrections \cite{BBE04}.

\subsection{FCNC Processes in the Littlest Higgs Model}
\label{LHIGGS}

The little Higgs models \cite{LH1}-\cite{LH5} offer 
an attractive and a rather simple solution to  the gauge hierarchy problem.
In these models
the electroweak Higgs boson is regarded as a pseudo-goldstone boson of a 
certain global 
symmetry that is broken spontaneously at a scale 
$\Lambda \sim 4 \pi f \sim \mathcal{O}\left(\textrm{10 TeV}\right)$, 
 much higher than the vacuum expectation value $v$ of the standard Higgs 
doublet. The Higgs field remains then light, 
being protected by 
the approximate global symmetry from acquiring quadratically divergent 
contributions to its mass at the one-loop level. 
On the diagramatic level the new heavy particles present in these models
allow to cancel,
analogously to supersymmetric 
particles, the quadratic divergencies in question.
Reviews of the little Higgs models can be found in \cite{LHREV}.

One of the simplest models of this type is the ``Littlest Higgs'' model 
\cite{LH4} (LH) in which, in addition to the Standard Model (SM) particles, 
new charged heavy vector bosons ($W_H^\pm$), a neutral heavy vector 
boson ($Z_H$), a heavy photon ($A_H$), a heavy top quark ($T$) and a 
triplet of heavy Higgs scalars ($\Phi^{++}$, $\Phi^{+}$, $\Phi^{0}$) 
are present. The details of this model have been 
worked out in \cite{Logan} and the constraints from various processes, 
in particular from electroweak precision observables and direct new particles
searches, have been extensively discussed in \cite{Logan}-\cite{PHEN6}. 
It has been found that 
except for the heavy photon $A_H$, that could still be as ``light'' as 
$500~\textrm{GeV}$, the masses of the remaining particles are constrained 
to be significantly larger than $1~ \textrm{TeV}$.

The new particles can also contribute to FCNC processes.
In \cite{BPU04}, in collaboration with Anton Poschenrieder and Selma Uhlig,
the $K^{0}-\bar{K}^{0}$, $B_{d,s}^{0}-\bar{B}_{d,s}^{0}$ 
mixing mass differences $\Delta M_K$, $\Delta M_{d,s}$ and the 
CP-violating parameter $\varepsilon_{K}$ have been calculated in the LH model. 
We have found that
even for $f/v$ as low as $5$, the enhancement of 
$\Delta M_{d}$ for the $t-T$ mixing parameter $x_L\le 0.8$
amounts to at most $20\%$. Similar comments apply to $\Delta M_s$
and $\varepsilon_{K}$.
The correction to $\Delta M_{K}$ is negligible.
Our results have recently been confirmed in \cite{IND1}. 
Only for $x_L\ge 0.85$ significantly larger effects could be found 
\cite{BPU04,DEC05}.
Significant effects could be present in $D^0-\bar D^0$ mixing \cite{QUAD}, 
where 
in contrast to  processes involving external down quarks, FCNC transitions 
are already present at the tree level.

Concerning FCNC decay processes, the corrections to  $B\to X_s\gamma$ 
have been found to be small \cite{BSG}, while the analysis of
$\klpn$ in \cite{IND2} resulted in  a large enhancement of the relevant
branching ratio. 

In \cite{BPU05} we have extended our study of FCNC processes in the LH model 
to
the rare decays $\kpn$, $\klpn$, $B_{s,d}\to\mu^+\mu^-$ and $B\to
X_{s,d}\nu\bar\nu$. Preliminary results appeared in \cite{DEC05}. 
Our results differ from those of \cite{IND2} in that we find only 
 very small corrections to $\kpn$ and $\klpn$. On the other hand a significant 
enhancement of $Br(B_{s}\to\mu^+\mu^-)$ has been identified.

The details of this analysis will appear in \cite{BPU05}. Although no 
dramatic modifications of the SM expectations for FCNC decays have been 
found, the analysis is interesting from the technical point of view and 
can be considered as a symphony of new particle contributions to rare 
decays. Indeed, this analysis involving
many  diagrams, in particular the $Z^0$, $Z_H$ and $A_H$ penguins 
with the heavy $T$, $W_H$, 
$\Phi^{\pm}$
 and the ordinary quarks exchanges has a certain 
beauty in view of only three new parameters involved.
\boldmath
\subsection{Upper Bounds on rare $K$ and $B$ Decays from MFV}
\unboldmath
Very recently a  detailed analysis of several branching ratios
for rare $K$ and $B$ decays in MFV models has been performed in 
\cite{MFV05}. Using the presently available information on the UUT, 
summarized in (\ref{UUTfit}), and from the measurements of $Br(B\to
X_s\gamma)$, $Br(B\to X_sl^+l^-)$ and $Br(\kpn)$, the upper bounds on various
branching ratios within the MFV scenario have been found. They are collected 
in Table~\ref{brMFV} together with the results within the SM. Moreover with 
$95\%$ probability one finds
\be
X(v)_{\rm max}=1.95, \qquad Y(v)_{\rm max}=1.43, \qquad Z(v)_{\rm max}=1.46,
\ee
to be compared with $X=1.54$, $Y=0.99$ and $Z=0.69$ in the SM.
Finally, anticipating that the leading role in constraining this kind of
physics will eventually be taken over by $\kpn$, $\klpn$ and
$B_{s,d}\to\mu^+\mu^-$, that are dominated by the function $C(v)$, reference 
\cite{MFV05} provides plots for several branching ratios as functions of 
$C(v)$.

\begin{table*}[t]
\small{
\begin{center}
\begin{tabular}{|c|c|c|c|c|}
\hline
{Branching Ratios} &  MFV (95\%) &  SM (68\%) &  SM (95\%) & exp
 \\ \hline
$Br(\kpn)\times 10^{11}$ & $< 11.9$ & $8.3 \pm 1.2$ &  $[6.1,10.9]$
& $(14.7^{+13.0}_{-8.9})$ \cite{E949}
\\ \hline
$Br(\klpn)\times 10^{11}$  & $< 4.59$ &  $3.08 \pm 0.56$ &  $[2.03,4.26]$ &
 $ < 5.9 \cdot10^{4}$  \cite{KTeV00X}
\\ \hline
$Br(\kmm)_{\rm SD}\times 10^{9} $ & $< 1.36$ & $0.87 \pm 0.13$ &  $[0.63,1.15]$ & -
\\ \hline
$Br(B\to X_s\nu\bar\nu)\times 10^{5}$ & $<5.17$ &  $3.66 \pm 0.21$ &  $[3.25,4.09]$
&  $<64 $ 
\cite{Barate:2000rc}
\\ \hline
$Br(B\to X_d\nu\bar\nu)\times 10^{6}$ &  $<2.17$ & $1.50 \pm 0.19$ &  $[1.12,1.91]$
& -
\\ \hline
$Br(B_s\to \mu^+\mu^-)\times 10^{9}$ &  $< 7.42$ & $3.67 \pm 1.01$ &  $[1.91,5.91]$
& $<2.7\cdot 10^{2}$  \cite{Herndon:2004tk}
\\ \hline
$Br(B_d\to \mu^+\mu^-)\times 10^{10}$ &  $< 2.20$ & $1.04 \pm 0.34$ &  $[0.47,1.81]$
& $<1.5 \cdot 10^3$ \cite{Herndon:2004tk}
\\ \hline
\end{tabular}
\caption[]{ Upper bounds for rare decays in MFV models at $95 \%$
  probability, the corresponding values in the SM (using inputs from
  the UUT analysis) and the available experimental information. See
  \cite{MFV05} for details.  }
\label{brMFV}
\end{center}
}
\end{table*}

The main message from from \cite{MFV05} is the following one:

The existing constraints coming from $K^+\to\pi^+\nu\bar\nu$, $B\to
X_s\gamma$ and $B\to X_s l^+l^-$ do not allow within the MFV scenario
of \cite{UUT} for substantial departures of the branching ratios for
all rare $K$ and $B$ decays from the SM estimates. This is evident
{}from Table~\ref{brMFV}.

This could be at first sight a rather pessimistic message. On the other 
hand it implies that finding practically any branching ratio enhanced 
by more than a factor of two with respect to the SM will automatically 
signal either the presence of new CP-violating phases or new operators, 
strongly suppressed within the SM, at work.
 In particular, recalling that in most extensions of the SM the
  decays $K\to\pi\nu\bar\nu$ are governed by the single $(V-A)\otimes
  (V-A)$ operator, the violation of the upper bounds on at least one
  of the $K\to\pi\nu\bar\nu$ branching ratios, will either signal the
  presence of new complex weak phases at work or new contributions
  that violate the correlations between the $B$ decays and $K$ decays
  present in models with MFV.

\subsection{Final Comments on MFV}
Assuming that the MFV scenario will survive future tests, the next step
will be to identify the correct model in this class. Clearly, direct searches
at high energy colliders can rule out or identify specific extensions of the
SM. But also FCNC processes can play an important role in this context,
provided the theoretical and experimental uncertainties in some of them will
be sufficiently decreased. In this case, by studying simultaneously several 
branching ratios it should be in principle possible to select the correct MFV 
models by just identifying the pattern of enhancements and suppressions
relative to the SM that is specific to a given model. If this pattern 
is independent of the values of the parameters defining the model, no 
detailed quantitative analysis of the enhancements and suppressions 
is required in order to rule it out. As an example the
distinction between the MSSM with MFV and the models with one universal 
extra dimension should be straightforward:
\begin{itemize}
\item In the MSSM with MFV the branching ratios for $\kpn$, $\klpn$,
  $B\to X_d\nu\bar\nu$ and $B_d\to\mu^+\mu^-$ are generally suppressed
  relative to the SM expectations, while those governed by $V_{ts}$
  like $B\to X_s\nu\bar\nu$, $B_s\to\mu^+\mu^-$ and $B\to X_s\gamma$
  can be enhanced or suppressed depending on the values of parameters
  involved~\cite{BRMSSM}.
\item In the model with one universal extra dimension analyzed 
in~\cite{BSW02,BPSW}, branching 
ratios for essentially all rare decays are enhanced, the enhancement
being stronger for the decays governed by $V_{ts}$ than for those 
where $V_{td}$ 
is involved. A prominent exception 
is the suppression of $B\to X_{s,d}\gamma$ \cite{BPSW,Deshpande}.
\end{itemize} 
Finally, if MFV will be confirmed, and some new particles
  will be observed, the rare processes discussed in this work will
  constitute a most powerful tool to probe the spectrum of the NP
  model, which might not be entirely accessible via direct studies at
  the LHC.

\section{New Weak Phases}\label{NPHASE}
\setcounter{equation}{0}
\subsection{Preliminaries}
We will now consider the class C of Section~\ref{CDECAYS}.
In this class of models the dominant operators are as in the MFV models 
(class A) but the
master functions become now complex quantities. If the new weak phases
are large, the deviations from the SM can be spectacular as we
will see below.

Let us consider first $\Delta F=2$ transitions.
In the MFV scenario discussed in the previous section, the NP effects enter 
universally
in $K^0-\bar K^0$,  $B_d^0-\bar B_d^0$ and $B_s^0-\bar B_s^0$ mixings 
through the
single real function $S(v)$, implying strong correlations between new physics
effects in the $\Delta F=2$ observables
of 
$K$ and $B$ decays. When new complex weak phases are present, the situation
could be more involved with $S(v)$ replaced  by
\be
S_K(v)=|S_K(v)|e^{i\theta_K}, \quad 
S_d(v)=|S_d(v)|e^{i\theta_d}, \quad
S_s(v)=|S_s(v)|e^{i\theta_s},
\ee
for $K^0-\bar K^0$,  $B_d^0-\bar B_d^0$ and $B_s^0-\bar B_s^0$ mixing,
respectively. If these three functions are different from each other, 
some universal properties found in the MFV models A are lost. In addition the mixing
induced CP asymmetries in $B$ decays will not measure the angles of the UT
but only sums of these angles and of $\theta_i$. 
An example is given in (\ref{mast}).
Yet, within each class of 
$K$, $B_d$ and $B_s$ decays, the NP effects of this sort will be
universal. 
Scenarios of this type have been considered for instance in 
\cite{WPDELTAF2,Laplace,FLISMA,AI01,Berg,BSU}.

New weak phases could enter also decay amplitudes. As now these effects enter
in principle differently in each decay, the situation can be very involved
with many free parameters, no universal effects and little predictive power.

Here I will only discuss one scenario, discussed first in 
\cite{BRS}--\cite{Buchalla:2000sk}
and recently in the context of a simultaneous analysis of prominent 
non--leptonic
$B$ decays like $B\to\pi\pi$, $B\to\pi K$, $B\to \psi K_S$ and $B\to\phi K_S$
and equally prominent rare decays like $K\to\pi\nu\bar\nu$,
$K_L\to\pi^0e^+e^-$, $B_{s,d}\to\mu^+\mu^-$, $B\to X_{s,d} e^+e^-$ and $\epe$
in \cite{BFRS-BIG,BFRS-PRL}. 
It is the scenario of enhanced $Z^0$ penguins with a
large complex weak 
phase
in which the only modification with respect to the MFV models 
is the replacement 
in the $Z^0$ penguin function
$C(v)\to |C(v)|e^{i\theta_C}$ that makes the master functions $X(v)$, $Y(v)$
and $Z(v)$ of Section~4 to be complex quantities:
\be
X(v)=|X(v)|e^{i\theta_X}, \qquad Y(v)=|Y(v)|e^{i\theta_Y}, \qquad 
Z(v)=|Z(v)|e^{i\theta_Z}.
\ee
The magnitudes and phases of these three functions are correlated with each
other as they depend only on $|C(v)|e^{i\theta_C}$ and other smaller
contributions that can be set to their SM values.
This new analysis has been motivated
by an interesting  experimental situation in $B\to\pi\pi$ and $B\to\pi K$
decays that we will summarize below.
\subsection{Basic Strategy}
The present studies of non-leptonic two-body $B$ decays 
and of rare $K$ and $B$ decays are very important as they will teach us 
both about the non-perturbative aspects of QCD and about the perturbative 
electroweak physics at very short distances. For the analysis of these
modes, it is essential to have a strategy available that could clearly 
distinguish between non-perturbative QCD effects and short-distance 
electroweak effects. A strategy that in the case of deviations from 
the SM expectations would allow us transparently to identify a possible 
necessity for modifications in our understanding of hadronic effects and 
for a change of the SM picture of electroweak flavour-changing 
interactions at short-distance scales.

In \cite{BFRS-BIG,BFRS-PRL}, we have developed a strategy
that allows us to address these questions in a systematic manner. It
encompasses non-leptonic $B$ and $K$ decays and rare $K$ and $B$ decays 
but has been at present used primarily for the analysis of 
$B\to\pi\pi$ and $B\to\pi K$ systems and rare $K$ and $B$ decays. In what 
follows I will 
summarize the basic ingredients of our strategy 
and list the most important results. The basic concepts can be found in 
\cite{BFRS-PRL,BFRS-BIG} and an update has been presented 
in \cite{BFRS-UPDATE}. The discussion presented below is entirely based 
on these papers and borrows a lot from the talk in \cite{SUSY04}.
In order to illustrate our strategy in explicit terms, we shall 
consider a simple extension of the SM in which new physics (NP) 
enters dominantly through enhanced $Z^0$ penguins involving a 
CP-violating weak phase, as advertised above. As we will see below, 
this choice is 
dictated by the pattern of the data on the $B\to\pi K$ observables 
and the great predictivity of this scenario.  
Our strategy consists of three interrelated 
steps, and has the following logical structure:

\vspace*{0.3truecm}

\noindent
{\bf Step 1:}

\noindent
Since $B\to\pi\pi$ decays and the usual analysis of the unitarity 
triangle (UT) are only insignificantly affected by electroweak (EW) 
penguins, the $B\to\pi\pi$ system can be 
described as in the SM. Employing the $SU(2)$ isospin flavour symmetry
of strong interactions and the information on $\gamma$ from the
UT fits, we may extract the hadronic parameters of the $B\to\pi\pi$ system, 
and find large 
non-factorizable contributions, which are in particular reflected by 
large CP-conserving strong phases. Having these parameters at hand, we 
may then also predict the direct and mixing-induced CP asymmetries
of the $B_d\to\pi^0\pi^0$ channel. See Table~\ref{asymtab}.
 A future measurement of one of these
observables allows a determination of $\gamma$.

\vspace*{0.3truecm}

\noindent
{\bf Step 2:}

\noindent
If we use the $SU(3)$ flavour symmetry and plausible dynamical 
assumptions, we can determine the hadronic $B\to\pi K$ parameters 
from the hadronic parameters of 
the $B\to\pi\pi$ system found in Step 1.
Subsequently, this allows us
 to calculate the $B\to\pi K$ 
observables in the SM. Interestingly, we find agreement with the pattern 
of the $B$-factory data for those observables where EW penguins play only 
a minor r\^ole. On the other hand, the observables receiving significant
EW penguin contributions do {\it not} agree with the experimental picture, 
thereby suggesting NP in the EW penguin sector. Indeed, a detailed analysis
shows \cite{BFRS-BIG,BFRS-PRL,BFRS-UPDATE}
that we may describe all the currently available data through moderately 
enhanced EW penguins with a large CP-violating NP phase around $-90^\circ$.
A future test of this scenario will be provided by the CP-violating 
$B_d\to \pi^0 K_{\rm S}$ observables, which we may predict. 
See Table~\ref{asymtab}.
Moreover, 
we may obtain valuable insights into $SU(3)$-breaking effects, which 
support our working assumptions, and may also determine the UT angle 
$\gamma$, in remarkable agreement with the UT fit of Section 6.  

\vspace*{0.3truecm}

\noindent
{\bf Step 3:}

\noindent
In turn, the modified EW penguins with their large CP-violating 
NP phase have important implications for rare $K$ and $B$ decays.
Interestingly, several predictions differ significantly from the SM 
expectations and should easily be identified once the data improve. 
Similarly, we may explore specific NP patterns in other non-leptonic 
$B$ decays such as $B_d\to\phi K_{\rm S}$.

\begin{figure}
\begin{center}
\includegraphics[width=11.5cm,angle=-90]{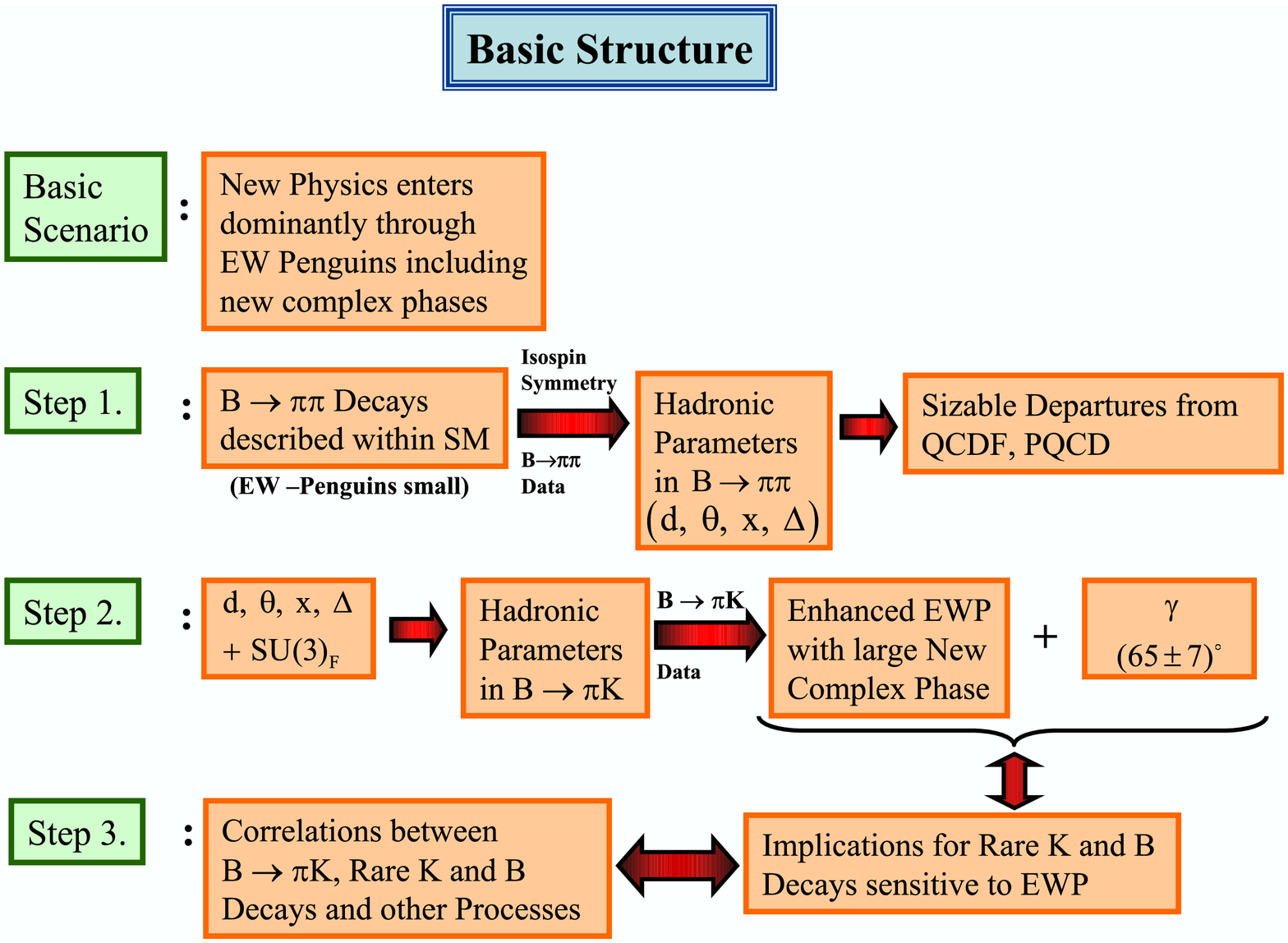}
\end{center}
\caption{Outline of the strategy of \cite{BFRS-BIG,BFRS-PRL}\label{chart}}
\end{figure}

\vspace*{0.6truecm}

A chart of the three steps in question is given in Fig.\ \ref{chart}.
Before going into the details it is important to emphasize that our strategy is
valid both in the SM and all SM extensions in which NP enters
predominantly through the EW penguin sector. This means that even if
the presently observed deviations from the SM in the $B\to\pi K$ sector would
diminish with improved data, our strategy would still be useful in correlating
the phenomena in $B\to\pi\pi$, $B\to\pi K$ and rare $K$ and $B$ decays within
the SM. If, on the other hand, the observed deviations from the SM in
$B\to\pi\pi$ decays would not be attributed to the modification in hadron
dynamics but to NP contributions, our approach should be properly
generalized.

\boldmath
\subsection{$B\to\pi\pi$ decays}
\unboldmath
The central quantities for our analysis of the $B\to\pi\pi$ decays are the
ratios
\begin{eqnarray}
R_{+-}^{\pi\pi}&\equiv&2\left[\frac{\mbox{Br}(B^+\to\pi^+\pi^0)
+\mbox{Br}(B^-\to\pi^-\pi^0)}{\mbox{Br}(B_d^0\to\pi^+\pi^-)
+\mbox{Br}(\bar B_d^0\to\pi^+\pi^-)}\right]
\frac{\tau_{B^0_d}}{\tau_{B^+}}\label{Rpm-def}\\
R_{00}^{\pi\pi}&\equiv&2\left[\frac{\mbox{Br}(B_d^0\to\pi^0\pi^0)+
\mbox{Br}(\bar B_d^0\to\pi^0\pi^0)}{\mbox{Br}(B_d^0\to\pi^+\pi^-)+
\mbox{Br}(\bar B_d^0\to\pi^+\pi^-)}\right]\label{R00-def}
\end{eqnarray}
of the CP-averaged $B\to\pi\pi$ branching ratios, and the CP-violating 
observables provided by the time-dependent rate asymmetry 
\begin{eqnarray}
\lefteqn{\frac{\Gamma(B^0_d(t)\to \pi^+\pi^-)-\Gamma(\bar B^0_d(t)\to 
\pi^+\pi^-)}{\Gamma(B^0_d(t)\to \pi^+\pi^-)+\Gamma(\bar B^0_d(t)\to 
\pi^+\pi^-)}}\nonumber\\
&&={\cal A}_{\rm CP}^{\rm dir}(B_d\to \pi^+\pi^-)\cos(\Delta M_d t)+
{\cal A}_{\rm CP}^{\rm mix}(B_d\to \pi^+\pi^-)
\sin(\Delta M_d t)\label{rate-asym}\,.
\end{eqnarray}
The current status of the $B\to\pi\pi$ data together with the relevant
references can be found in Table \ref{tab:Bpipi-input}.
\begin{table}
\begin{center}
\begin{tabular}{|c||c|c|}
\hline
Quantity & Input & Exp.\ reference
\\ \hline
 \hline
$\mbox{Br}(B^\pm\to\pi^\pm\pi^0)/10^{-6}$ & $5.5\pm0.6$ &
\cite{BaBar-Bpi0pi0, Chao:2003ue}
\\ \hline 
$\mbox{Br}(B_d\to\pi^+\pi^-)/10^{-6}$ & $4.6\pm0.4$ & 
\cite{Aubert:2002jb, Chao:2003ue} \\ \hline
$\mbox{Br}(B_d\to\pi^0\pi^0)/10^{-6}$ & $1.51\pm0.28$ & 
\cite{BaBar-Bpi0pi0, Belle-Bpi0pi0-new}
\\ \hline
 \hline
$R_{+-}^{\pi\pi}$ & $2.20\pm0.31$ & 
\\ \hline
$R_{00}^{\pi\pi}$ & $0.67\pm0.14$ & 
\\ \hline
 \hline
${\cal A}_{\rm CP}^{\rm dir}(B_d\to \pi^+\pi^-)$ & $-0.37\pm 0.11$ & 
\cite{BaBar-CP-Bpipi, Belle-CP-Bpipi}
\\ \hline
${\cal A}_{\rm CP}^{\rm mix}(B_d\to \pi^+\pi^-)$ & $+0.61\pm0.14$ & 
\cite{BaBar-CP-Bpipi, Belle-CP-Bpipi}
\\ \hline
\end{tabular}
\end{center}
\caption[]{The current status of the $B\to\pi\pi$ input data for our
strategy, with averages taken from \cite{HFAG}. For the 
evaluation of $R_{+-}^{\pi\pi}$, we have used the life-time ratio
$\tau_{B^+}/\tau_{B^0_d}=1.086\pm0.017$ \cite{PDG}.}\label{tab:Bpipi-input}
\end{table}
The so-called ``$B\to\pi\pi$ puzzle'' is reflected in a surprisingly 
large value of ${Br}(B_d\to\pi^0\pi^0)$ and a somewhat small
value of ${Br}(B_d\to\pi^+\pi^-)$, which results in
large values of both $R_{00}^{\pi\pi}$ and $R_{+-}^{\pi\pi}$. For instance,
the central values calculated within QCD factorization (QCDF) \cite{Be-Ne} 
give $R_{+-}^{\pi\pi}=1.24$ and $R_{00}^{\pi\pi}=0.07$, although in the
scenario ``S4'' of \cite{Be-Ne} values 2.0 and 0.2, respectively, can
be obtained. As  pointed out in \cite{BFRS-PRL}, these data
indicate important non-factorizable contributions rather than NP
effects, and can be perfectly accommodated in the SM. The same applies
to the NP scenario considered in \cite{BFRS-PRL,BFRS-BIG}, in which
the EW penguin contributions to $B\to\pi\pi$ are marginal.

In order to address this issue in explicit terms, we use the isospin
symmetry to find
\begin{eqnarray}
\sqrt{2}A(B^+\to\pi^+\pi^0)&=&-[\tilde T+\tilde C] = 
-[T+C]\label{B+pi+pi0}\\
A(B^0_d\to\pi^+\pi^-)&=&-[\tilde T + P]\label{Bdpi+pi-}\\
\sqrt{2}A(B^0_d\to\pi^0\pi^0)&=&-[\tilde C - P].\label{Bdpi0pi0}
\end{eqnarray}
The individual amplitudes of (\ref{B+pi+pi0})--(\ref{Bdpi0pi0}) can
be expressed as
\begin{eqnarray}
P&=&\lambda^3 A({\cal P}_t-{\cal P}_c)\equiv\lambda^3 A {\cal P}_{tc}
\label{P-def}\\
\tilde T &=&\lambda^3 A R_b e^{i\gamma}\left[{\cal T}-\left({\cal 
P}_{tu}-{\cal E}\right)\right]\label{T-tilde}\\
\tilde C &=&\lambda^3 A R_b e^{i\gamma}\left[{\cal C}+\left({\cal P}_{tu}-
{\cal E}\right)\right],\label{C-tilde}
\end{eqnarray}
where $\lambda$, $A$ and $R_b$ have been defined in Section 3.
The ${\cal P}_q$ describe the strong 
amplitudes of QCD penguins with internal $q$-quark exchanges 
($q\in\{t,c,u\}$), including annihilation and exchange penguins,
while ${\cal T}$ and ${\cal C}$ are the strong amplitudes of 
colour-allowed and colour-suppressed tree-diagram-like topologies, 
respectively, and ${\cal E}$ denotes the strong amplitude of an 
exchange topology. The amplitudes $\tilde T$ and $\tilde C$ differ from
\begin{equation}
T =\lambda^3 A R_b e^{i\gamma}{\cal T}, \quad
C =\lambda^3 A R_b e^{i\gamma}{\cal C}
\end{equation}
through the $({\cal P}_{tu}-{\cal E})$ pieces, which may play
an important r\^ole \cite{PAP0}. Note that these terms contain 
also the ``GIM penguins'' with internal up-quark exchanges, 
whereas their ``charming penguin'' counterparts enter in $P$ 
through ${\cal P}_{c}$, as can be seen in (\ref{P-def}) 
\cite{PAP0,c-pen,BFM,BPRS}. 

In order to characterize the dynamics of the $B\to\pi\pi$ system, we
introduce four hadronic parameters $d$, $\theta$, $x$ and $\Delta$ 
through
\begin{equation}\label{d-theta-def}
d e^{i\theta}=
-\left|\frac{P}{\tilde T}\right|e^{i(\delta_P-\delta_{\tilde T})}\,,
\quad
x e^{i\Delta}=
\left|\frac{\tilde C}{\tilde T}\right|e^{i(\delta_{\tilde C}-
\delta_{\tilde T})}\,,
\end{equation}
with $\delta_i$ being strong phases.
Using this parametrization, we have
\begin{equation}\label{Rpipi-gen}
R_{+-}^{\pi\pi}=F_1(d,\theta,x,\Delta;\gamma), \quad
R_{00}^{\pi\pi}=F_2(d,\theta,x,\Delta;\gamma),
\end{equation}
\begin{equation}\label{CP-Bpipi-gen}
{\cal A}_{\rm CP}^{\rm dir}(B_d\to \pi^+\pi^-)=
G_1(d,\theta;\gamma), \quad
{\cal A}_{\rm CP}^{\rm mix}(B_d\to \pi^+\pi^-)=
G_2(d,\theta;\gamma,\phi_d),
\end{equation}
with explicit expressions for $F_1$, $F_2$, $G_1$ and $G_2$
given in \cite {BFRS-BIG}. Taking then as the input
\begin{equation}\gamma=(65\pm7)^\circ\,,\quad
\phi_d=2\beta=\left(46.5^{+3.2}_{-3.0}\right)^\circ
\end{equation}
and the data for $R_{+-}^{\pi\pi}$, $R_{00}^{\pi\pi}$,
${\cal A}_{\rm CP}^{\rm dir}$ and ${\cal A}_{\rm CP}^{\rm mix}$
of Table~\ref{tab:Bpipi-input}, 
we obtain a set of four equations with the four unknowns
$d$, $\theta$, $x$ and $\Delta$. Interestingly, as demonstrated in
\cite{BFRS-BIG,BFRS-UPDATE}, a unique solution for these
parameters can be found
\begin{equation} \label{dthetaxdelta}
d=0.51^{+0.26}_{-0.20}, \quad \theta=\left(140^{+14}_{-18}\right)^\circ, \quad 
x=1.15^{+0.18}_{-0.16}, \quad \Delta=-\left(59^{+19}_{-26}\right)^\circ.
\end{equation}
The large values of the strong phases $\theta$ and $\Delta$ and the
large values of $d$ and $x$ signal departures from the picture of
QCDF. Going back to (\ref{T-tilde}) and (\ref{C-tilde}) we observe
that these effects can be attributed to the enhancement of the
$({\mathcal{P}}_{tu}-{\cal E})$ terms that in turn suppresses 
${\tilde T}$ and enhances ${\tilde C}$.
In this manner, ${Br}(B_d\to\pi^+\pi^-)$ and
${Br}(B_d\to\pi^0\pi^0)$ are suppressed and enhanced,
respectively. 

With the hadronic parameters at hand, we can predict the direct
and mixing-induced CP asymmetries of the $B_d\to\pi^0\pi^0$ channel.
They are given in Table~\ref{asymtab}. As seen there
these predictions, while still subject to large uncertainties, have
been confirmed by the most recent data. On the other hand, as
illustrated in \cite{BFRS-BIG}, a future precise measurement of
${\cal A}_{\rm CP}^{\rm dir}(B_d\to \pi^0\pi^0)$ or
${\cal A}_{\rm CP}^{\rm mix}(B_d\to \pi^0\pi^0)$ allows a theoretically
clean determination of $\gamma$.

The large non-factorizable effects found in 
\cite{BFRS-PRL} have been discussed at length in
\cite{BFRS-BIG,BFRS-UPDATE}, and have been confirmed in 
\cite{BPRS,ALP-Bpipi,CGRS,He:2004ck,ML04}. For the following discussion, 
the most important outcome of this analysis are the values of the 
hadronic parameters $d$, $\theta$, $x$ and $\Delta$ in (\ref{dthetaxdelta}).
These quantities  allow us -- with the help of the $SU(3)$ flavour 
symmetry -- to determine the corresponding hadronic parameters of the 
$B\to\pi K$ system.

\boldmath
\subsection{$B\to\pi K$ decays}
\unboldmath
The key observables for our discussion are the following ratios:
\begin{eqnarray}
R&\equiv&\left[\frac{Br(B_d^0\to\pi^- K^+)+
Br(\bar B_d^0\to\pi^+ K^-)}{Br(B^+\to\pi^+ K^0)+
Br(B^-\to\pi^- \bar K^0)}
\right]\frac{\tau_{B^+}}{\tau_{B^0_d}}\label{R-def}\\
R_{\rm c}&\equiv&2\left[\frac{Br(B^+\to\pi^0K^+)+
Br(B^-\to\pi^0K^-)}{Br(B^+\to\pi^+ K^0)+
Br(B^-\to\pi^- \bar K^0)}\right]\label{Rc-def}\\
R_{\rm n}&\equiv&\frac{1}{2}\left[
\frac{Br(B_d^0\to\pi^- K^+)+
Br(\bar B_d^0\to\pi^+ K^-)}{Br(B_d^0\to\pi^0K^0)+
Br(\bar B_d^0\to\pi^0\bar K^0)}\right],\label{Rn-def}
\end{eqnarray}
where the current status of the relevant branching ratios and the
corresponding values of the $R_i$ is summarized in 
Table \ref{tab:BpiK-input1}.
\begin{table}
\begin{center}
\begin{tabular}{|c||c|c|}
\hline
Quantity & Data & Exp. reference
\\ \hline
 \hline
$Br(B_d\to\pi^\mp K^\pm)/10^{-6}$ & $18.2\pm0.8$ & 
\cite{Aubert:2002jb, Chao:2003ue}
\\ \hline
$Br(B^\pm\to\pi^\pm K)/10^{-6}$ & $24.1\pm1.3$ &
\cite{BaBar-BdKK-obs, Chao:2003ue}
\\ \hline
$Br(B^\pm\to\pi^0K^\pm)/10^{-6}$ & $12.1\pm0.8$ & 
\cite{BaBar-Bpi0pi0, Chao:2003ue}
\\ \hline
$Br(B_d\to\pi^0K)/10^{-6}$ & $11.5\pm1.0$ & 
\cite{BaBar-BK0pi0, Chao:2003ue}
\\ \hline
 \hline
$R$ & $0.82\pm0.06$ & $0.91\pm0.07$
\\ \hline
$R_{\rm c}$ & $1.00\pm0.08$ & $1.17\pm0.12$
\\ \hline
$R_{\rm n}$ & $0.79\pm0.08$ & $0.76\pm0.10$
\\ \hline
\end{tabular}
\end{center}
\caption[]{The current status of the CP-averaged $B\to\pi K$ 
branching ratios, with averages taken from \cite{HFAG}. We also give the
values of the ratios $R$, $R_{\rm c}$ and $R_{\rm n}$ introduced
in (\ref{R-def}), (\ref{Rc-def}) and (\ref{Rn-def}), where  
$R$ refers again to $\tau_{B^+}/\tau_{B^0_d}=1.086\pm0.017$ 
\cite{PDG}. In the last column we also give the values of
$R_i$ at the time of the analyses in \cite{BFRS-PRL,BFRS-BIG}.
}\label{tab:BpiK-input1}
\end{table}
The so-called ``$B\to\pi K$ puzzle'', which was already pointed out in
\cite{BF-neutral2}, is reflected in the small value of
$R_{\rm n}$ that is significantly lower than $R_{\rm c}$.
We will return to this below.

In order to analyze this issue, we neglect for simplicity the
colour-suppressed EW penguins and use the $SU(3)$ flavour symmetry 
to find:
\begin{eqnarray}
A(B^0_d\to\pi^-K^+)&=&P'\left[1-re^{i\delta}e^{i\gamma}
\right]\label{Bdpi-K+}\\
A(B^{\pm}_d\to\pi^{\pm} K^0)&=&-P'\label{B+pi-K0}\\
\sqrt{2}A(B^+\to\pi^0K^+)&=&P'\left[1-\left(e^{i\gamma}-qe^{i\phi}\right)
r_{\rm c}e^{i\delta_{\rm c}}\right]\label{B+pi0K+-SM}\\
\sqrt{2}A(B^0_d\to\pi^0K^0)&=&-P'\left[1+\rho_{\rm n}e^{i\theta_{\rm n}}
e^{i\gamma}-qe^{i\phi}r_{\rm c}e^{i\delta_{\rm c}}\right].\label{B0pi0K0-SM}
\end{eqnarray}
Here, $P'$ is a QCD penguin amplitude that does not concern us as it cancels 
in the ratios $R_i$ and in the CP asymmetries considered. The parameters $r$, 
$\delta$, $\rho_{\rm n}$, $\theta_{\rm n}$, $r_{\rm c}$ and $\delta_{\rm c}$ 
are of hadronic origin. If they were considered as completely free, the 
predictive power of (\ref{Bdpi-K+})--(\ref{B0pi0K0-SM}) would be rather low. 
Fortunately, using the $SU(3)$ flavour symmetry, they can be related to the 
parameters $d$, $\theta$, $x$ and $\Delta$ 
in (\ref{dthetaxdelta}).
Explicit expressions for these relations 
can be found in \cite{BFRS-BIG}. In 
this manner, the values of $r$, $\delta$, $\rho_{\rm n}$, $\theta_{\rm n}$, 
$r_{\rm c}$ and $\delta_{\rm c}$ can be found.
The specific numerical values for these parameters are not of particular 
interest here and can be found in \cite{BFRS-UPDATE}. It suffices to say 
that they also signal large non-factorizable hadronic effects. 

The most important recent experimental result concerning the $B\to\pi K$
system is the observation of direct CP violation in $B_d\to\pi^\mp K^\pm$
decays \cite{BaBar-CP-dir-obs,Belle-CP-dir-obs}. This phenomenon is 
described by the following rate asymmetry:
\begin{equation}
{\cal A}_{\rm CP}^{\rm dir}(B_d\to\pi^\mp K^\pm)\equiv
\frac{Br(B^0_d\to\pi^-K^+)-
Br(\bar B^0_d\to\pi^+K^-)}{Br(B^0_d\to\pi^-K^+)+
Br(\bar B^0_d\to\pi^+K^-)}=+0.113\pm 0.019,
\end{equation}
where the numerical value is the average compiled in \cite{HFAG}. Using 
the values of $r$ and $\delta$ as determined above and (\ref{Bdpi-K+}), 
we obtain 
${\cal A}_{\rm CP}^{\rm dir}(B_d\to\pi^\mp K^\pm)=+0.127^{+0.102}_{-0.066}$,
which is in nice accordance with the experimental result.  
Following the lines of \cite{RF-Bpipi,Fl-Ma}, 
we may determine the angle $\gamma$ of the UT by complementing 
the CP-violating $B_d\to\pi^+\pi^-$ observables with either the 
ratio of the CP-averaged branching ratios $Br(B_d\to\pi^\mp K^\pm)$ 
and $Br(B_d\to\pi^+\pi^-)$ or the direct CP-asymmetry 
${\cal A}_{\rm CP}^{\rm dir}(B_d\to\pi^\mp K^\pm)$. 
These avenues, where the latter is theoretically more favourable, 
yield the following results: 
\begin{equation}
\label{gamma}
\left.\gamma\right|_{\rm Br}=
\left(63.3\,^{+7.7}_{-11.1}\right)^\circ, \quad
\left.\gamma\right|_{{\cal A}_{\rm CP}^{\rm dir}}= 
\left(66.6\,^{+11.0}_{-11.1}\right)^\circ, 
\end{equation}
which are nicely consistent with each other. Moreover, these ranges
are in remarkable accordance with the results of Section~\ref{UT-Det}. 
A similar extraction of $\gamma$
can be found in \cite{Wu:2004xx}.

Apart from ${\cal A}_{\rm CP}^{\rm dir}(B_d\to\pi^\mp K^\pm)$,
two observables are left that are only marginally affected by
EW penguins: the ratio $R$ introduced in (\ref{R-def}) and
the direct CP asymmetry of $B^\pm\to\pi^\pm K$ modes. These observables
may be affected by another hadronic parameter $\rho_{\rm c}
e^{i\theta_{\rm c}}$, which is expected to play a minor r\^ole and
was neglected in (\ref{B+pi-K0}) and (\ref{B+pi0K+-SM}). In this
approximation, the direct $B^\pm\to\pi^\pm K$ CP asymmetry vanishes --
in accordance with the experimental value of $+0.020\pm0.034$ -- and 
the new experimental value of $R=0.82\pm0.06$, which is on the lower 
side, can be converted into $\gamma\leq (64.9^{+4.8}_{-4.2})^\circ$ 
with the help of the bound derived in \cite{FM}. On the other hand,
if we use the values of $r$ and $\delta$ as determined above, we
obtain
\begin{equation}\label{R-pred}
R=0.943^{+0.028}_{-0.021},
\end{equation}
which is sizeably larger than the experimental value. The nice agreement 
of the data with our prediction of 
${\cal A}_{\rm CP}^{\rm dir}(B_d\to\pi^\mp K^\pm)$, which 
is independent of $\rho_{\rm c}$, suggests that this parameter is 
actually the origin of the deviation of $R$. In fact, as discussed
in detail in \cite{BFRS-UPDATE}, the emerging signal for 
$B^\pm\to K^\pm K$ decays, which provide direct access to
$\rho_{\rm c}$, shows that our value of $R$ in (\ref{R-pred}) 
is shifted towards the experimental value through this parameter,
thereby essentially resolving this small discrepancy. 
A nice related discussion can be found in \cite{FR-I,FR-II}.

It is important to emphasize that we could accommodate all the 
$B\to\pi\pi$ and $B\to\pi K$ data so far nicely in the SM. Moreover,
as discussed in detail in \cite{BFRS-BIG,BFRS-UPDATE}, there are
also a couple of internal consistency checks of our working 
assumptions, which work very well within the current 
uncertainties.

Let us now turn our attention to (\ref{B+pi0K+-SM}) and (\ref{B0pi0K0-SM}). 
The only variables in these formulae that we did
not discuss so far are the parameters $q$ and $\phi$ that parametrize
the EW penguin sector. The fact that EW penguins cannot be neglected
here is related to the simple fact that a $\pi^0$ meson can be emitted
directly in these colour-allowed EW penguin topologies, while the
corresponding emission with the help of QCD penguins is 
colour-suppressed. In the SM, the parameters $q$ and $\phi$ can be 
determined with the help of the $SU(3)$ flavour symmetry of strong 
interactions \cite{NRBOUND}, yielding
\begin{equation}\label{q-SM}
q=0.69 \times\left[\frac{0.086}{|V_{ub}/V_{cb}|}\right], 
\qquad \phi=0^\circ.
\end{equation}
In this manner, predictions for $R_{\rm c}$ and $R_{\rm n}$ can be made
\cite{BFRS-PRL,BFRS-BIG}, which read as follows \cite{BFRS-UPDATE}: 
\begin{equation}\label{Rn-Rc-pred}
\left.R_{\rm c}\right|_{\rm SM}=1.14 \pm 0.05 ,\,\quad
\left. R_{\rm n}\right|_{\rm SM}=1.11^{+0.04}_{-0.05}\,.
\end{equation}
Comparing with the experimental results in Table~\ref{tab:BpiK-input1},
we observe that  there is only a marginal discrepancy in the
case of $R_{\rm c}$, whereas the value of $R_{\rm n}$ in 
(\ref{Rn-Rc-pred}) is definetely too large. The ``$B\ \to \pi K$'' 
puzzle is seen here in explicit terms.

\begin{figure}
\begin{center}
\includegraphics[width=12cm]{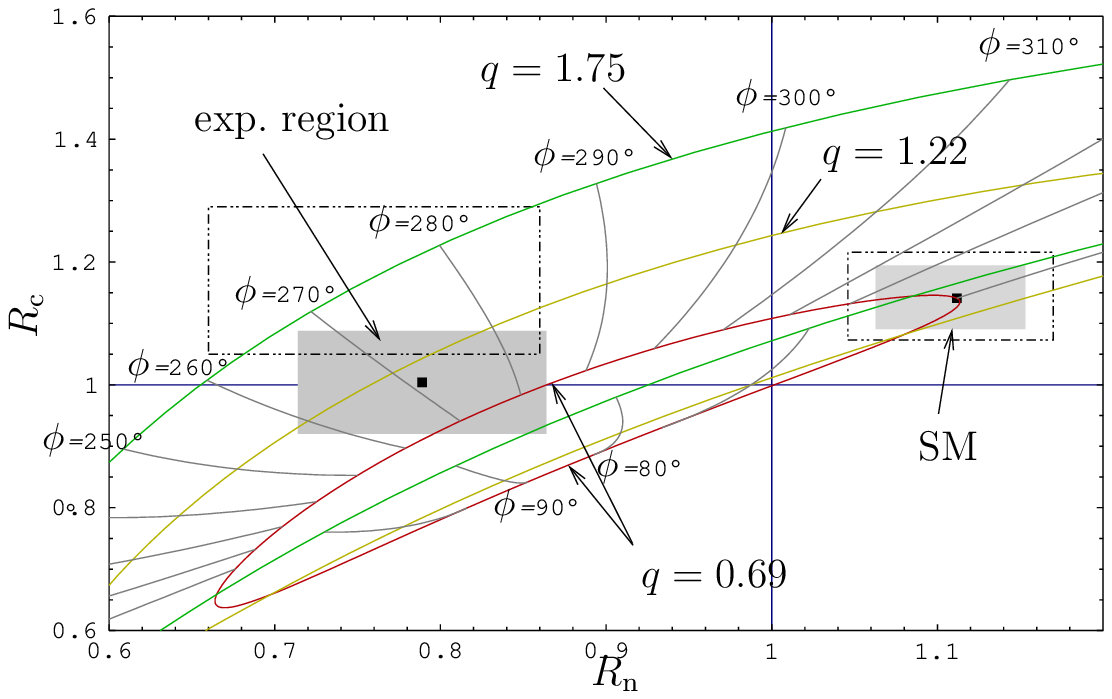}
\end{center}
\caption{The situation in the $R_{\rm n}$--$R_{\rm c}$ plane
\cite{BFRS-UPDATE}.  We show contours
for values of $q=0.69$, $q=1.22$ and $q=1.75$, with
$\phi \in [0^\circ,360^\circ]$. 
The experimental ranges for $R_{\rm c}$ and $R_{\rm n}$ and those predicted 
in the SM are indicated in grey, the dashed lines serve as a reminder of
the corresponding ranges in \cite{BFRS-BIG}.\label{fig:Rn-Rc}}
\end{figure}

The disagreement of the SM with the data can be resolved in the
scenario of enhanced EW penguins carrying a non-vanishing phase
$\phi$. Indeed, using the measured values of $R_{\rm n}$ and 
$R_{\rm c}$, we find \cite{BFRS-UPDATE}:
\begin{equation}\label{q-det}
q = 1.08\,^{+0.81}_{-0.73} , \quad \, \phi =
-(88.8^{+13.7}_{-19.0})^\circ \,. 
\end{equation}
We observe that -- while $q$ is consistent with the SM estimate
within the errors -- the large phase $\phi$ is a spectacular signal
of possible NP contributions. It is useful to consider the 
$R_{\rm n}$--$R_{\rm c}$ plane, as we have done in Fig.~\ref{fig:Rn-Rc}. 
There we show contours corresponding to different values of $q$, and 
indicate the experimental and SM ranges. 

We close this section with a list of predictions for the CP asymmetries
in the $B \to \pi \pi$ and $B\to \pi K$ systems, which are summarized
in Table~\ref{asymtab}.

\begin{table}[hbt]
\begin{center}
\begin{tabular}{|c||c|c|}
\hline
  Quantity &   Our Prediction &  Experiment
 \\ \hline
  ${\cal A}_{\rm CP}^{\rm dir}(B_d\!\to\!\pi^0 \pi^0)$ 
 &   $-0.28^{+0.37}_{-0.21}$ \rule{0em}{1.05em} & $-0.28 \pm 0.39$ \\\hline
  ${\cal A}_{\rm CP}^{\rm mix}(B_d\!\to\!\pi^0 \pi^0)$ 
 &  $-0.63^{+0.45}_{-0.41}$ \rule{0em}{1.05em} &
 $-0.48_{-0.40}^{+0.48}$ \\\hline
${\cal A}_{\rm CP}^{\rm dir}(B_d\!\to\!\pi^{\mp} K^{\pm}) $ 
&   $0.127^{+0.102}_{-0.066}$ \rule{0em}{1.05em} &  $0.113\pm0.019$ \\
\hline
${\cal A}_{\rm CP}^{\rm dir}(B^\pm\!\to\!\pi^0 K^\pm) $ 
&   $0.10^{+0.25}_{-0.19}$ \rule{0em}{1.05em} &  $-0.04 \pm 0.04$ \\ \hline
  ${\cal A}_{\rm CP}^{\rm dir}(B_d\!\to\!\pi^0 K_{\rm S})$ 
&   $0.01^{+0.15}_{-0.18}$ \rule{0em}{1.05em} &
$0.09 \pm 0.14$ \\ \hline 
  ${\cal A}_{\rm CP}^{\rm mix}(B_d\!\to\!\pi^0 K_{\rm S})$ 
&  $-0.98^{+0.04}_{-0.02} $ \rule{0em}{1.05em} &  $-0.34^{+0.29}_{-0.27}$
\\ \hline
\end{tabular}
\caption{Compilation of our predictions for the CP-violating 
$B\to\pi\pi$, $\pi K$ asymmetries.}
\label{asymtab}
\end{center}
\end{table}

\boldmath
\subsection{Implications for Rare $K$ and $B$ Decays}
\unboldmath
The rates for rare $K$ and $B$ decays are sensitive functions
of the EW penguin contributions. 
We have discussed it in detail in Section 8 in the context of
$K\to\pi\nu\bar\nu$ decays.
In a simple scenario in
which NP enters the EW penguin sector predominantly through $Z^0$
penguins and the local operators in the effective Hamiltonians for
rare decays are the same as in the SM\footnote{See
 \cite{Barger:2004hn} for a discussion of the $B \to \pi K$ system in a
 slightly different scenario involving an additional $Z^{'}$ boson.},
there is a strict relation \cite{BFRS-I,BFRS-PRL,BFRS-BIG} between the
EW penguin effects  in the $B\to\pi K$ system and the corresponding
effects in rare  $K$ and $B$ decays. The  $Z^0$-penguin
function $C(x_t)$  of the SM generalizes to $C(v)$ with 
\begin{equation}
C(v)=|C(v)|e^{i\theta_C}=2.35 \bar{q} e^{i\phi}-0.82, \qquad
\bar{q}=q\left|\frac{V_{ub}/V_{cb}}{0.086}\right| \,. 
\end{equation}
In turn, the functions $X$, $Y$, $Z$ that govern rare decays in the
scenarios in question become explicit functions of $q$ and $\phi$:
\begin{eqnarray}\label{Xfunc}
X=|X|e^{i\theta_X}&=&2.35 \bar{q}e^{i\phi}-0.09,
\\
Y=|Y|e^{i\theta_Y}&=&2.35 \bar{q}e^{i\phi}-0.64,
\\ \label{Zfunc}
Z=|Z|e^{i\theta_Z}&=&2.35 \bar{q}e^{i\phi}-0.94.
\end{eqnarray}

If the phase $\phi$ was zero (the case considered in \cite{BFRS-I}), 
the functions $X$, $Y$, $Z$ would remain to be real quantities as in 
the SM and the MFV model but the enhancement of $q$ would imply 
enhancements of 
$X$, $Y$, $Z$ as well. As in the scenario considered $X$, $Y$, $Z$ are 
not independent of one another, it is sufficient to constrain one of 
them from rare decays in order to see whether the enhancement of $q$ 
is consistent with the existing data on rare decays. It turns out that 
the data on inclusive $B\to X_s l^+l^-$ decays 
\cite{Kaneko:2002mr,Abe:2004sg} are presently most
powerful to constrain $X$, $Y$, $Z$, but 
as demonstrated very recently in \cite{MFV05} their impact is considerably 
increased when combined with data on $B\to X_s\gamma$ and $\kpn$. One 
finds then in particular $X_{\rm max}=1.95$ and $Y_{\rm max}=1.43$ to be 
compared with $1.54$ and $0.99$ in the SM. Correspondingly as seen in 
Table~\ref{brMFV}, the enhancements of rare branching ratios within the MFV
scenario of \cite{UUT} by more than a factor of two over the SM expectations
are no longer possible.

The situation changes drastically if $\phi$ is required to be
non-zero, in particular when its value is in the ball park
of $-90^\circ$ as found above. In this case, $X$, $Y$
and $Z$ become complex quantities, as seen in (\ref{Xfunc})--(\ref{Zfunc}), 
with the phases $\theta_i$ in the ball park of $-90^\circ$ 
\cite{BFRS-PRL,BFRS-BIG}:
\be\label{rXrYrZ}
 \theta_X= -(86\pm 12)^\circ, \qquad
 \theta_Y= -(100\pm 12)^\circ, \qquad 
 \theta_Z=-(108\pm 12)^\circ .
\ee
Actually, the data for the $B\to\pi K$ decays used in our first analysis 
\cite{BFRS-BIG} were such that $q=1.75^{+1.27}_{-0.99}$ and 
$\phi=-(85^{+11}_{-14})^\circ$ were required, implying
$|X|\approx|Y|\approx|Z|\approx4.3^{+3.0}_{-2.4}$, barely consistent
with the data. Choosing $|Y|=2.2$ as high as possible but still
consistent with the data on $B\to X_s l^+l^-$ at the time of the analysis 
in \cite{BFRS-BIG}, we found
\begin{equation} 
\label{qwithRD}
\bar{q}=0.92^{+0.07}_{-0.05}\,, \qquad \phi=-(85^{+11}_{-14})^{\circ}\,, 
\end{equation}
 which has been already taken into account in obtaining the values in 
(\ref{rXrYrZ}).
This in turn made us expect that the experimental values
$R_{\rm c}=1.17\pm0.12$ and $R_{\rm n}=0.76\pm0.10$ known at the time of the
analysis in \cite{BFRS-BIG} could change (see Fig.\ \ref{fig:Rn-Rc}) 
once the data improve. Indeed, our expectation \cite{BFRS-BIG}, 
\begin{equation}
R_{\rm c}=1.00^{+0.12}_{-0.08} \qquad R_{\rm n}=0.82^{+0.12}_{-0.11},
\end{equation}
has been confirmed by the most recent results in Table
\ref{tab:BpiK-input1}, 
making the overall description of the $B\to \pi \pi$, $B\to \pi K$ and 
rare decays within our approach significantly better with respect to our 
previous analysis.

The very recent analysis of MFV in \cite{MFV05}, indicating allowed values 
for $Y$ below 2.2, will have some indirect impact also on our analysis that
goes beyond MFV. However, a preliminary analysis indicates that the impact 
on our $B\to\pi K$ results is rather insignificant. In particular the phase 
$\phi$ remains very large. We will return elsewhere to these modifications 
once the data on $B\to\pi\pi$ and $B\to\pi K$ improve.

There is a characteristic pattern of modifications of
branching ratios relative to the case of $\phi=0^\circ$ and 
$\theta_i=0^\circ$:
\begin{itemize}
\item  The formulae for the branching ratios proportional to $|X|^2$, like 
$Br(B\to X_s\nu\bar\nu)$, and  $|Y|^2$,
like $Br(B_{d,s}\to \mu^+\mu^-)$, remain unchanged relative to
the case $\phi=0^\circ$, except that the correlation between $|X|$ and 
$|Y|$ for $\phi\not=0$ differs from the one in the MFV models. 
\item  In CP-conserving transitions in which in addition to
top-like contributions also charm contribution plays some
r\^ole, the {\it constructive} interference between top and charm
contributions in the SM becomes {\it destructive} or very small if 
the new phases $\theta_i$ are large, thereby compensating 
for the enhancements of $X$, $Y$ and $Z$. In particular, 
$Br(K^+\to \pi^+\nu\bar\nu)$ turns out to be rather
close to the SM estimates, and the short-distance part of 
$Br(K_{\rm L}\to \mu^+\mu^-)$ is even smaller than in the SM.
\item Not surprisingly, the most spectacular impact of large
phases $\theta_i$ is seen in CP-violating quantities sensitive to EW
penguins.
\end{itemize}

In particular, one finds \cite{BFRS-PRL,BFRS-BIG}
\be
\frac{Br(\klpn)}{Br(\klpn)_{\rm SM}}=
\left|\frac{X}{X_{\rm SM}}\right|^2
\left[\frac{\sin(\beta-\theta_X)}{\sin(\beta)}\right]^2, 
\ee
with the two factors on the right-hand side in the ball park of $2$ and
$5$, respectively. Consequently, $Br(\klpn)$ can be enhanced over the
SM prediction even by an order of magnitude and is expected
to be roughly by a factor of $4$ larger than 
$Br(\kpn)$. 
We would like to emphasize that this pattern is only moderately affected 
by the results in \cite{MFV05} as the maximal value of $|X|$ used in 
\cite{BFRS-BIG} is only slightly higher than the upper bound found in 
\cite{MFV05}.

\begin{figure}[hbt]
\vspace{0.10in}
\centerline{
\epsfysize=8.0cm
\epsffile{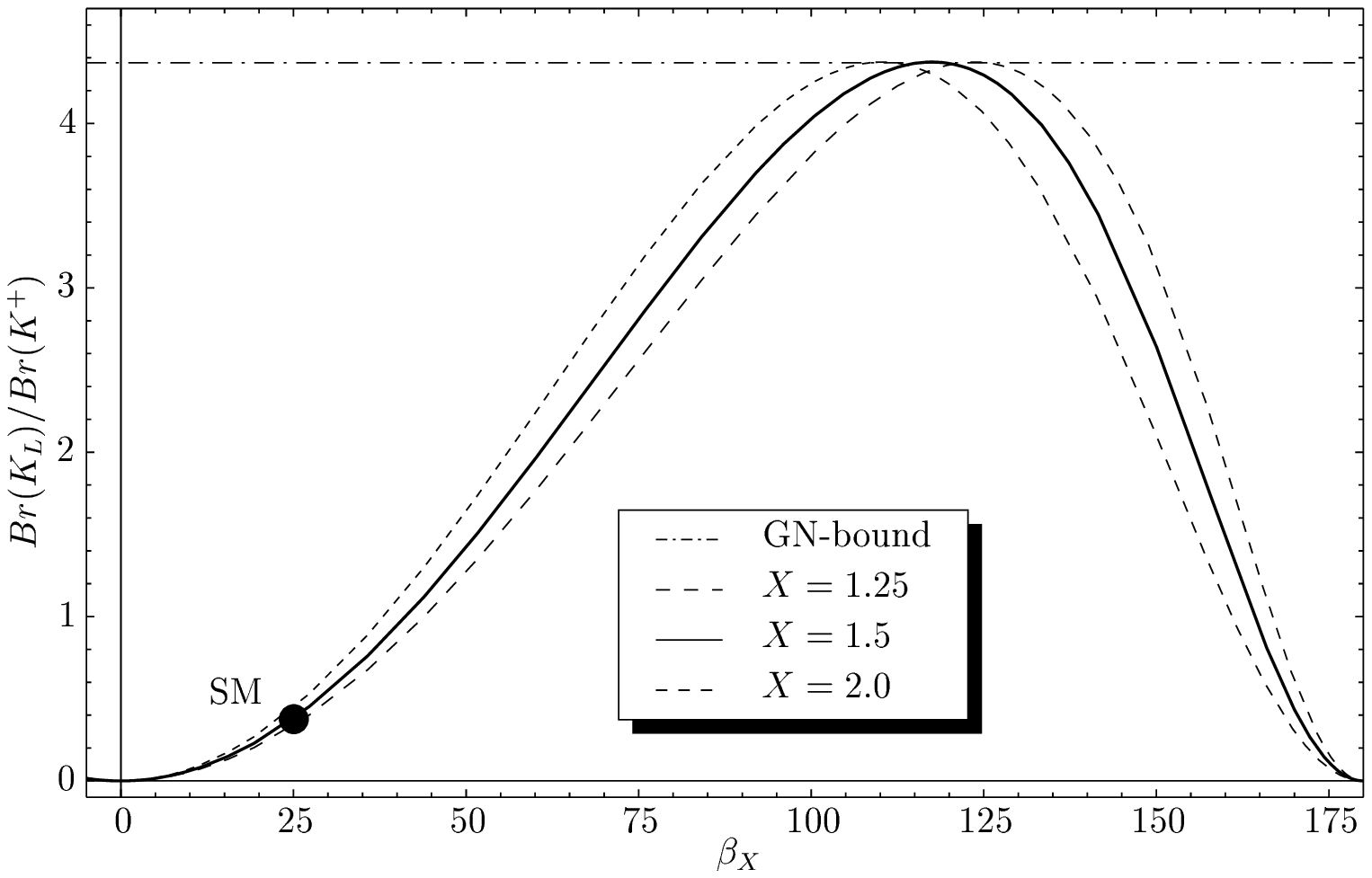}
}
\vspace{0.08in}
\caption{The ratio of $\klpn$ and $\kpn$ branching ratios as a function of 
$\beta_X$ for different calues of $|X|$ \cite{BSU}. The horizontal line 
represents the bound in (\ref{NRBOUND}).
}\label{fig:Brratio}
\end{figure}

In the SM and most MFV models the pattern is totally 
different with
$Br(\klpn)$ smaller than $Br(\kpn)$
by a factor of $2$--$3$ \cite{BSU,BB4,BB96,BF01}. On the other hand a
recent analysis shows that a pattern of $Br(K \to \pi \nu \bar{\nu})$
expected in our NP 
scenario can be obtained in a general MSSM \cite{Buras:2004qb}.
In Fig.~\ref{fig:Brratio} we show
the ratio of $\klpn$ and $\kpn$ branching ratios as a function of 
$\beta_X$ for different values of $|X|$ \cite{BSU}.
It is clear from this plot that  accurate measurements of both branching
ratios will give a precise value of the new phase $\theta_X$ without 
essentially any theoretical uncertainties.

\begin{figure}[htbp]
\begin{center}
\begin{tabular}{cc}
\epsfig{file=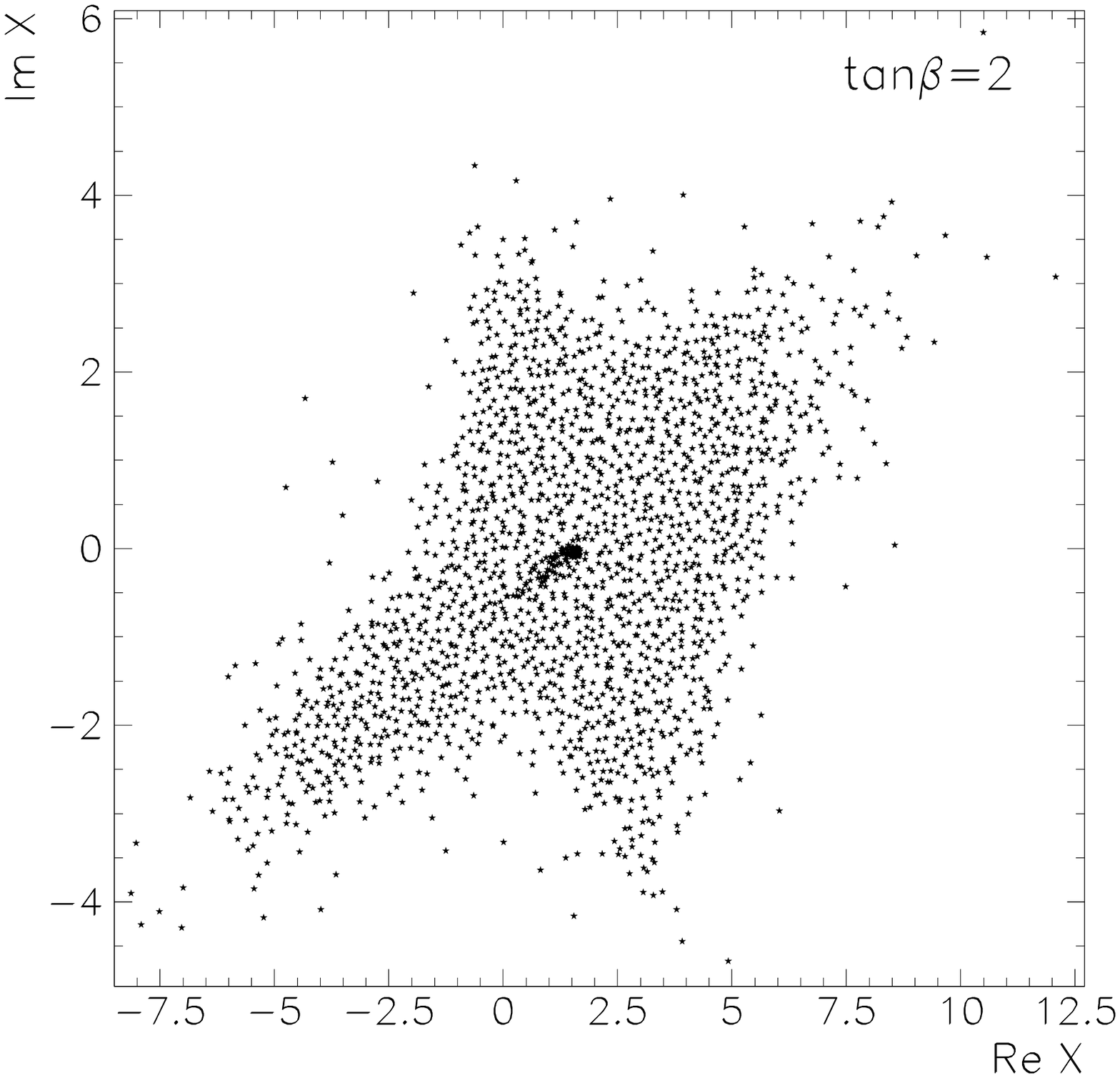,width=0.48\linewidth}
&
\epsfig{file=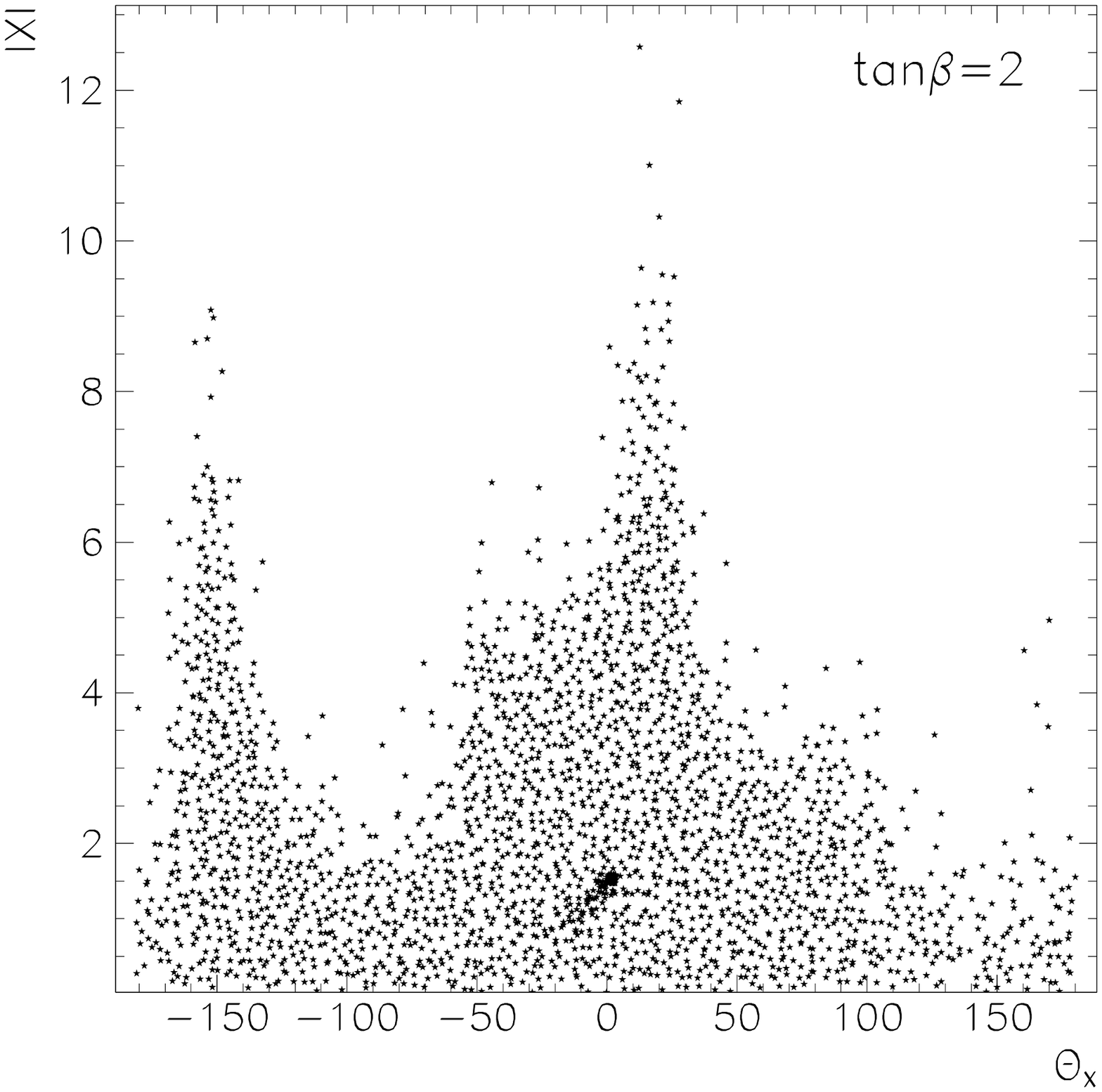,width=0.48\linewidth}
\\
\epsfig{file=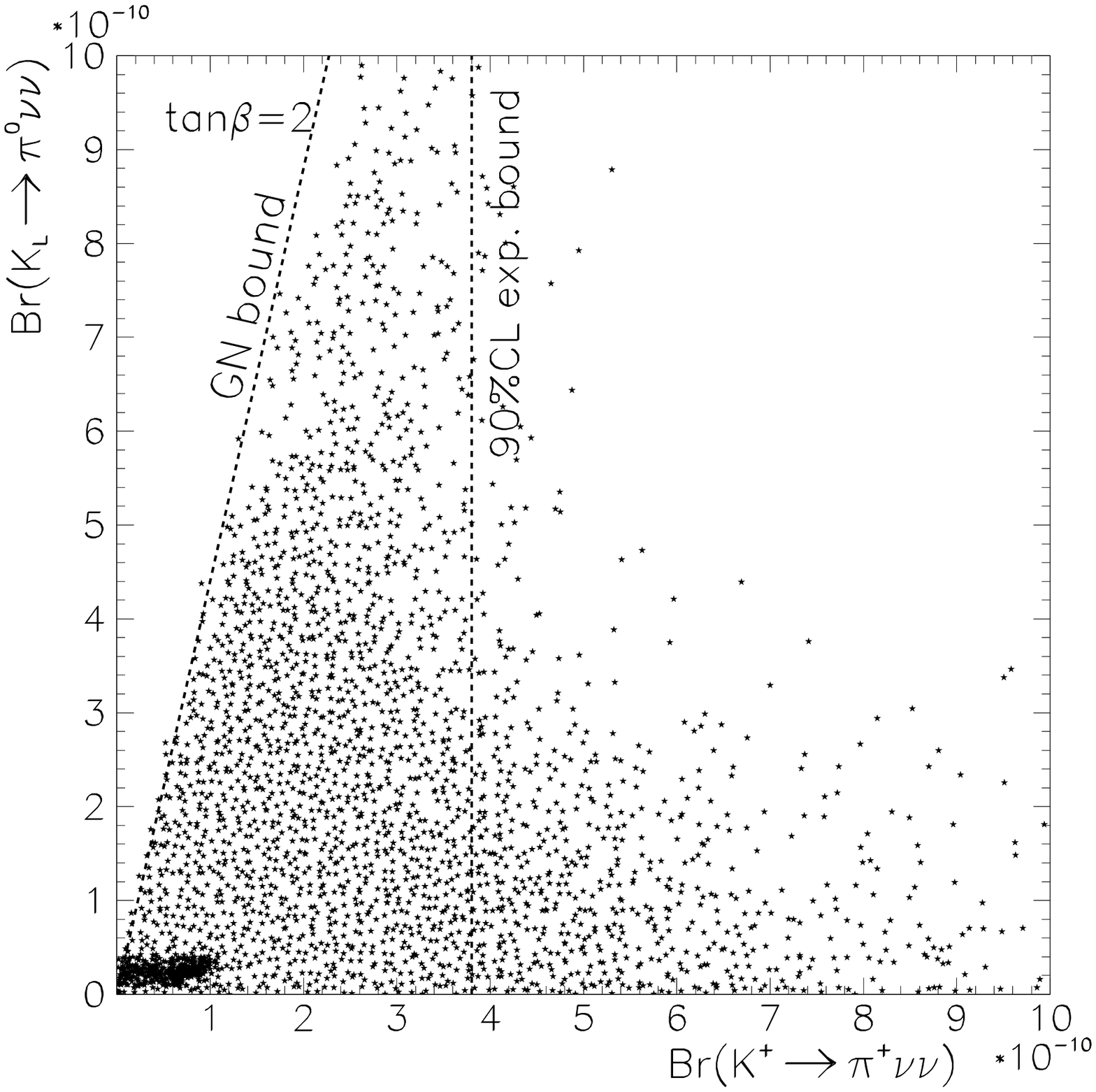,width=0.48\linewidth}
&
\epsfig{file=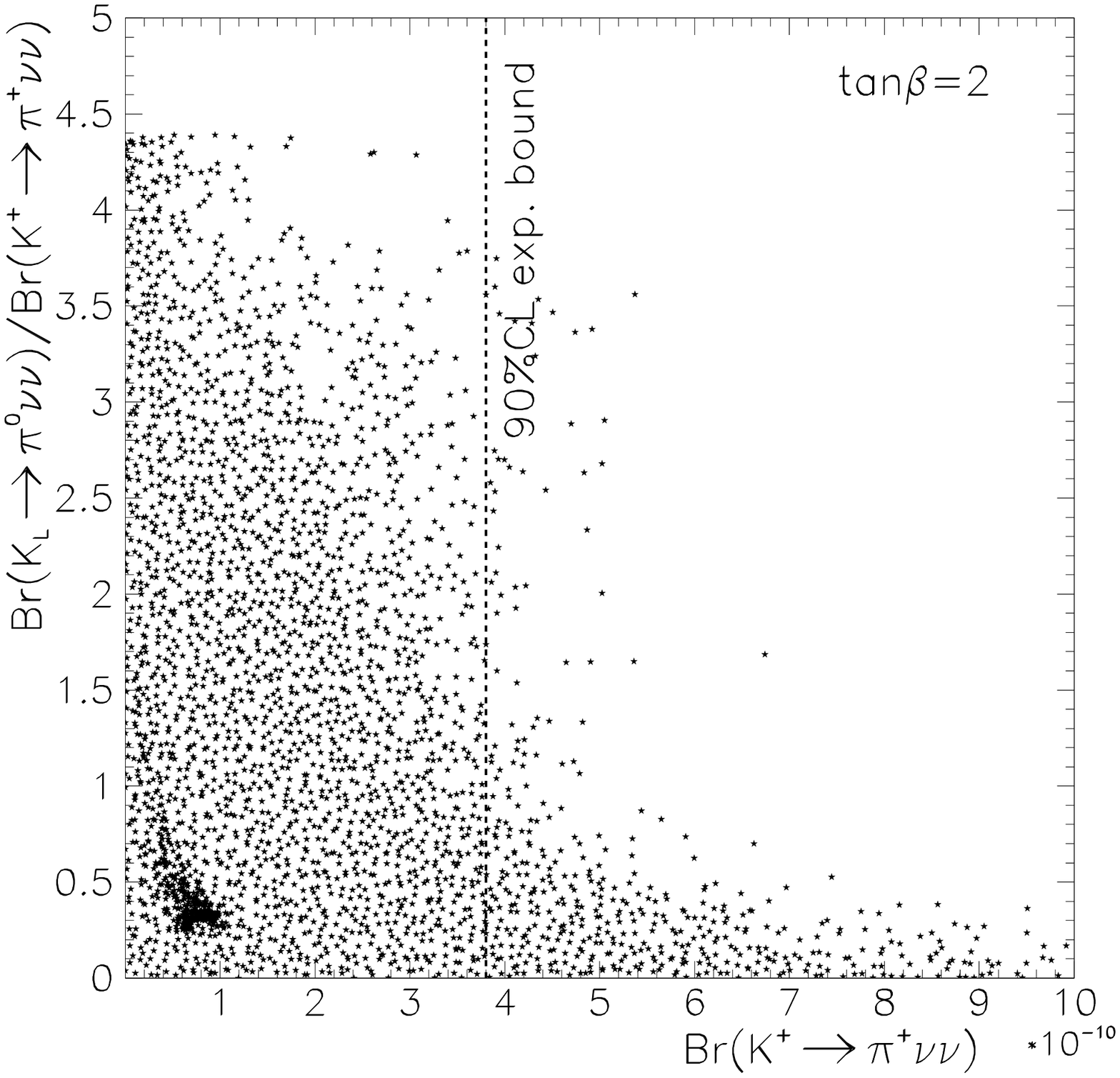,width=0.48\linewidth}
\\
\end{tabular}
\caption{Distributions of $X$, $Br(K_L\to \pi^0 \nu 
\bar \nu)$ and $Br(K^+\to \pi^+ \nu \bar \nu)$ for $\tan\beta=2$ in 
the 63-parameter scan \cite{Buras:2004qb}.}
\label{fig:fxtb2}
\end{center}
\end{figure}

The result in a general MSSM is shown in Fig.~\ref{fig:fxtb2} 
\cite{Buras:2004qb}.
The points in this figure for which both branching ratios are large and 
$Br(\klpn)\ge Br(\kpn)$ correspond to large complex phases in 
the non-diagonal terms $\delta_{LL}^{12}$, 
$\delta_{ULR}^{13}$, $\delta_{ULR}^{23}$
in the squark mass matrix .

\begin{figure}
\vspace*{0.3truecm}
\begin{center}
\includegraphics[width=10cm]{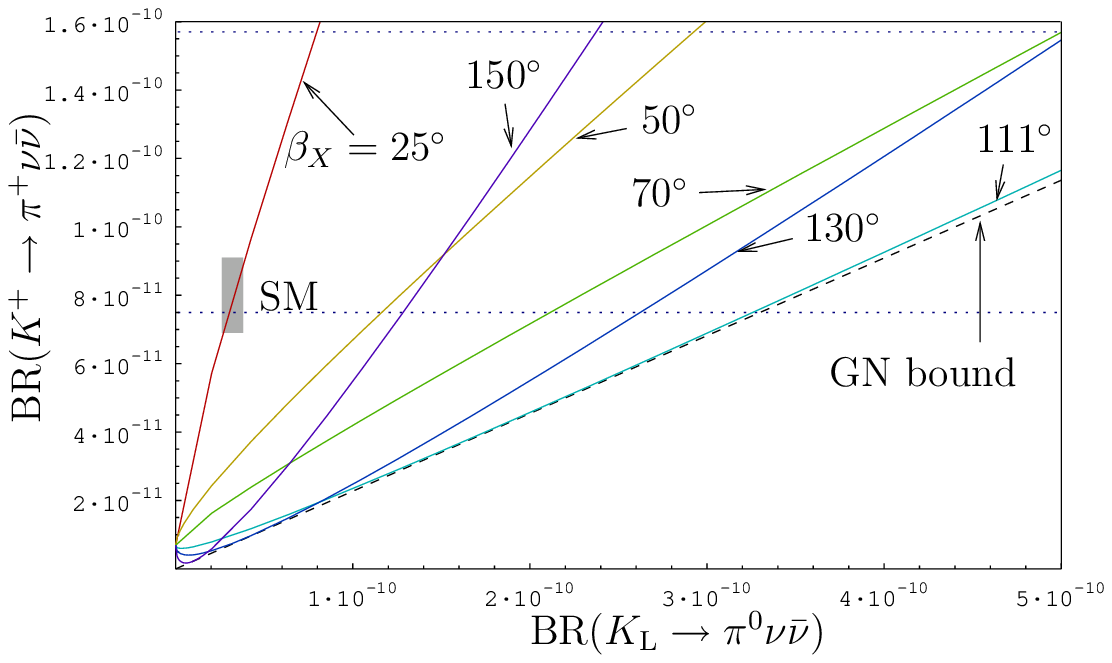}
\end{center}
\caption{${Br}(\kpn)$ as a function of ${Br}(\klpn)$
for various values of $\beta_X=\beta-\theta_X$ \cite{BFRS-BIG}. 
The dotted horizontal 
lines indicate 
the lower part of the experimental range
 and the grey area the SM prediction. We also show the 
bound of \cite{GRNIR}.  \label{fig:KpKl}}
\end{figure}

We note that $Br(\klpn)$ is predicted to be rather close to 
its model-independent
upper bound~\cite{GRNIR} in (\ref{NRBOUND}).
Moreover, another  spectacular 
implication of these findings is a strong violation of the relation 
\cite{BB4,BB96}
\be 
(\sin 2 \beta)_{\pi \nu\bar\nu}=(\sin 2 \beta)_{\psi K_{\rm S}},
\ee 
which is valid in the SM and any model with MFV. Indeed, we find
\cite{BFRS-PRL,BFRS-BIG}
\be
(\sin 2 \beta)_{\pi \nu\bar\nu}=\sin 2(\beta-\theta_X) =
-(0.69^{+0.23}_{-0.41}),
\ee
in striking disagreement with $(\sin 2 \beta)_{\psi K_{\rm S}}= 
0.725\pm0.037$. Fig.~\ref{fig:KpKl} shows what happens when
$\beta_X=\beta-\theta_X$ is varied.

Even if eventually the departures from the SM and MFV pictures
could turn out to
be smaller than estimated here, finding 
$Br(\klpn)$ to be larger than $Br(\kpn)$
would be a clear signal of new complex phases at work. For a more
general discussion of the $K \to \pi \nu \bar{\nu}$ decays beyond the
SM, see \cite{BSU}.

Similarly, as seen in Table~\ref{rare}, interesting enhancements are
found in $K_{\rm L}\to\pi^0 l^+l^-$ \cite{BFRS-BIG,Isidori:2004rb,CSMITH} 
and the forward--backward CP asymmetry 
in $B\to X_sl^+l^-$ as discussed in \cite{BFRS-BIG}.
The impact of the very recent analysis in \cite{MFV05} is to moderately
suppress the enhancements seen in this table. The exception is 
$Br(B_s\to\mu^+\mu^-)$ for which values higher than $1.0 \cdot 10^{-8}$ are
not possible any longer also in our scenario. 
Further implications for rare decays in our scenario can be found in 
\cite{CGC}.

\begin{table}[hbt]
\begin{center}
\begin{tabular}{|c||c|c|c|}
\hline
 Decay & SM prediction &  Our scenario & 
Exp.\ bound \quad(\mbox{90\% {\rm C.L.}}) 
 \\ \hline
$K^+ \to \pi^+ \bar \nu \nu$  & $ (7.8 \pm 1.2) \cdot 10^{-11}$ 
&  $(7.5 \pm 2.1)\cdot 10^{-11}$ & $(14.7^{+13.0}_{-8.9})\cdot 10^{-11} $ 
\cite{Anisimovsky:2004hr} \\ \hline
 $K_{\rm L} \to \pi^0 \bar \nu \nu$   & $ (3.0 \pm 0.6)\cdot 10^{-11}$
&  $ (3.1\pm 1.0)\cdot 10^{-10}$ &  $ < 5.9 \cdot10^{-7}$ 
\cite{KTeV00X} \\ \hline
 $K_{\rm L} \to \pi^0  e^+ e^-$ &  $ (3.7^{+1.1}_{-0.9})\cdot 10^{-11}$  &
 $(9.0\pm 1.6)\cdot 10^{-11}$  &   $<2.8\cdot 10^{-10}$ 
\cite{Alavi-Harati:2003mr} \\ \hline
$K_{\rm L} \to \pi^0  \mu^+ \mu^-$ &  $ (1.5 \pm 0.3)\cdot 10^{-11}$  &
 $(4.3\pm 0.7)\cdot 10^{-11}$  &   $<3.8\cdot 10^{-10}$ 
\cite{Alavi-Harati:2000hs} \\ \hline  
$B \to X_s\bar \nu \nu$  &  $(3.5\pm 0.5)\cdot 10^{-5}$  
&  $ \approx 7\cdot 10^{-5}$ &  $<6.4\cdot 10^{-4}$ 
\cite{Barate:2000rc} \\ \hline
  $B_s \to \mu^+ \mu^-$  &   $(3.42 \pm 0.53)\cdot 10^{-9}$ & 
  $\approx 17\cdot 10^{-9}$ &  $<5.0\cdot 10^{-7}$ 
\cite{Herndon:2004tk,Abazov:2004dj} \\ \hline
\end{tabular}
\end{center}
\caption[]{\small Predictions for various rare decays in the scenario
considered compared with the SM expectations and experimental
bounds. For a theoretical update on $K_{\rm L} \to \pi^0  e^+ e^-$ and 
a discussion of $K_{\rm L} \to \pi^0  \mu^+ \mu^-$, see \cite{Isidori:2004rb}.
\label{rare}}
\end{table}

As emphasized above, the new data on $B\to \pi K$ improved the overall
description of $B\to \pi\pi$, $B\to \pi K$ and rare decays within our
approach.  However, the most interesting question is whether 
the large negative values of $\phi$ and $\theta_i$ will be reinforced 
by the future more accurate data. This would be a very spectacular
signal of NP!

We have also explored the implications for the decay $B_d\to\phi K_{\rm S}$ 
\cite{BFRS-BIG}. Large hadronic uncertainties preclude a precise 
prediction, but assuming that the sign of the cosine 
of a strong phase agrees with factorization, we find that 
$(\sin 2 \beta)_{\phi K_{\rm S}}\ge(\sin 2 \beta)_{\psi K_{\rm S}}$.
This pattern contradicted the first data on 
$(\sin 2 \beta)_{\phi K_{\rm S}}$ in (\ref{gaphi}) but the most recent data 
in (\ref{gaphinew}) are fully consistent with our expectations.
A future confirmation of $(\sin 2 \beta)_{\phi K_{\rm S}}>(\sin 2 \beta)_{\psi
K_{\rm S}}$ could be 
another signal of enhanced CP-violating $Z^0$ penguins with a large weak 
complex phase at work. A very recent analysis of the $B\to\pi K$ decays 
and the correlation with $B\to\phi K_S$ in supersymmetric theories can 
be found in \cite{Khalil:2005qg}.

\subsection{Outlook}
We have presented a strategy for analyzing $B\to \pi \pi $, $B\to \pi K$ 
decays and rare $K$ and $B$ decays. Within a simple NP scenario of enhanced  
CP-violating EW penguins considered here, the NP contributions enter 
significantly only in certain $B\to\pi K$ decays and rare $K$ and $B$ decays, 
while the $B\to\pi\pi$ system is practically 
unaffected by these contributions and can be described within the SM. 
The confrontation of this strategy with the most recent data on 
$B \to \pi \pi$ and $B \to \pi K$ modes from the BaBar and Belle 
collaborations is very encouraging. 
In particular, our earlier predictions for the direct CP asymmetries of 
$B_d \to \pi^0 \pi^0$ and $B_d \to \pi^{\mp} K^{\pm}$ have been confirmed 
within the theoretical and experimental uncertainties, and the shift in the
experimental values of $R_{\rm c}$ and $R_{\rm n}$ took place as
expected. 

It will be exciting to follow the experimental progress on
$B\to\pi\pi$ and $B\to\pi K$ decays and the corresponding efforts in rare
decays. In particular, new messages from BaBar and Belle that 
the present central values of $R_{\rm c}$ and $R_{\rm n}$ have been confirmed 
at a high confidence level, a slight increase of $R$ and a message 
from KEK \cite{JPAR} in the next two
years that the decay $K_{\rm L}\to\pi^0\nu\bar\nu$ has been observed would 
give a strong support to the NP scenario considered here.

\boldmath
\section{Shopping List}\label{SHOP}
\unboldmath
Flavour physics and CP violation have been with us for almost 50 years but
only in the present decade we can expect to be able to test the CKM picture
of flavour and CP violation at a satisfactory level. There is of course 
hope that at a certain level of precision some deviations from the SM 
expectations will be observed, signalling new physics contributions of the
MFV type or even beyond it.

Here is my shopping list for the rest of the decade, that I present here not
necessarily in a chronological order.

\begin{itemize}
\item
There has been an impressive progress on the determination of $\vcb$ and 
$|V_{ub}|$ in the last years but both determinations, in particular the one
of $|V_{ub}|$, will hopefully be improved.
\item
The angle $\beta$ in the UT is already known with a high precision. Yet, it
is important to measure it more accurately as the corresponding corner of the UT
is placed far from its apex, where the main action in the
$(\bar\varrho,\bar\eta)$ takes place. Even more importantly, one should find 
out whether the solution for $\beta$ chosen by the SM is chosen by nature
and whether new phases in the $B^0_d-\bar B^0_d$ mixing could pollute this
measurement. 
\item
An important mile stone in the physics of flavour violation will be 
undoubtedly
the measurement of $\Delta M_s$ that in combination with $\Delta M_d$ will 
offer a rather clean measurement of $\vtd$ and indirectly of the angle
$\gamma$ in the UT.
\item
Another important message will be the measurement of the angle $\gamma$ by
means of the strategies discussed in Section~\ref{STRATEGIES}. 
At Tevatron the U-spin
strategies seem to be promising. Also the $B\to\pi\pi$ and
the $B\to\pi K$ systems offer interesting results. But of course these 
extractions of $\gamma$ will be surpassed one day by very clean measurements 
of this angle in $B\to DK$ decays.
\item
A precise determination of $\gamma$ without hadronic and new physics
uncertainties that is available by means of the tree level $B\to DK$ decays, 
will be an important mile stone in the tests of the KM picture of CP
violation. Combined with hopefully precise value of $R_b$, following from 
$\vub$, it will allow the construction of the reference UT (RUT) \cite{refut}. 
This triangle 
together with $\vus$ and $\vcb$ will give us the true CKM matrix without 
essentially any new physics pollution. First steps in this direction have
been made in \cite{UTfit}.
\item
These results will be confronted one day with the accurate measurements of
$Br(\kpn)$ and $Br(\klpn)$. To this end a number of golden relations will
allow us to test the SM and MFV scenarios. 
We have listed them in Section~\ref{KPNN}.
It should be emphasized that 
although theoretically very clean, the decays $K\to\pi\nu\bar\nu$, in
contrast to tree level decays relevant for the RUT, are sensitive to new
physics contributions. Consequently the determination of the UT by means of
these decays is subject to new physics uncertainties. But precisely the
comparison of the UT from $K\to\pi\nu\bar\nu$ with the RUT will teach us
about  new physics in a clean environment. 
\item
Of course the simultaneous comparison of various determinations of the UT 
with the RUT will allow us not only to discover possible new physics 
contributions but also to identify the type of this new physics. 
\item
An important issue is a better clarification of the
 measurement of 
$(\sin2\beta)_{\phi K_S}$. The confirmation of the significant departure of 
$(\sin2\beta)_{\phi K_S}$ from the already accurate value of 
$(\sin2\beta)_{\psi K_S}$, would be a clear signal of new physics that 
cannot be accomodated within classes A and B.
Also in a particular scenario of class C, discussed in the previous Section, 
one finds 
$(\sin2\beta)_{\phi K_S}>(\sin2\beta)_{\psi K_S}$. 
\item
Also very important are the measurements of $Br(B_{d,s}\to \mu^+\mu^-)$. 
The possible enhancements of $Br(B_{d,s}\to \mu^+\mu^-)$ by 
 factors as high as $100$ in some versions of supersymmetric models, are 
the largest 
enhancements in the field of $K$
and $B$ decays, that are still consistent with all available data. 
A recent review can be found in \cite{Kolda04}.
The correlation of the measurement of $Br(B_{d,s}\to \mu^+\mu^-)$ with the
one of $\Delta M_s$ may teach us something about the nature of new physics
contributions. 
Finding $\Delta M_s$ below $(\Delta M_s)_{\rm SM}$ and $Br(B_{d,s}\to
\mu^+\mu^-)$ well above the SM expectations would be a nice confirmation 
of a SUSY scenario with a large $\tan\beta$ that 
has been considered in \cite{BCRS}.
\item
The improved measurements of several $B\to\pi\pi$ and $B\to\pi K$ observables
are very important in order to see whether the theoretical approaches like 
QCDF \cite{BBNS1}, PQCD \cite{PQCD} and SCET \cite{SCET} in addition to their
nice theoretical structures are also
phenomenologically useful. On the other hand, independently of the outcome of
these measurements, the pure phenomenological strategy
\cite{BFRS-BIG,BFRS-PRL} presented in Section~\ref{NPHASE},
 will be useful in correlating the experimental results for $B\to\pi\pi$ and
$B\to\pi K$ with those for rare $K$ and $B$ decays, $B_d\to\phi K_S$ and 
$\epe$.
\item
Assuming that the future more accurate data on $B\to\pi\pi$ and
$B\to\pi K$ will not modify significantly the presently observed pattern in
these decays, the scenario of enhanced $Z^0$ penguins with a new large
complex weak phase will remain to be an attractive possibility. While the 
enhancement of $Br(\klpn)$ by one order of magnitude would be very welcome to
our experimental colleagues and $(\sin2\beta)_{\pi\nu\bar\nu}<0$
 would be a very spectacular signal of NP, 
even more moderate departures of this sort
 from the SM and the MFV expectations could be easily identified in the very
clean $K\to\pi\nu\bar\nu$ decays as clear signals of NP. 
This is seen in Figs.~\ref{fig:Brratio} and \ref{fig:KpKl}.
\item
The improved measurements of $Br(B\to X_sl^+l^-)$ and $Br(\kpn)$ in the
coming years will efficiently bound the possible enhancements of
$Z^0$ penguins, at least within the scenarios A--C discussed here.
A very recent analysis in \cite{MFV05}, that we briefly summarized 
in Section 9, demonstrates very clearly that the existing constraints 
on $Z^0$-penguins within the MFV scenario are already rather strong.
\item
Also very important is an improved measurement of $Br(B\to X_s\gamma)$ as
well as the removal of its sensitivity to $\mu_c$ in $m_c(\mu_c)$ through a
NNLO calculation. This would increase the precision on the MFV correlation   
between $Br(B\to X_s\gamma)$ and the zero $\hat s_0$ in the forward--backward
asymmetry $A_{\rm FB}(\hat s)$ in $B\to X_s l^+l^-$ \cite{BPSW}. 
A $20\%$ suppression of 
$Br(B\to X_s\gamma)$ with respect to the SM accompanied by a downward shift
of $\hat s_0$ would be an interesting confirmation of the correlation in
question and consistent with the effects of universal extra dimensions with 
a low compactification scale of order few hundred GeV. On the other hand 
finding no zero in $A_{\rm FB}(\hat s)$ would
likely point towards flavour violation beyond the MFV.
\item
Finally, improved bounds and/or measurements of processes not existing or
very strongly suppressed in the SM, like various electric dipole moments and
FCNC transitions in the charm sector will be very important in the search for
new physics. The same applies to $\mu\to e\gamma$ and generally lepton
flavour violation.
\end{itemize}

We could continue this list for ever, in particular in view of the
expected progress at Belle and BaBar, charm physics at Cornell,
experimental program at LHCb and the planned rare $K$ physics 
experiments.
But the upper bound on the maximal number of pages for these lectures has
been saturated which is a clear signal that I should 
conclude here.
The conclusion is not unexpected:
in this decade, it will be very 
exciting to follow the development in this field  and
to monitor the values of various observables provided by our experimental
colleagues by using the strategies presented here and other strategies 
found in the rich literature. 
\vskip1.8truecm

\noindent
{\bf Acknowledgements}\\
\noindent
I would like to thank the organizers for inviting me to such a wonderful 
school and the students for questions and table tennis.
I also thank my collaborators for the time we spent together. 
 Special 
thanks go to Monika Blanke for a careful reading of the manuscript 
and comments.
The discussions with Stefan Recksiegel during the final stages of 
this work are truely appreciated. 
This work has been supported in part by the
Bundesministerium f\"ur
Bildung und Forschung under the contract 05HT4WOA/3 and by the German-Israeli
Foundation under the contract G-698-22.7/2002.


\renewcommand{\baselinestretch}{0.95}

\vfill\eject


\begin{thebibliography}{999}
\bibitem{WARN} 
The numerical value on the r.h.s of (\ref{100}) being proportional to
$\lambda^{-10}$ is very sensitive to $\lambda$ but as $A\propto \lambda^{-2}$
and $P_c(\varepsilon)\propto \lambda^{-4}$ the constraint (\ref{100}) depends
only quadratically on $\lambda$.
\end{thebibliography}
\end{document}